\documentclass[rmp,aps,10pt,twocolumn,superscriptaddress,preprintnumbers]{revtex4-1}
\usepackage{amsmath,amssymb}
\usepackage[version=4]{mhchem}
\usepackage{tablefootnote}
\usepackage{graphicx}
\usepackage{url}
\usepackage{color}
\usepackage{natbib}
\usepackage{textcase}
\usepackage[dvipsnames]{xcolor}
\usepackage[colorlinks=true,breaklinks=true,linktocpage=true]{hyperref}
\hypersetup{allcolors=[rgb]{0.0 0.0 1.0}}

\makeatletter
\def\NAT@sort{\z@}
\makeatother

\usepackage{multirow}

\newcommand{\rprime}{\vec{r}^{\,\prime}}

\begin{document}

\preprint{MPP-2019-205}

\title{Grand Unified Neutrino Spectrum at Earth:\\
Sources and Spectral Components}

\author{Edoardo Vitagliano}
\affiliation{Max-Planck-Institut f\"ur Physik
 (Werner-Heisenberg-Institut), \hbox{F\"ohringer Ring 6, 80805 M\"unchen, Germany}}
\affiliation{\hbox{Department of Physics and Astronomy, University of California},
\hbox{Los Angeles, California, 90095-1547, USA}}

\author{Irene Tamborra}
\affiliation{Niels Bohr International Academy \& DARK,
  \hbox{Niels Bohr Institute, Blegdamsvej 17, 2100 Copenhagen, Denmark}}

\author{Georg Raffelt}
\affiliation{Max-Planck-Institut f\"ur Physik
 (Werner-Heisenberg-Institut), \hbox{F\"ohringer Ring 6, 80805 M\"unchen, Germany}}

\date{October 25, 2019; revised July 4, 2020}

\begin{abstract}
We briefly review the dominant neutrino fluxes at Earth from different
sources and present the {\em Grand Unified Neutrino Spectrum\/}
ranging from meV to PeV energies.  For each energy band and source,
we discuss both theoretical expectations and experimental data. This
compact review should be useful as a brief reference to those
interested in neutrino astronomy, fundamental particle physics,
dark-matter detection, high-energy astrophysics, geophysics, and other
related topics.
\end{abstract}

\maketitle

\renewcommand{\baselinestretch}{0.995}\normalsize
\tableofcontents
\renewcommand{\baselinestretch}{1.0}\normalsize


\section{Introduction}
\label{sec:introduction}

In our epoch of multi-messenger astronomy, the Universe is no longer
explored with electromagnetic radiation alone, but in addition to
cosmic rays, neutrinos and gravitational waves are becoming crucial
astrophysical probes. While the age of gravitational-wave detection
has only begun \cite{Abbott:2016nmj}, neutrino astronomy has
evolved from modest beginnings in the late 1960s with first detections
of atmospheric \cite{Achar:1965ova,Reines:1965qk} and solar neutrinos
\cite{Davis:1968cp} to a main-stream effort. Today, a vast array of
experiments observes the neutrino sky over a large range of
energies~\cite{Koshiba:1992yb, Cribier:2016wmg, Becker:2007sv, Spiering:2012xe,
  Gaisser:2017lkd}.

When observing distant sources, inevitably one also probes the
intervening space and the propagation properties of the radiation,
providing tests of fundamental physics.  Examples include
time-of-flight limits on the masses of photons \cite{Tanabashi:2018,
  Wei:2018pyh, Goldhaber:2008xy}, gravitons
\cite{Goldhaber:2008xy,deRham:2016nuf} and neutrinos
\cite{Tanabashi:2018, Loredo:1988mk, Loredo:2001rx, Beacom:2000ng,
  Nardi:2003pr, Ellis:2012ji, Lu:2014zma}, photon or graviton mixing
with axion-like particles \cite{Tanabashi:2018, Raffelt:1987im,
  Meyer:2013pny, Meyer:2016wrm, Liang:2018mqm, Galanti:2018myb,
  Conlon:2017ofb}, the relative propagation speed of different types
of radiation \cite{Longo:1987ub, Stodolsky:1987vd, Laha:2018hsh,
  Ellis:2018ogq}, tests of Lorentz invariance violation
\cite{Laha:2018hsh, Ellis:2018ogq, Liberati:2009pf, Liberati:2013xla,
  Lang:2017wpe}, or the Shapiro time delay in gravitational potentials
\cite{Longo:1987gc, Krauss:1987me, Pakvasa:1988gd, Desai:2016nqu, Wang:2016lne, Shoemaker:2017nqv,
  Wei:2017nyl, Boran:2018ypz}.

Neutrinos are special in this regard because questions about their
internal properties were on the table immediately after the
first observation of solar neutrinos.  The daring interpretation of
the observed deficit in terms of flavor oscillations
\cite{Gribov:1968kq}, supported by atmospheric neutrino measurements~\cite{Fukuda:1998mi}, eventually proved correct \cite{Aharmim:2009gd, Esteban:2016qun, Capozzi:2018ubv, deSalas:2017kay}.  Today this effect is a standard
ingredient to interpret neutrino measurements from practically any
source.  While some parameters of the neutrino mixing matrix remain to
be settled (the mass ordering and the CP-violating phase), it is
probably fair to say that in neutrino astronomy today the focus is on
the sources and less on properties of the radiation. Of course, there
is always room for surprises and new discoveries.

One major exception to this development is the cosmic neutrino
background (CNB) that has never been directly detected and where the
question of the absolute neutrino mass scale, and the Dirac
vs.\ Majorana question, is the main unresolved issue. Here neutrinos
are a hybrid between radiation and dark matter. If neutrinos were
massless, the CNB today would be blackbody radiation with
$T_\nu=1.95~{\rm K}=0.168~{\rm meV}$, whereas the minimal neutrino
mass spectrum implied by flavor oscillations is $m_1=0$, $m_2=8.6$,
and $m_3=50$~meV, but all masses could be larger in the form of a
degenerate spectrum and the ordering could be inverted in the form
$m_3<m_1<m_2$.  One may actually question if future CNB measurements
are part of traditional neutrino astronomy or the first case of
dark-matter astronomy.

\begin{figure*}
\includegraphics[width=0.69\textwidth]{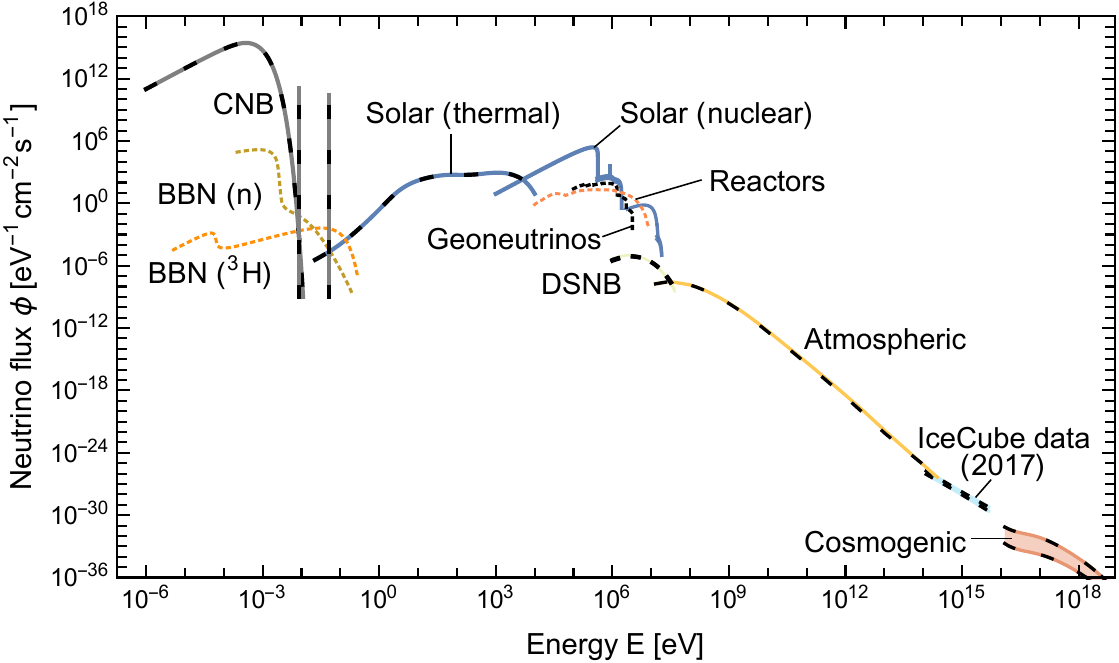}
\vskip6pt
\includegraphics[width=0.69\textwidth]{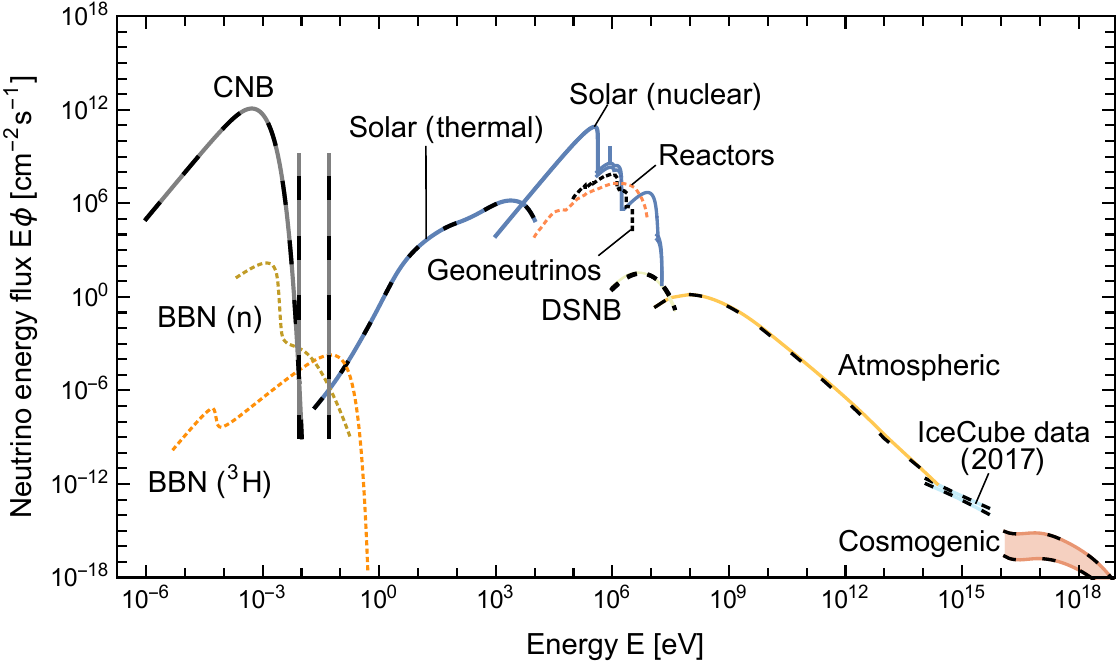}
\vskip-4pt
\caption{Grand Unified Neutrino Spectrum (GUNS) at Earth, integrated over directions and
summed over flavors. Therefore, flavor conversion between source and detector does not affect this plot.
Solid lines are for neutrinos, dashed or dotted lines for
antineutrinos, superimposed dashed and solid lines for sources of both
$\nu$ and $\overline\nu$. The fluxes from BBN, the Earth, and reactors
encompass only antineutrinos, the Sun emits only neutrinos, whereas
all other components include both. The CNB is shown for
a minimal mass spectrum of $m_1=0$, $m_2=8.6$, and $m_3=50$~meV,
producing a blackbody spectrum plus two monochromatic lines
of nonrelativistic neutrinos with energies corresponding to
$m_2$ and~$m_3$. See Appendix~\ref{sec:GUNS} for an exact description of the individual curves.
{\it Top panel:} Neutrino flux $\phi$ as a function of energy; line sources in units of ${\rm cm}^{-2}~{\rm s}^{-1}$. {\it Bottom panel:} Neutrino energy flux $E \times \phi$ as a function of energy; line sources in units of
${\rm eV}~{\rm cm}^{-2}~{\rm s}^{-1}$.}
\label{fig:GUNS0}
\vskip-4pt
\end{figure*}

The large range of energies and the very different types of sources
and detectors makes it difficult to stay abreast of the developments
in the entire field of neutrino astronomy. One first entry
to the subject is afforded by a graphical representation and brief explanation
of what we call the
{\em Grand Unified Neutrino Spectrum}\footnote{We borrow this terminology
  from the seminal {\em Grand Unified Photon Spectrum}
  of \textcite{Ressell:1989rz}.}
(GUNS), a single plot of the neutrino and antineutrino
background at Earth from the CNB in the meV range
to the highest-energy cosmic neutrinos at
PeV ($10^{15}~{\rm eV}$) energies \cite{Koshiba:1992yb, Haxton:2000xb,Cribier:2016wmg, Becker:2007sv, Spiering:2012xe,
  Gaisser:2017lkd}.
As our main result we here produce an updated version
of the GUNS plots shown in Fig.~\ref{fig:GUNS0}. The
top panel shows the
neutrino flux $\phi$ as a function of energy, while the energy flux $E \times \phi$ is
shown in the bottom~panel.

Our initial motivation for this task came from the low-energy part
which traditionally shows a gap between solar neutrinos and the CNB,
the latter usually depicted as blackbody radiation.  However, the
seemingly simple task of showing a new component ---
the keV thermal neutrino flux from the Sun
and the neutrinos from $\beta$ decays of primordial elements
--- in the context of the full
GUNS quickly turned into a much bigger project because one is forced to
think about \hbox{all components}.

While our brief review can be minimally thought of as an updated and
annotated version of the traditional GUNS plot, ideally it serves as a
compact resource for students and researchers to get a first sense in
particular of those parts of the spectrum where they are no immediate
experts. One model for our work could be the format of the
mini reviews provided in the
{\em Review of Particle Physics\/} \cite{Tanabashi:2018}.
In addition, we provide the input of what
exactly went on the plot in the form of tables or analytic formulas
(see Appendix~\ref{sec:GUNS}).

Astrophysical and terrestrial neutrino fluxes can be modified by any number of nonstandard effects, including mixing with hypothetical sterile neutrinos~\cite{Mention:2011rk, Abazajian:2012ys}, large nonstandard interactions~\cite{Davidson:2003ha, Antusch:2008tz, Biggio:2009nt, Ohlsson:2012kf}, spin-flavor oscillations by large nonstandard magnetic dipole moments~\cite{Raffelt:1990pj, Haft:1993jt, Giunti:2014ixa}, decay and annihilation into majoron-like bosons \cite{Schechter:1981cv, Gelmini:1983ea, Beacom:2002vi, Beacom:2002cb, Denton:2018aml, Funcke:2019grs, Pakvasa:2012db, Pagliaroli:2015rca, Bustamante:2016ciw}, for the CNB large primordial asymmetries and other novel early-universe phenomena~\cite{Pastor:2008ti, Arteaga:2017zxg}, or entirely new sources such as dark-matter decay~\cite{Barger:2001ur, Halzen:2010yj, Fan:2013faa, Feldstein:2013kka, Agashe:2014yua, Rott:2014kfa, Kopp:2015bfa, Boucenna:2015tra, Chianese:2016opp, Cohen:2016uyg, Chianese:2019kyl, Esmaili:2013gha, Bhattacharya:2014vwa, Higaki:2014dwa, Fong:2014bsa, Murase:2015gea} and annihilation in the Sun or Earth~\cite{Srednicki:1986vj, Silk:1985ax, Ritz:1987mh, Kamionkowski:1991nj, Cirelli:2005gh}.  We will usually not explore such topics and rather stay in a minimal framework which of course includes normal flavor conversion.

In the main text we walk the reader through the GUNS plots of Fig.~\ref{fig:GUNS0} and briefly review the different components approximately in increasing order of energy.  In Sec.~\ref{sec:CNB} we begin with the CNB, discussing primarily the impact of neutrino masses. In
Fig.~\ref{fig:GUNS0} we show a minimal example where the smallest neutrino mass vanishes, providing the traditional blackbody radiation, and two mass components which are nonrelativistic today.

In Sec.~\ref{sec:BBN} we turn to neutrinos from the big-bang
nucleosynthesis (BBN) epoch that form a small
but dominant contribution at energies just above the CNB.
This very recently recognized flux derives from neutron and triton
decays, $n\to p+e^-+\overline\nu_e$ and $^3{\rm H}\to{}^3{\rm He}+e^-+\overline\nu_e$,
that are left over from BBN.

In  Sec.~\ref{sec:Solar-Nuclear} we turn to the Sun,
which is especially bright in neutrinos because of its proximity,
beginning with the traditional
MeV-range neutrinos from nuclear reactions that produce
only~$\nu_e$. We continue in Sec.~\ref{sec:Thermal-Solar}
with a new contribution in the keV range of
thermally produced fluxes that are equal for $\nu$ and $\overline\nu$.
In both cases, what exactly arrives at Earth depends
on flavor conversion, and for MeV energies also whether the Sun is observed
through the Earth or directly (day-night effect).

Nuclear fusion in the Sun produces only $\nu_e$, implying that the
MeV-range $\overline\nu_e$ fluxes, of course also modified by oscillations,
are of terrestrial origin from nuclear fission.  In Sec.~\ref{sec:Geo}
we consider geoneutrinos that predominantly come from natural
radioactive decays of potassium, uranium and thorium.  In
Sec.~\ref{sec:Reactors} we turn to nuclear power reactors. Both fluxes
strongly depend on location so that their contributions to the GUNS
are not universal.

In Sec.~\ref{sec:snnu} we turn to the 1--100~MeV range where neutrinos
from the next nearby stellar collapse, which could be an exploding or
failed supernova, is one of the most exciting if rare targets.
However, some of the most interesting information is in the detailed
time profile of these few-second bursts.  Moreover, the range of
expected distances is large and the signal depends on the viewing
angle of these very asymmetric events.  Therefore, such sources fit
poorly on the GUNS and are not shown in Fig.~\ref{fig:GUNS0}.  On the
other hand, the diffuse supernova neutrino background (DSNB) from all
past collapsing stellar cores in the Universe dominates in the 10--50
MeV range (Sec.~\ref{sec:DSNB}). If the CNB is all hot dark matter,
the DSNB is actually the largest neutrino radiation component in the
Universe. It may soon be detected by the upcoming JUNO and
gadolinium-enhanced Super-Kamiokande experiments, opening a completely
new frontier.

Beyond the DSNB begins the realm of high-energy neutrinos.  Up to
about $10^{14}~{\rm eV}$ atmospheric neutrinos rule supreme
(Sec.~\ref{sec:ATM}). Historically they were the first ``natural''
neutrinos to be observed in the 1960s as mentioned earlier, and the
observed up-down asymmetry by the Super-Kamiokande detector led to the
first incontrovertible evidence for flavor conversion in
1998. Today, atmospheric neutrinos are still being used for
oscillation physics. Otherwise they are mainly a background to
astrophysical sources in this energy range.

In Sec.~\ref{sec:HE} we turn to the range beyond atmospheric
neutrinos.  Since 2013, the IceCube observatory at the South Pole has
reported detections of more than 100 high-energy cosmic neutrinos with
energies $10^{14}$--$10^{16}~{\rm eV}$, an achievement that marks the beginning of galactic and extra-galactic neutrino astronomy. The sources of this apparently diffuse flux remain uncertain.
At yet larger energies, a diffuse ``cosmogenic neutrino flux''
may exist as a result of possible cosmic-ray interactions at extremely high energies.

We conclude in Sec.~\ref{sec:Conclusions} with a brief summary and discussion
of our results. We also speculate about possible developments
in the foreseeable future.

\section{Cosmic Neutrino Background}
\label{sec:CNB}

The cosmic neutrino background (CNB), a relic from the early universe
when it was about 1~sec old, consists today of about $112~{\rm
  cm}^{-3}$ neutrinos plus antineutrinos per flavor. It is the largest
neutrino density at Earth, yet it has never been measured.  If
neutrinos were massless, the CNB would be blackbody radiation at
$T_\nu=1.945~{\rm K}=0.168~{\rm meV}$. However, the mass differences
implied by flavor oscillation data show that at least two mass
eigenstates must be nonrelativistic today, providing a dark-matter
component instead of radiation.  The CNB and its possible detection is
a topic tightly interwoven with the question of the absolute scale of
neutrino masses and their Dirac vs.\ Majorana nature.

\subsection{Standard properties of the CNB}

Cosmic neutrinos \cite{Dolgov:2002wy, Hannestad:2006zg,
  lesgourgues2013neutrino, Lesgourgoues:2017} are a thermal relic from
the hot early universe, in analogy to the cosmic microwave background
(CMB).  At cosmic temperature $T$ above a few MeV, photons, leptons
and nucleons are in thermal equilibrium, so neutrinos follow a
Fermi-Dirac distribution. If the lepton-number asymmetry in neutrinos
is comparable to that in charged leptons or to the primordial baryon
asymmetry, i.e., of the order of $10^{-9}$, their chemical potentials
are negligibly small.

The true origin of primordial particle asymmetries remains unknown,
but one particularly attractive scenario is leptogenesis,
which is directly connected to the origin of neutrino masses
\cite{Fukugita:1986hr,Buchmuller:2005eh,Davidson:2008bu}. There
exist many variations of leptogenesis, but its generic structure suggests
\hbox{sub-eV} neutrino Majorana masses. In this sense, everything that exists
in the universe today may trace its fundamental origin to neutrino
Majorana masses.

Much later in the cosmic evolution, at $T\sim 1~{\rm MeV}$, neutrinos
freeze out in that their interaction rates become slow compared to the
Hubble expansion, but they continue to follow a Fermi-Dirac
distribution at a common $T$ because, for essentially massless
neutrinos, the distribution is kinematically cooled by cosmic
expansion. Around $T\sim 0.1~{\rm MeV}$, electrons and positrons
disappear, heating photons relative to neutrinos. In the adiabatic
limit, one finds that afterwards $T_\nu=(4/11)^{1/3}\,T_\gamma$. Based
on the present-day value $T_{\rm CMB}=2.725~{\rm K}$ one finds
$T_\nu=1.945~{\rm K}$ today.

The radiation density after $e^+e^-$ disappearance is provided by
photons and neutrinos and, before the latter become nonrelativistic, is usually expressed~as
\begin{equation}
  \rho_{\rm rad}=\left[1+N_{\rm eff}\,\frac{7}{8}\,
    \left(\frac{4}{11}\right)^{4/3}\right]\rho_{\gamma}\,,
\end{equation}
where $N_{\rm eff}$, the effective number of thermally excited
neutrino degrees of freedom, is a way to parameterize $\rho_{\rm
  rad}$. The standard value is $N_{\rm eff}=3.045$
\cite{deSalas:2016ztq}, where the deviation from $3$ arises from
residual neutrino heating by $e^+e^-$ annihilation and other small
corrections.  Both big-bang
nucleosynthesis and cosmological data, notably of the CMB angular
power spectrum measured by Planck, confirm $N_{\rm eff}$ within $\sim
10\%$ errors \cite{Cyburt:2015mya, Ade:2015xua, Aghanim:2018eyx,
  Lesgourgoues:2017}.

While leptogenesis in the early universe is directly connected to
the origin of neutrino masses, they play no role in the
subsequent cosmic evolution. In particular, \hbox{sub-eV} masses are too small
for helicity-changing collisions to have any practical
effect. If neutrino masses are of Majorana type and thus
violate lepton number, any primordial asymmetry would remain
conserved, i.e., helicity plays the role of lepton number and
allows for a chemical potential.  In the Dirac case, the
same reasoning implies that the sterile partners will not be thermally
excited. Therefore, the standard CNB will be the same for both types
of neutrino masses~\cite{Long:2014zva, Balantekin:2018azf}.

Leptogenesis is not proven and one may speculate about large
primordial neutrino-antineutrino asymmetries in one or all flavors. In
this case flavor oscillations would essentially equilibrate the
neutrino distributions before or around thermal freeze-out at $T\sim
1~{\rm MeV}$ so that, in particular, the $\nu_e$ chemical potential
would be representative of that for any flavor
\cite{Dolgov:2002ab,Castorina:2012md}.  It is strongly constrained by
big-bang nucleosynthesis and its impact on $\beta$ equilibrium through
reactions of the type $p+e^-\leftrightarrow n+\nu_e$. Moreover, a
large neutrino asymmetry would enhance $N_{\rm eff}$. Overall, a
neutrino chemical potential, common to all flavors, is
constrained by $|\mu_\nu/T|\alt 0.1$
\cite{Castorina:2012md,Oldengott:2017tzj}, allowing at most for a
modest modification of the radiation density in the CNB.

\subsection{Neutrinos as hot dark matter}
\label{sec:NuMasses}

Flavor oscillation data reveal the squared-mass differences discussed
in Appendix~\ref{sec:MassMatrix}. They imply a minimal neutrino mass
spectrum
\begin{equation}\label{eq:mass-spectrum}
  m_1=0,
  \quad
  m_2=8.6~{\rm meV},
  \quad
  m_3=50~{\rm meV}\,,
\end{equation}
that we will use as our reference case for plotting the GUNS.  While
normal mass ordering is favored by global fits, it could also be
inverted ($m_3<m_1<m_2$) and there could be a common offset from
zero. The value of the smallest neutrino mass remains a key open question.

In view of $T_\nu=0.168~{\rm meV}$ for massless neutrinos, at
least two mass eigenstates are dark matter today. Indeed, cosmological
data provide restrictive limits on the hot dark matter fraction,
implying 95\% C.L.\ limits on $\sum m_\nu$ in the range
0.11--$0.68~{\rm eV}$, depending on the used data sets and
cosmological model \cite{Ade:2015xua, Aghanim:2018eyx,
  Lesgourgoues:2017}.  Near-future surveys should be sensitive enough
to actually provide a lower limit \cite{Lesgourgoues:2017,Brinckmann:2018owf},
i.e., a neutrino-mass detection perhaps even on the level of the minimal mass
spectrum of Eq.~(\ref{eq:mass-spectrum}).

Ongoing laboratory searches for neutrino masses include, in
particular, the KATRIN experiment \cite{Arenz:2018kma,Aker:2019uuj} to measure the
electron endpoint spectrum in tritium $\beta$ decay. The
neutrino-mass sensitivity reaches to about 0.2~eV for the common mass
scale, i.e., a detection would imply a significant tension with
cosmological limits and thus point to a nonstandard CNB or other
issues with standard cosmology. In the future, Project~8, an experiment based on cyclotron radiation emission spectroscopy, could reach a sensitivity down to 40~meV~\cite{Esfahani:2017dmu}.
\begin{figure}[!b]
\includegraphics[scale=0.56]{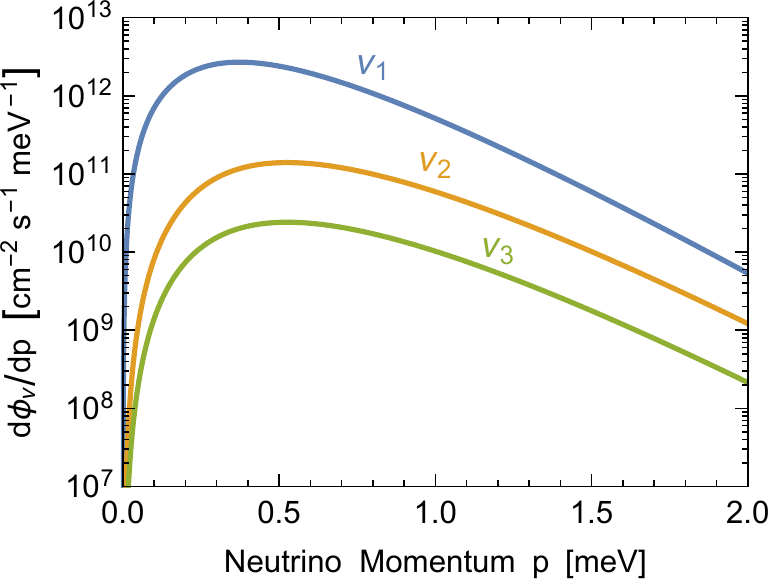}
\vskip-6pt
\caption{Isotropic $\nu$ or $\overline\nu$ differential flux today, $d\Phi_{\nu}/dp$,
  for neutrinos with mass as given in Eq.~(\ref{eq:p-flux}).
  The different curves correspond to our reference mass spectrum
  of Eq.~(\ref{eq:mass-spectrum}).
\label{fig:nucosmo}}
\end{figure}

\begin{figure*}
\hbox to\textwidth{\includegraphics[scale=0.46]{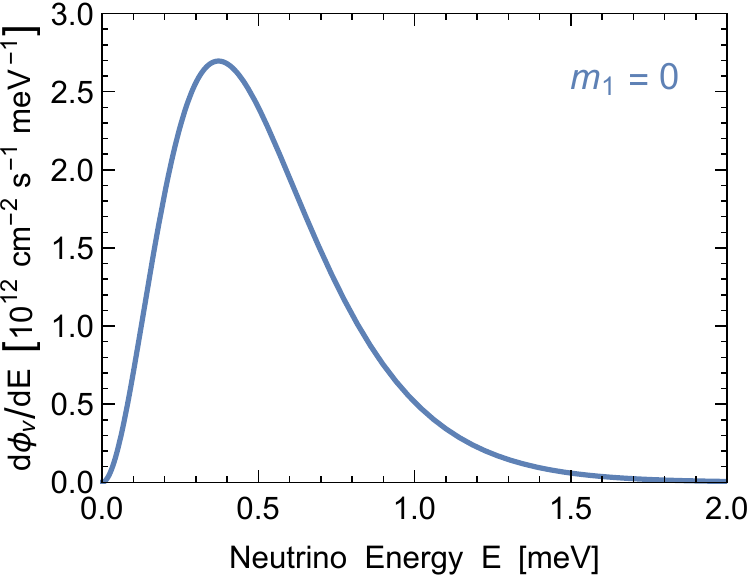}\hfil
  \includegraphics[scale=0.46]{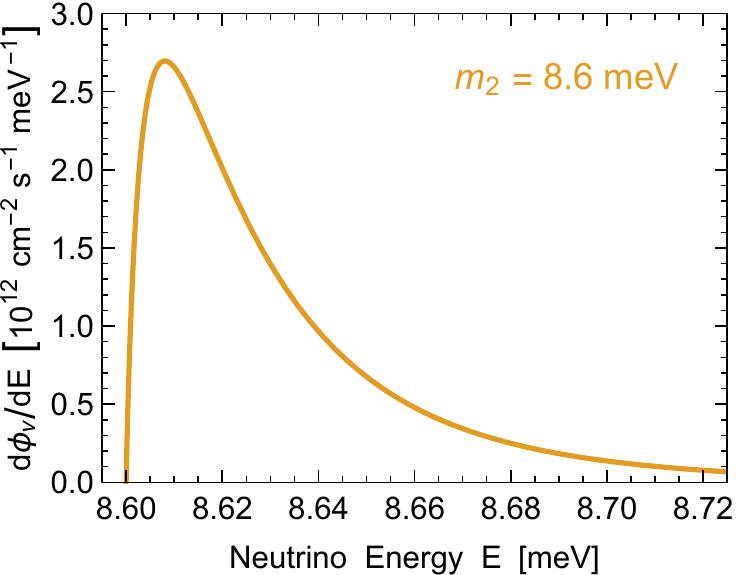}\hfil
  \includegraphics[scale=0.46]{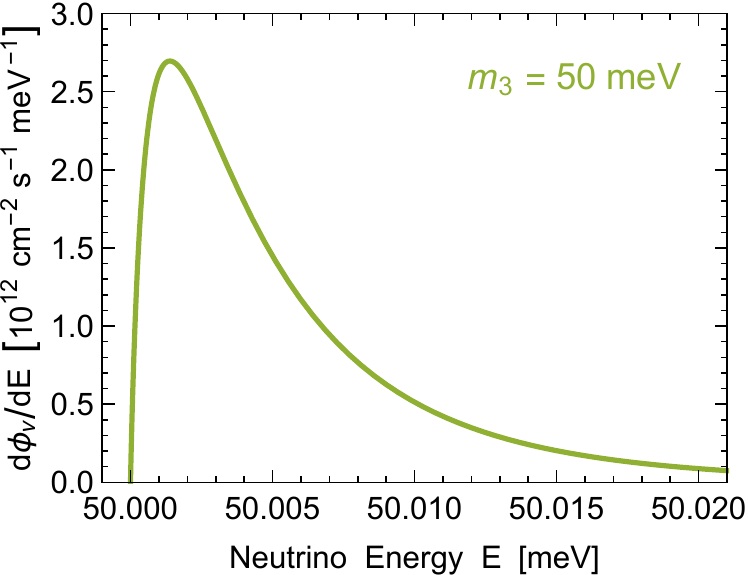}}
\caption{Neutrino differential flux $d\Phi_{\nu}/dE$ according to Eq.~(\ref{eq:E-flux}) for our reference
 mass spectrum of Eq.~(\ref{eq:mass-spectrum}). The maximum flux} does not depend on $m_{\nu}$ and is
 $2.70\times10^{12}~{\rm cm}^{-2}~{\rm s}^{-1}~{\rm meV}^{-1}$.
 \label{fig:enecosmo}
\vskip6pt
\end{figure*}

\subsection{Spectrum at Earth}
\label{sec:SpectrumEarth}

Which neutrino spectrum would be expected at Earth and should be shown
on the GUNS plot?  For neutrinos with mass, not the energy
but the momentum is redshifted by cosmic expansion,
so the phase-space occupation at redshift $z$
for free-streaming neutrinos is
\begin{equation}
	f_{\nu}(p) = \frac{1}{e^{p/T_z} + 1}\,,
\end{equation}
where $T_z=T_\nu (1+z)$ and $T_\nu=1.945~{\rm K}$ is today's
temperature of hypothetical massless neutrinos. The present-day number
density for one species of $\nu$ or $\overline\nu$, differential relative
to momentum, is therefore
\begin{equation}\label{eq:phi-p}
\frac{dn_\nu}{dp} = \frac{1}{2\pi^2}\,\frac{p^2}{e^{p/T_\nu} + 1}\,.
\end{equation}
Integration provides $n_\nu=56~{\rm cm}^{-3}$ as mentioned earlier.

Expressed as an isotropic flux, perhaps for a detection
experiment, requires the velocity $p/E$ with
$E=\sqrt{p^2+m_{i}^2}$, where $m_{i}$ is one of the mass
eigenstates $i=1$, 2 or 3.
So the isotropic differential flux today is
\begin{equation}\label{eq:p-flux}
\frac{d\Phi_{\nu}}{dp}=\frac{p}{E}\frac{dn_\nu}{dp}
= \frac{1}{2\pi^2}\,\frac{p^3}{\sqrt{p^2+m_{i}^2}}\,\frac{1}{e^{p/T_\nu} + 1}\,.
\end{equation}
In Fig.~\ref{fig:nucosmo} we show this flux for our reference mass
spectrum given in Eq.~(\ref{eq:mass-spectrum}).

On the other hand, for plotting the GUNS, the spectrum in terms of energy
is more useful. In this case we need to include a Jacobian
$dp/dE=E/p$ that cancels the velocity factor so that
\begin{equation}\label{eq:E-flux}
\frac{d\Phi_{\nu}}{dE}=\frac{p}{E}\frac{dn_\nu}{dE}
= \frac{1}{2\pi^2}\,
\frac{E^2-m_{i}^2}{e^{\sqrt{E^2-m_{i}^2}\big/T_\nu} + 1}\,.
\end{equation}
The maximum of this function does not depend
on $m_{i}$ and is
$2.70\times10^{12}~{\rm cm}^{-2}~{\rm s}^{-1}~{\rm meV}^{-1}$.
We show the energy spectrum for our reference neutrino masses in
Fig.~\ref{fig:enecosmo} and notice that for larger masses it is
tightly concentrated at $E\agt m_{i}$.  Traditional GUNS plots
\cite{Becker:2007sv, Spiering:2012xe} apply only to massless
neutrinos.

These results ignore that the Earth is located in the gravitational
potential of the Milky Way. Beginning with the momentum distribution
of Eq.~(\ref{eq:phi-p}) we find for the average of the velocity
$v=p/E$
\begin{equation}\label{eq:CNB-velocity}
  \langle v\rangle=\frac{2700\,\zeta_5}{7\pi^4}\,\frac{T}{m}
  +{\cal O}\left(\frac{T}{m}\right)^3
  \approx 4.106\,\,\frac{T}{m}\,.
\end{equation}
For $T=0.168~{\rm meV}$ and $m=50~{\rm meV}$ this is $\langle
v\rangle=1.38\times10^{-2}$, significantly larger than the galactic
virial velocity of about $10^{-3}$.  Therefore, gravitational
clustering is a small effect \cite{Ringwald:2004np, deSalas:2017wtt}
and our momentum and energy distributions remain approximately valid
if neutrino masses are as small as we have assumed.

One CNB mass eigenstate of $\nu_i$ plus $\bar\nu_i$
contributes at Earth a number and energy density of
\begin{eqnarray}
\label{eq:CNB-density}
  n_{\nu\overline\nu} &=& 112~{\rm cm}^{-3}\ ,
  \\[0.8ex]
  \rho_{\nu\overline\nu} &=& \begin{cases} 59.2~{\rm meV}~{\rm cm}^{-3}
    \kern2.3em\quad\hbox{for $m_\nu\ll T_\nu$}\ ,
    \label{eq:CNB-density-b}\\
  112~{\rm meV}~{\rm cm}^{-3}~\frac{m_\nu}{{\rm meV}}
  \kern0.3em\quad\hbox{for $m_\nu\gg T_\nu$}\ ,
  \end{cases}
\end{eqnarray}
ignoring small clustering effects in the galaxy. Here $T_\nu=1.95~{\rm K}=0.168~{\rm meV}$ as explained earlier.

The CNB consists essentially of an equal mixture of all flavors, so the probability for finding a random CNB $\nu$ or $\overline\nu$ in any of the mass eigenstates is equal to 1/3. Put another way, if the neutrino distribution is uniform among flavors and thus their flavor matrix is proportional to the unit matrix, this is true in any basis.

\subsection{Detection perspectives}

Directly measuring the CNB remains extremely challenging \cite{Ringwald:2009bg, Vogel:2015vfa, Li:2017fpz}.  Those ideas based on the electroweak potential on electrons caused by the cosmic
neutrino sea \cite{Stodolsky:1974aq}, an ${\cal O}(G_{\rm F})$ effect,
depend on the net lepton number in neutrinos which today we know
cannot be large as explained earlier and also would be washed out in
the limit of nonrelativistic neutrinos. Early proposals based on the use of the neutrino wind~\cite{Opher:1974drq,Lewis:1979mu} had been found to be not viable, as there is no net acceleration~\cite{Cabibbo:1982bb}.

At ${\cal O}(G_{\rm F}^2)$ one can also consider mechanical forces on
macroscopic bodies by neutrino scattering and the annual modulation
caused by the Earth's motion in the neutrino wind \cite{Duda:2001hd, Hagmann:1999kf}, but the experimental realization of such ideas seems implausible with the available Cavendish-like balance technology.  The results are not encouraging also for similar concepts based on interferometers \cite{Domcke:2017aqj}.

Another idea for the distant future is radiative atomic emission
of a neutrino pair \cite{Yoshimura:2014hfa}. The CNB affects
this process by Pauli phase-space blocking.

Extremely high-energy  neutrinos, produced as cosmic-ray
secondaries or from ultra-heavy particle decay or cosmic strings,
would be absorbed by the CNB, a resonant process if the CM
energy matches the $Z^0$ mass \cite{Weiler:1982qy}. For now there is
no evidence for neutrinos in the required energy range beyond
$10^{20}~{\rm eV}$ so that absorption dips cannot yet be looked for~\cite{Ringwald:2009bg}.

Perhaps the most realistic approach uses inverse $\beta$ decay
\cite{Weinberg:1962zza,Cocco:2007za,Long:2014zva,Arteaga:2017zxg,Lisanti:2014pqa,Akhmedov:2019oxm},
notably on tritium, $\nu_e+{\rm H}^3\to {\rm He}^3+e^-$, which is
actually pursued by the PTOLEMY project \cite{Betts:2013uya,Baracchini:2018wwj}. However,
our reference scenario with the mass spectrum given in
Eq.~(\ref{eq:mass-spectrum}) is particularly difficult because $\nu_3$
has the smallest $\nu_e$ admixture of all mass eigenstates. On the
other hand, if the mass spectrum is inverted and/or quasi degenerate,
the detection opportunities may be more realistic.  Such an experiment
may also be able to distinguish Dirac from Majorana neutrinos
\cite{Long:2014zva} and place constraints on nonstandard neutrino
couplings \cite{Arteaga:2017zxg}.  Moreover, polarization of the
target might achieve directionality \cite{Lisanti:2014pqa}.

The properties of the CNB, the search for the neutrino mass scale, and
the Dirac vs.\ Majorana question, remain at the frontier of particle
cosmology and neutrino physics.  Moreover, while neutrinos are but a
small dark-matter component, detecting the CNB would be a first step
in the future field of dark-matter astronomy.

\section[Neutrinos from Big-Bang Nucleosynthesis]{\kern-1pt Neutrinos from Big-Bang Nucleosynthesis}
\label{sec:BBN}

During its first few minutes, the universe produces the observed light
elements. Subsequent decays of neutrons ($n\to p+e+\overline\nu_e$) and
tritons ($^3{\rm H}\to{}^3{\rm He}+e+\overline\nu_e$) produce a very small
$\overline\nu_e$ flux, which however dominates the GUNS in the gap between
the CNB and thermal solar neutrinos roughly for
$E_\nu=10$--100~meV. While a detection is currently out of the
question, it would provide a direct observational window to primordial
nucleosynthesis.

\subsection{Primordial nucleosynthesis}

Big-bang nucleosynthesis of the light elements is one of the pillars
of cosmology \cite{Alpher:1948ve, Alpher:1950zz, Steigman:2007xt,
  Iocco:2008va, Cyburt:2015mya} and historically has led to a
prediction of the CMB long before it was actually detected
\cite{Gamow:1946eb,1948Natur.162..774A, Alpher:1988kr}. In the early universe,
protons and neutrons are in $\beta$ equilibrium, so their relative
abundance is $n/p=\exp(-\Delta m/T)$ with $\Delta m=1.293~{\rm MeV}$
their mass difference. Weak interactions freeze out about 1~s after
the big bang when $T\approx1~{\rm MeV}$, leaving
$n/p\approx1/6$. Nuclei form only 5~min later when $T$ falls below
60~keV and the large number of thermal photons no longer keeps nuclei
dissociated. Neutrons decay, but their lifetime of 880~s leaves about
$n/p\approx1/7$ at that point. Subsequently most neutrons end up in
$^4$He, leaving the famous primordial helium mass fraction of~25\%.

\begin{figure}[!b]
\includegraphics[width=0.90\columnwidth]{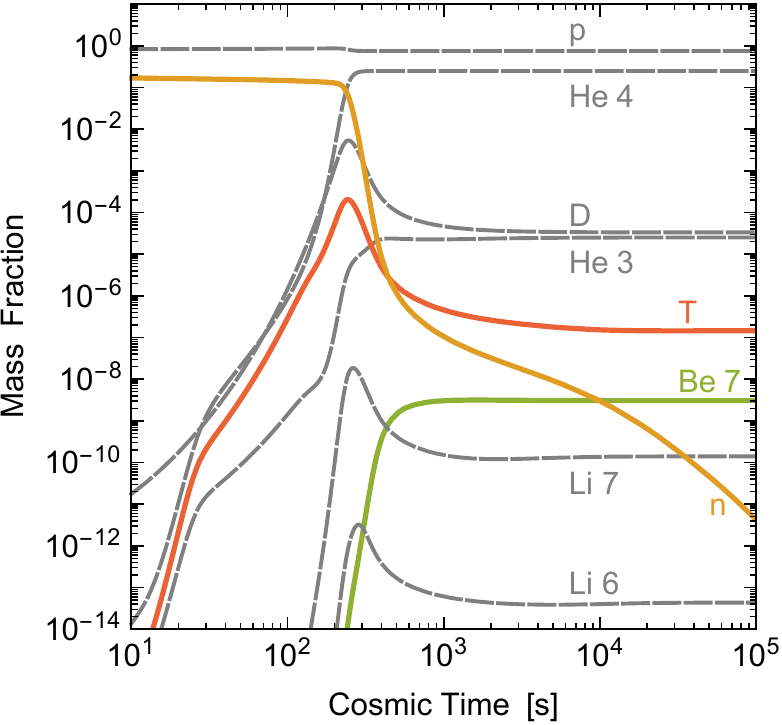}
\caption{Evolution of light-element abundances in the early universe as indicated at the lines.
  Colored (solid) lines are neutrons (n)  and the unstable isotopes tritium (T) and beryllium ($^7$Be) that produce $\overline{\nu}_e$ and that do not survive until today.
  Plot was adapted from \href{http://cococubed.asu.edu/code\_pages/net\_bigbang.shtml}{http://cococubed.asu.edu /code\_pages/net\_bigbang.shtml}, where $\eta=6.23\times10^{-10}$ and
  $H_0=70.5~{\rm km}~{\rm s}^{-1}~{\rm Mpc}^{-1}$ was
  used.}\label{fig:bbn-elements}
\end{figure}

In detail, one has to solve a nuclear reaction network in the
expanding universe and finds the evolution of light isotopes as shown
in Fig.~\ref{fig:bbn-elements}, where neutrons and the unstable isotopes are
shown in color. Besides the nuclear-physics input,
the result depends on the cosmic baryon fraction $\eta=n_B/n_\gamma$.
With $\eta=6.23\times10^{-10}$ that
was chosen in Fig.~\ref{fig:bbn-elements} and the
density $n_\gamma=411~{\rm cm}^{-3}$ of CMB photons, the baryon
density is $n_B=2.56\times10^{-7}~{\rm cm}^{-3}$. The 95\%
C.L.\ range for $n_B$ is 2.4--2.7 in these units
\cite{Tanabashi:2018}. Of particular interest are the unstable
but long-lived isotopes tritium (T) and $^7$Be for which
Fig.~\ref{fig:bbn-elements} shows final mass fractions
$1.4\times10^{-7}$ and $3.1\times10^{-9}$, corresponding to
\begin{subequations}\label{eq:BBN-densities}
  \begin{eqnarray}
  n_{\rm T}&=&1.2\times10^{-14}~{\rm cm}^{-3},\\
  n_{\rm ^7Be}&=&1.1\times10^{-16}~{\rm cm}^{-3}
\end{eqnarray}
\end{subequations}
in terms of a present-day number density.

\begin{figure}[!b]
\includegraphics[width=0.90\columnwidth]{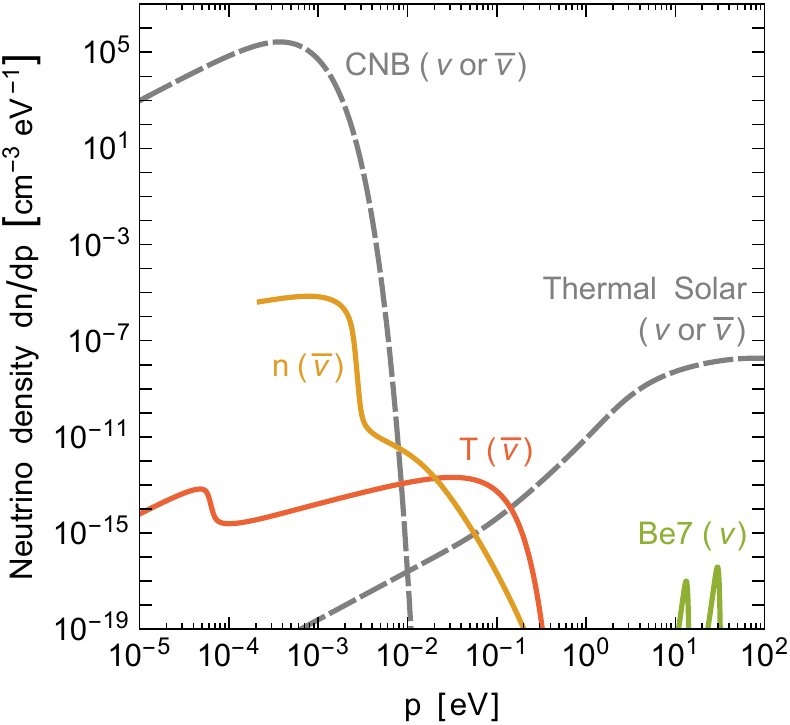}
\caption{Density of low-energy neutrinos, taken to be massless ($p=E$).
  The CNB and thermal solar neutrinos include all flavors, but the lines are only for either
  $\nu$ or $\overline\nu$. Colored (solid) lines are BBN neutrinos:
  $\overline\nu_e$~from $n$ and tritium decay and $\nu_e$ from $^7$Be electron
  capture.}\label{fig:bbn-nu-dens}
\end{figure}

\subsection{Neutrinos from decaying light isotopes}

The isotopes shown in color in Fig.~\ref{fig:bbn-elements} are $\beta$
unstable and thus produce a small cosmic $\overline\nu_e$ or $\nu_e$
density which is  much smaller than the CNB density given in
Eq.~\eqref{eq:CNB-density}, but shows up at larger energies because of
less redshifting due to late decays \cite{Khatri:2010ed,
  Ivanchik:2018fxy, Yurchenko:2019uxu}. Ignoring for now the question of neutrino masses
and flavor conversion, the resulting present-day number densities are
shown in Fig.~\ref{fig:bbn-nu-dens} in comparison with the CNB
(Sec.~\ref{sec:CNB}) and the low-energy tail of thermal solar
neutrinos (Sec.~\ref{sec:Thermal-Solar}). These two sources produce
$\nu\overline\nu$ pairs of all flavors, so their number density is equal
for $\nu$ and $\overline\nu$. In Fig.~\ref{fig:bbn-nu-dens} we show the
all-flavor $\nu$ density of these sources, equal to that for
$\overline\nu$, to compare with either the $\nu$ or $\overline\nu$ density of
BBN neutrinos. The low-energy tail of traditional solar $\nu_e$ from
nuclear reactions (Sec.~\ref{sec:Solar-Nuclear}) and of the $\overline\nu_e$
geoneutrino (Sec.~\ref{sec:Geo}) and reactor fluxes
(Sec.~\ref{sec:Reactors}) are all very much smaller than the solar
thermal $\nu$ or $\overline\nu$ flux.  One concludes that the BBN neutrinos
($\overline\nu_e$) from later neutron ($n$) and tritium decays produce
the dominant density in the valley between the CNB and thermal solar
neutrinos around neutrino energies of 10--200~meV. Of course, a
detection of this flux is out of the question with present-day
technology.

\bigskip

\noindent{\em Beryllium recombination.}---Considering the individual sources in more detail, we begin with
$^7$Be which emerges with a much larger abundance than
$^7$Li. Eventually it decays to $^7$Li by electron capture, producing
$\nu_e$ of 861.8~keV (89.6\%) or 384.2~keV (10.4\%), analogous to the
solar $^7$Be flux (Sec.~\ref{sec:Solar-Nuclear}). However, the electrons captured
in the Sun are free, so their average energy increases
by a thermal amount of a few keV (Table~\ref{table:sun}). In the
dilute plasma of the early universe, electrons are captured from bound
states, which happens only at around 900~years (cosmic redshift $z_{\rm
  rec}\approx29,200$) when $^7$Be atoms form.  The kinetics of $^7$Be
recombination and decay was solved by \textcite{Khatri:2010ed}
who found $z_{\rm rec}$ to be larger by about 5000 than implied by the
Saha equation. The present-day energies of the lines are
$13.1~{\rm eV}=384.2~{\rm keV}/(z_{\rm rec}+1)$ and
$29.5~{\rm eV}=861.8~{\rm keV}/(z_{\rm rec}+1)$, each with a full
width at half maximum of 7.8\%,
given by the redshift profile of $^7$Be recombination,
i.e., 1.0 and 2.3~eV.

The $^7$Be lines in Fig.~\ref{fig:bbn-nu-dens} were extracted
from Fig.~5 of \textcite{Khatri:2010ed} with two modifications. The
integrated number densities in the lines should be 10.4~:~89.6
according to the branching ratio of the $^7$Be decay, whereas in
\textcite{Khatri:2010ed} the strength of the lower-energy line is
reduced by an additional factor $(384.2/861.8)^2$ which we have
undone.\footnote{We thank Rishi Khatri for confirming this issue
which was caused at the level of plotting by a multiplication
with $384.2/861.8$ instead of $861.8/384.2$ to convert
the high-energy line to the low-energy one. The formula for the redshifted
lines given in their Sec.~4 is correct.}
Moreover, we have multiplied both lines with
a factor 5.6 to arrive at
the number density $n_{\rm Be7}$ of Eq.~\eqref{eq:BBN-densities}.
Notice that \textcite{Khatri:2010ed} cite
a relative $^7$Be number density at the end of BBN
of around $10^{-10}$, whereas their cited literature
and also our Fig.~\ref{fig:bbn-elements} shows about 5--6 times
more.

\bigskip

\noindent{\em Tritium decay.}---BBN produces a tritium (T or $^3$H) abundance given
in Eq.~\eqref{eq:BBN-densities} which later decays with a lifetime of 17.8~years
by $^3{\rm H}\to{}^3{\rm He}+e+\overline\nu_e$, producing the same number density of $\overline\nu_e$ with a spectral shape given by Eq.~\eqref{eq:allowed-spectrum}
with $E_{\rm max}=18.6$~keV. During radiation domination, a cosmic age of 17.8~years
corresponds to a redshift of $2\times10^5$, so an energy of 18.6~keV is today 90~meV, explaining the $\overline\nu_e$ range shown in Fig.~\ref{fig:bbn-nu-dens}.

This spectrum was taken from Fig.~2 of \textcite{Ivanchik:2018fxy}. Pre-asymptotic
tritium (i.e.\ the population existing at the onset of BBN, identified by the spike in Fig.~\ref{fig:bbn-elements}) was also included, producing the low-energy step-like feature.
The isotropic flux shown by \textcite{Ivanchik:2018fxy} was multiplied with a factor $2/c$ to obtain our number density.\footnote{We thank \textcite{Ivanchik:2018fxy} for providing a data file for this curve and for explaining the required factor. They define the flux of an isotropic
gas by the number of particles passing through a 1~cm$^2$ disk per sec according to their Eq.~(7) and following text, providing a factor $c/4$. Then they apply a factor of~2
to account for neutrinos passing from both sides. See our Appendix~\ref{sec:Units} for our definition of an isotropic flux.} Our integrated $\overline\nu_e$ density then corresponds well to the tritium density in Eq.~\eqref{eq:BBN-densities}.

\bigskip

\noindent{\em Neutron decay.}---After weak-interaction freeze-out near 1~sec, neutrons decay with
a lifetime of 880~s, producing $\overline\nu_e$ with a spectrum given by
Eq.~\eqref{eq:allowed-spectrum} with $E_{\rm max}=782~{\rm keV}$. The
short lifetime implies that there is no asymptotic value around the
end of BBN. Notice also that the late $n$ evolution shown in
Fig.~\ref{fig:bbn-nu-dens} is not explained by decay alone that would
imply a much faster decline, i.e., residual nuclear reactions provide
a late source of neutrons.  The $\overline\nu_e$ number density shown in
Fig.~\ref{fig:bbn-nu-dens} was obtained from \textcite{Ivanchik:2018fxy} with the same prescription that we used
for tritium.

\subsection{Neutrinos with mass}

The cross-over region between CNB, BBN, and solar neutrinos shown in Fig.~\ref{fig:bbn-nu-dens} is at energies where sub-eV neutrino masses become important. For the purpose of illustration we use the minimal masses in normal ordering of Eq.~\eqref{eq:mass-spectrum} with 0, 8.6, and 50~meV. Neutrinos reaching Earth will have decohered into mass eigenstates, so one needs to determine the three corresponding spectra.

The CNB consists essentially of an equal mixture of all flavors, so the probability
for finding a random CNB neutrino or antineutrino in any of the mass
eigenstates is
\begin{equation}\label{eq:CNB-prob}
  P_{i}^{\rm CNB}=\frac{1}{3}
  \quad\hbox{for}\quad i=1,2,3.
\end{equation}
The flavor density matrix is essentially proportional to the unit matrix
from the beginning and thus is the same in any basis. Flavor conversion has no effect.

On the other hand, the BBN neutrinos are produced in the $e$ flavor, so their flavor content will change with time. Flavor evolution in the early universe can involve many complications in that the matter effect at $T\agt1$~MeV is dominated by a thermal term~\cite{Notzold:1987ik}. Moreover, neutrinos themselves are an important background medium, leading to collective flavor evolution~\cite{Kostelecky:1993yt, Duan:2010bg}.

However, the BBN neutrinos are largely produced after BBN is complete at $T\alt 60$~MeV. Scaling the present-day baryon density of $2.5\times10^{-7}~{\rm cm}^{-3}$ to the post-BBN epoch provides a matter density of the order of $10^{-5}~{\rm g}~{\rm cm}^{-3}$, very much smaller than the density of Sun or Earth, so the matter or neutrino backgrounds are no longer important. For the purpose of flavor evolution of MeV-range neutrinos we are in vacuum and the mass-content of the original states does not evolve. So we may use the best-fit probabilities $P_{ei}$ of finding a $\nu_e$ or $\overline\nu_e$ in any of the mass eigenstates given in the top row of Eq.~\eqref{eq:probability-matrix},
\begin{equation}\label{eq:BBN-prob}
  P_{1}^{\rm BBN}=0.681,
  \quad
  P_{2}^{\rm BBN}=0.297,
  \quad
  P_{3}^{\rm BBN}=0.022.
\end{equation}
Notice that here we have forced the numbers to add up to unity to correct for rounding errors.

Thermal solar neutrinos emerge in all flavors, but not with equal probabilities~\cite{Vitagliano:2017odj}. For very low energies, the mass-eigenstate probabilities
are (see text below Eq.~\ref{eq:sol-lowE-flux})
\begin{equation}\label{eq:Sun-prob}
  P_{1}^{\rm Sun}=0.432,
  \quad
  P_{2}^{\rm Sun}=0.323,
  \quad
  P_{3}^{\rm Sun}=0.245.
\end{equation}
For higher energies, these probabilities are plotted in the bottom
panel of Fig.~\ref{fig:Thermal-Solar}.

\begin{figure}
\includegraphics[width=0.90\columnwidth]{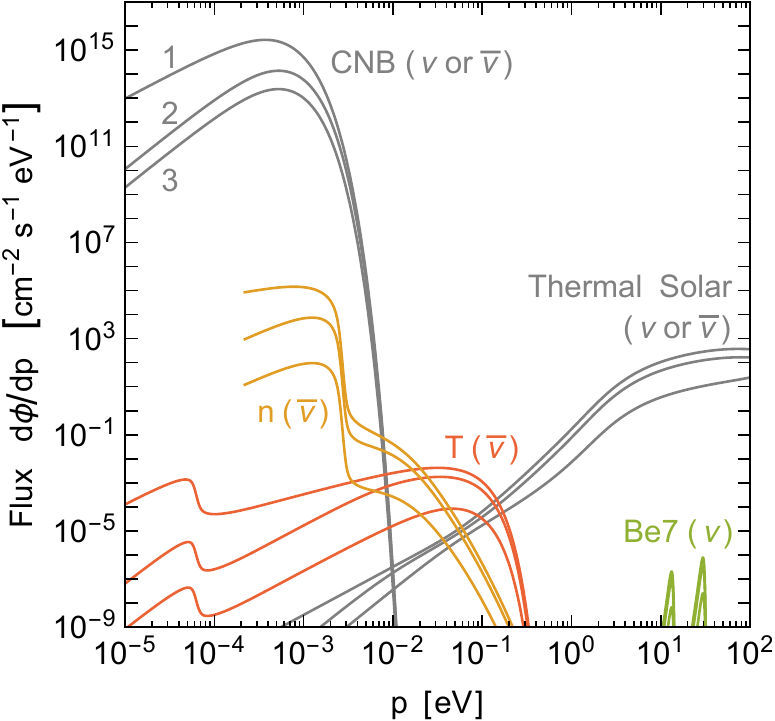}
\vskip10pt
\includegraphics[width=0.90\columnwidth]{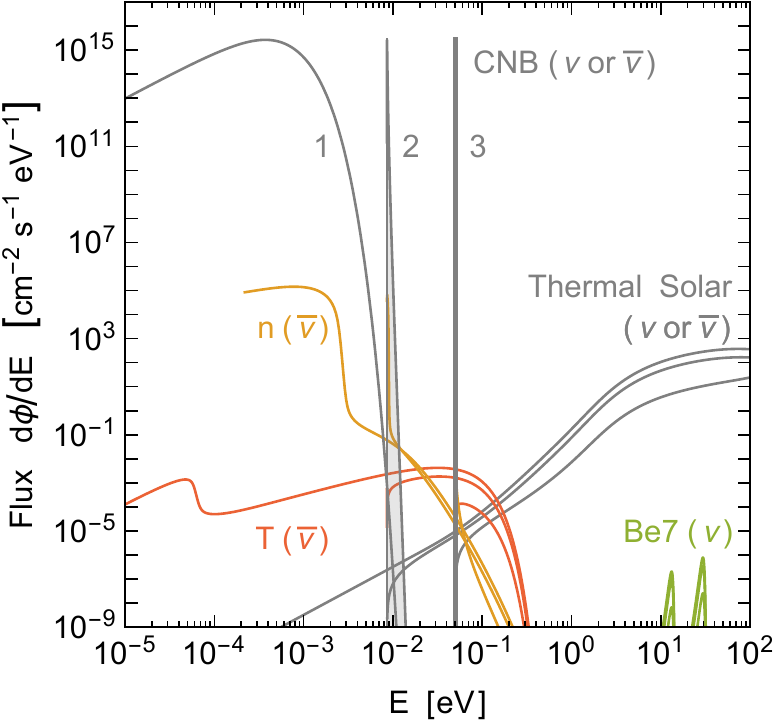}
  \caption{Flux densities of mass-eigenstate neutrinos for $m_i=0$, 8.6 and 50~meV as indicated
  at the curves, using the probabilities of Eqs.~(\ref{eq:CNB-prob}--\ref{eq:Sun-prob}) and the spectra of Fig.~\ref{fig:bbn-nu-dens}. {\em Top:\/} $d\Phi/dp$ which includes a velocity factor $v_i=p/E_i$ for each mass state. {\em Bottom:\/} $d\Phi/dE$ showing sharp lines at
  $E=m_{2,3}$.}\label{fig:bbn-flux}
\end{figure}

The CNB and BBN neutrinos are produced with high energies and later
their momenta are redshifted by cosmic expansion. Therefore, their
comoving differential number spectrum $dn/dp$ as a function of $p$
remains unchanged. If we interpret the horizontal axis of
Fig.~\ref{fig:bbn-nu-dens} as $p$ instead of $E$ and the vertical axis
as $dn/dp$ instead of $dn/dE$, the CNB and BBN curves actually do not
change, except that we get three curves, one for each mass eigenstate,
with the relative amplitudes of Eqs.~\eqref{eq:CNB-prob}
and~\eqref{eq:BBN-prob}.

For thermal solar neutrinos, the same argument applies to
bremsstrahlung, which dominates at low energies, because the
spectrum is essentially determined by phase space alone
(Sec.~\ref{sec:SolarVeryLow}). At higher energies, where our assumed
small masses are not important, the mass would also enter in the
matrix element and one would need an appropriate evaluation of plasmon
decay.

For experimental searches, the flux may be a more
appropriate quantity. Multiplying the number density spectra of
Fig.~\ref{fig:bbn-nu-dens} for each $p$ with
the velocity $v_i=p/\sqrt{p^2+m_i^2}$ provides the mass-eigenstate
flux spectra $d\Phi/dp$ shown in Fig.~\ref{fig:bbn-flux} (top), in
analogy to Fig.~\ref{fig:nucosmo}.

For experiments considering the absorption of neutrinos,
e.g.\ inverse $\beta$ decay on tritium, the
energy $E$ is a more appropriate variable instead of the
momentum $p$, so we show $d\Phi/dE$ as a function of $E$
in Fig.~\ref{fig:bbn-flux} (bottom). Notice that the velocity
factor $v_i$ is undone by a Jacobian $E/p$, so for example the
maxima of the mass-eigenstate curves are the same for every
$m_i$ as discussed in Sec.~\ref{sec:SpectrumEarth} and
illustrated in Fig.~\ref{fig:enecosmo}. Relative to the massless
case of Fig.~\ref{fig:bbn-nu-dens}, the vertical axis is simply
scaled with a factor $c$, whereas the curves are compressed
in the horizontal direction by $p\to E=\sqrt{p^2+m_i^2}$.
Effectively one obtains narrow lines at the
non-vanishing neutrino masses that are vastly dominated
by the CNB. The integrated
fluxes of the three mass eigenstates in either $\nu$ or
$\overline\nu$ are
\begin{subequations}
\begin{eqnarray}
  \Phi_1 &=& 1.68\times10^{12}~{\rm cm}^{-2}~{\rm s}^{-1}, \\[0.8ex]
  \Phi_2 &=& 1.35\times10^{11}~{\rm cm}^{-2}~{\rm s}^{-1}, \\[0.8ex]
  \Phi_3 &=& 2.32\times10^{10}~{\rm cm}^{-2}~{\rm s}^{-1},
\end{eqnarray}
\end{subequations}
where we have used Eqs.~\eqref{eq:CNB-velocity}
and~\eqref{eq:CNB-density} of Sec.~\ref{sec:CNB}.

Note that we have assumed $m_1 = 0$ in this Section; a degenerate mass spectrum (i.e., $m_1 \gg T_\nu=0.168$~meV) would make the flux densities of all mass eigenstates similar to each other, they will all have a spike-like behavior, and they will be shifted to larger energies. In this case there is no neutrino radiation in the universe today, only neutrino hot dark matter.

\section{Solar Neutrinos from Nuclear Reactions}
\label{sec:Solar-Nuclear}

The Sun emits 2.3\% of its nuclear energy production in the form of MeV-range electron neutrinos. They arise from the effective fusion reaction $4p+2e^-\to{}^4{\rm He}+2\nu_e+26.73~{\rm MeV}$ that proceeds through several reaction chains and cycles.  The history of solar neutrino
measurements is tightly connected with the discovery of flavor
conversion and the matter effect on neutrino dispersion.  There is
also a close connection to precision modeling of the Sun, leading to a
new problem in the form of discrepant sound-speed profiles relative to
helioseismology. This issue may well be related to the photon
opacities and thus to the detailed chemical abundances in the solar
core, a prime target of future neutrino precision
experiments. Meanwhile, solar neutrinos are becoming a background to weakly interacting massive particle
(WIMP) dark-matter searches. In fact, dark-matter detectors in future
may double as solar neutrino observatories.

\subsection{The Sun as a neutrino source}

\begin{figure}[htb]
\includegraphics[width=0.90\columnwidth]{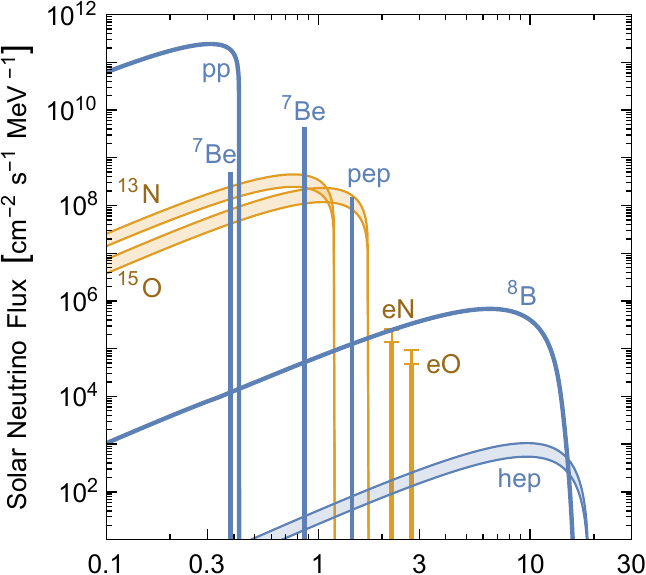}
\vskip2pt
\includegraphics[width=0.90\columnwidth]{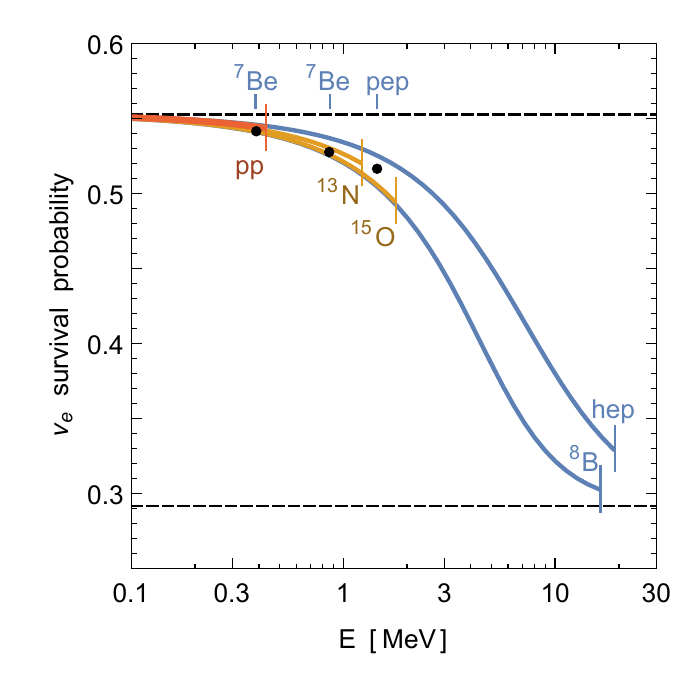}
\caption{Solar neutrinos from different source reactions. In blue (dark gray) pp-chain neutrinos (pp, $^7$Be, pep, $^8$B, hep), in orange (light gray) CNO neutrinos ($^{13}$N, $^{15}$O and the electron-capture lines eN and eO). {\em Top panel:\/} Differential fluxes, where line sources are in units of $\rm{cm}^{-2}~\rm{s}^{-1}$. pp-chain fluxes (except for hep) according to the measurements shown in Table~\ref{table:sun} where the uncertainties are too small to show. For the CNO and hep fluxes the range is bracketed by the lowest AGSS09 and highest GS98 predictions. $^{17}$F is a very small correction to the $^{15}$O flux and thus not shown. {\em Bottom panel:\/} Adiabatic $\nu_e$ survival probability due to flavor conversion (see Sec.~\ref{sec:FlavorConversion}) which depends on the radial distributions of the different production processes. For the eN and eO lines, these distributions have not been published. The
black dots show the survival probabilities of the three pp-chain lines from $^7$Be and pep.
The horizontal dashed lines show the survival probability for vanishing and infinite neutrino energy.}
  \label{sunnuclear}
\end{figure}

The Sun produces nuclear energy by hydrogen fusion to helium that proceeds
through the pp chains (exceeding 99\% for solar conditions) and the rest
through the CNO cycle \cite{Clayton:1983, Kippenhahn:2012, Bahcall:1987jc,
  Bahcall:1989ks, Robertson:2012ib, Serenelli:2016dgz}.  For every
produced ${}^4{\rm He}$ nucleus, two protons need to convert to
neutrons by what amounts to $p+e^-\to n+\nu_e$, i.e., two electrons
disappear in the Sun and emerge as $\nu_e$.
The individual $\nu_e$-producing reactions are listed in Table~\ref{table:sun}
(for more details see below)
and the expected flux spectra at Earth are shown in Fig.~\ref{sunnuclear}.

\begin{table*}
  \caption{Neutrino fluxes at Earth from different nuclear reactions in the Sun. Theoretical predictions from \textcite{Vinyoles:2016djt} for models with GS98 \cite{Grevesse:1998bj} and AGSS09 \cite{Asplund:2009fu} abundances. The predicted electron capture (EC)
    fluxes from the CNO cycle were obtained by scaling the
    $\beta^+$-decay fluxes \cite{Stonehill:2003zf}.
    The neutrino endpoint energy $E_{\rm max}$ and
    average $E_{\rm av}$ includes thermal energy of a few keV
    \cite{Bahcall:1997eg} except for the CNO-EC lines, where
    the given $E_{\rm av}$ is $E_{\rm max}+2m_e$ of the corresponding $\beta^+$ process.
    Observed fluxes with $1\sigma$ errors from the global analysis of
    \textcite{Bergstrom:2016cbh}.}\label{table:sun}
\vskip6pt
\begin{tabular*}{\textwidth}{@{\extracolsep{\fill}}|l|l|l|l|l|ll|ll|l|r|}
\hline
\hline
   \multirow{2}{*}{Channel} & \multirow{2}{*}{Flux}  & \multirow{2}{*}{Reaction} & $E_{\rm av}$& $E_{\rm max}$   &\multicolumn{6}{c|}{Flux at Earth}\\
   \cline{6-11}

           &         &          & MeV &  MeV    &\multicolumn{2}{l|}{GS98}&\multicolumn{2}{l|}{AGSS09}   & Observed & \multicolumn{1}{l|}{Units} \\
\hline
pp Chains ($\beta^+$)&$\Phi_{\rm pp}$ & $p+p\rightarrow d+ e^+ +\nu_e$ &0.267 & 0.423 & 5.98 & $\pm0.6\%$ & 6.03 & $\pm0.5\%$ & $5.971^{+0.62\%}_{-0.55\%}$ & $10^{10}~{\rm cm}^{-2}~{\rm s}^{-1}$ \\
 & $\Phi_{\rm B}$ & $\ce{^8B}\rightarrow \ce{^8Be^*}+ e^+ +\nu_e$ & $6.735\pm0.036$ &$\sim 15$ & 5.46 & $\pm12\%$ & 4.50 & $\pm12\%$ & $5.16^{+2.5\%}_{-1.7\%}$ & $10^{6}~{\rm cm}^{-2}~{\rm s}^{-1}$ \\
 & $\Phi_{\rm hep}$& $\ce{^3He}+p\rightarrow\ce{^4 He}+e^++\nu_e$ & 9.628 & 18.778   & 0.80 & $\pm30\%$ & 0.83 & $\pm30\%$ & $1.9^{+63\%}_{-47\%}$ & $10^{4}~{\rm cm}^{-2}~{\rm s}^{-1}$\\
\hline
pp Chains (EC)& \multirow{2}{*}{$\Phi_{\rm Be}$} &{$e^-+\ce{^7Be}\rightarrow\ce{^7Li}+\nu_e$} & 0.863 (89.7\%)& &
\multirow{2}{*}{4.93} &
\multirow{2}{*}{$\pm6\%$} &
\multirow{2}{*}{4.50} &
\multirow{2}{*}{$\pm6\%$}&
\multirow{2}{*}{$4.80^{+5.9\%}_{-4.6\%}$} &
\multirow{2}{*}{$10^{9}~{\rm cm}^{-2}~{\rm s}^{-1}$} \\
             &                 &{$e^-+\ce{^7Be}\rightarrow\ce{^7Li^*}+\nu_e$}& 0.386 (10.3\%)& &      &          &      &          &                     &                                   \\
& $\Phi_{\rm pep}$ & $p+e^- + p\rightarrow d +\nu_e$ & 1.445 & & 1.44 & $\pm1\%$ & 1.46 & $\pm0.9\%$ & $1.448^{+0.90\%}_{-0.90\%}$ & $10^{8}~{\rm cm}^{-2}~{\rm s}^{-1}$ \\
 \hline
CNO Cycle ($\beta^+$) & $\Phi_{\rm N}$  & $\ce{^13N}\rightarrow\ce{^13C}+e^++\nu_e$ & 0.706 & 1.198 & 2.78 & $\pm15\%$ & 2.04 & $\pm14\%$ & ${}<13.7$ & $10^{8}~{\rm cm}^{-2}~{\rm s}^{-1}$  \\
&   $\Phi_{\rm O}$  &  $\ce{^15O}\rightarrow\ce{^15N}+e^++\nu_e$ & 0.996 & 1.732& 2.05 & $\pm17\%$ & 1.44 & $\pm16\%$ & ${}<2.8$ & $10^{8}~{\rm cm}^{-2}~{\rm s}^{-1}$  \\
&      $\Phi_{\rm F}$  &  $\ce{^17F}\rightarrow\ce{^17O}+e^++\nu_e$ & 0.998 & 1.736  & 5.29 & $\pm20\%$ & 3.26 & $\pm18\%$ & ${}<8.5$ & $10^{6}~{\rm cm}^{-2}~{\rm s}^{-1}$ \\
\hline
CNO Cycle (EC)& $\Phi_{\rm eN}$ &$\ce{^13N}+e^-\rightarrow\ce{^13C}+\nu_e$ & 2.220 & &2.20 & $\pm15\%$ & 1.61 & $\pm14\%$ & \multicolumn{1}{c|}{---}& $10^5~{\rm cm}^{-2}~{\rm s}^{-1}$ \\
&    $\Phi_{\rm eO}$   & $\ce{^15O}+e^-\rightarrow\ce{^15N}+\nu_e$ & 2.754 &  &0.81 & $\pm17\%$  & 0.57 & $\pm16\%$ & \multicolumn{1}{c|}{---} & $10^5~{\rm cm}^{-2}~{\rm s}^{-1}$ \\
&    $\Phi_{\rm eF}$   & $\ce{^17F}+e^-\rightarrow\ce{^17O}+\nu_e$ & 2.758 &  &
3.11 & $\pm20\%$     &  1.91 & $\pm18\%$ &\multicolumn{1}{c|}{---} & $10^3~{\rm cm}^{-2}~{\rm s}^{-1}$ \\
 \hline
 \hline
\end{tabular*}
\end{table*}

All pp chains begin with $p+p\to d+e^++\nu_e$, the pp reaction, which on average releases 0.267~MeV as~$\nu_e$. Including other processes (GS98 predictions of Table~\ref{table:sun}) implies $\langle E_{\nu_e}\rangle=0.312~{\rm MeV}$. The solar luminosity without neutrinos
is $L_\odot=3.828\times10^{33}~{\rm erg}~{\rm s}^{-1}=2.39\times10^{39}~{\rm MeV}~{\rm s}^{-1}$,
implying a solar $\nu_e$ production of
\begin{equation}
L_{\nu_e}=2\times \frac{L_\odot}{26.73~\rm{MeV}-2\,
\langle E_{\nu_e}\rangle}=1.83\times10^{38}~{\rm s}^{-1},
\end{equation}%
where 26.73~MeV is the energy released per He fusion and $2$ the number of neutrinos per fusion. The average distance of $1.496\times10^{13}~{\rm cm}$ thus implies
a flux, number density, and energy density at Earth of
\begin{subequations}
  \begin{eqnarray}
    \Phi_\nu &=& 6.51\times10^{10}~{\rm cm}^{-2}~{\rm s}^{-1},\\
    n_\nu    &=& 2.17~{\rm cm}^{-3},\\
    \rho_\nu &=& 0.685~{\rm MeV}~{\rm cm}^{-3}.
  \end{eqnarray}
\end{subequations}
These numbers change by $\pm3.4\%$ in the course of the year due to the
ellipticity of the Earth's orbit, a variation confirmed
by the Super-Kamiokande detector~\cite{Fukuda:2001nj}.\looseness=-1

While this overall picture is robust, the flux spectra of those
reactions with larger $E_{\nu_e}$ are particulary important
for detection and flavor-oscillation physics, but
are side issues for overall solar physics. Therefore, details of
the production processes and of solar modeling are crucial
for predicting the solar neutrino spectrum.

\subsection{Production processes and spectra}

The proton-neutron conversion required for hydrogen burning proceeds
either as $\beta^+$ decay of the effective form $p\to n+e^++\nu_e$,
producing a continuous spectrum, or as electron capture (EC) $e^-+p\to
n+\nu_e$, producing a line spectrum. The nuclear MeV energies imply a
much larger final-state $\beta^+$ phase space than the initial-state
phase space occupied by electrons with keV thermal energies, so the
continuum fluxes tend to dominate \cite{Bahcall:1990tb}.

Line energies are larger $+2m_e$ relative to the continuum end point) and lines produce
a distinct detection signature \cite{Bellini:2013lnn, Agostini:2017ixy}.
The ${}^7{\rm Be}$ line is particularly important because
the nuclear energy is too small for $\beta^+$ decay. (Actually in
10\% of all cases it proceeds through an excited state of
${}^7{\rm Li}$, so there are two lines, together forming
the ${}^7{\rm Be}$ flux.)

We neglect ${}^3{\rm He}+e^-+p\to{}^4{\rm He}+\nu_e$, the heep flux
\cite{Bahcall:1990tb}.  On the other hand, we include the often
neglected lines from EC in CNO reactions, also called ecCNO processes
\cite{Stonehill:2003zf, Villante:2014txa}. Our
flux predictions come from scaling the continuum fluxes \cite{Vinyoles:2016djt} with
the ratios provided by \textcite{Stonehill:2003zf},
although these are based on a different solar model. This
inconsistency is small compared with the overall uncertainty of
the CNO fluxes.

The endpoint $E_{\rm max}$ of a continuum spectrum is given, in
vacuum, by the nuclear transition energy. However, for the reactions
taking place in the Sun one needs to include thermal kinetic energy
of a few keV. The endpoint and average energies
listed in Table~\ref{table:sun} include this effect according to
Bahcall \cite{Bahcall:1997eg}. For the same reason the EC lines are
slightly shifted and have a thermal width of a few keV
\cite{Bahcall:1993ej}, which is irrelevant in practice for
present-day experiments. The energies of the ecCNO lines were obtained
from the listed continuum endpoints by adding $2m_e$, which agrees
with \textcite{Stonehill:2003zf} except for
${}^{17}{\rm F}$, where they show 2.761 instead of 2.758~MeV.

\begin{figure}

\includegraphics[width=0.90\columnwidth]{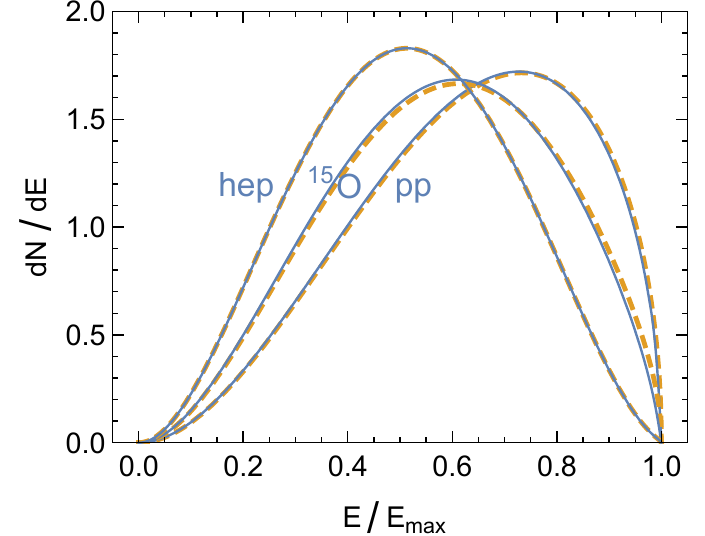}
\caption{Spectra of pp, $^{15}$O and hep neutrinos. The other
CNO spectra are similar to $^{15}$O.
{\em Solid:\/} Tabulated spectra according to \textcite{Bahcall:1997eg}.
{\em Dashed:\/} Allowed nuclear decay spectra according to Eq.~(\ref{eq:allowed-spectrum}).}
\label{fig:sun-spectra-1}
\end{figure}

\begin{figure}
\includegraphics[width=0.90\columnwidth]{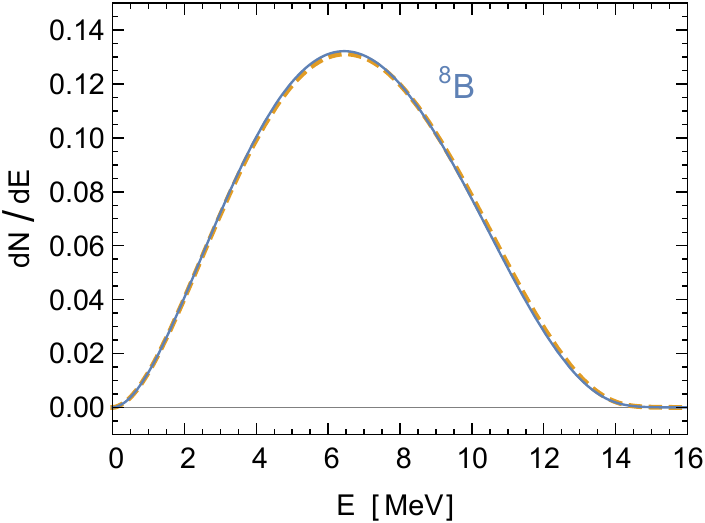}
\caption{Spectrum of $^{8}$B neutrinos.
{\em Solid:\/} According to \textcite{Bahcall:1996qv}.
{\em Dashed:\/} According to \textcite{Winter:2004kf}.}
\label{fig:B8-spectrum}
\end{figure}

Except for $^8$B, the continuum spectra follow from an allowed
nuclear $\beta$ decay, being dominated by the phase space of the final-state
$e^+$ and $\nu_e$. In vacuum and ignoring $e^+$ final-state interactions it is
\begin{equation}\label{eq:allowed-spectrum}
  \frac{dN}{dE}\propto E^2(Q-E)\sqrt{(Q-E)^2-m_e^2}\,,
\end{equation}
where $Q=E_{\rm max}+m_e$. In Fig.~\ref{fig:sun-spectra-1} (dashed
lines) we show these spectra in normalized form for the pp and hep
fluxes as well as $^{15}{\rm O}$, representative of the CNO
fluxes.  We also show the spectra (solid lines), where final-state
corrections and thermal initial-state distributions are included
according to \textcite{Bahcall:1997eg}. Notice that the spectra
provided on the late \href{http://www.sns.ias.edu/~jnb/}{John
  Bahcall's homepage} are not always exactly identical with those in
\textcite{Bahcall:1997eg}.

The ${}^8$B flux is the dominant contribution in many solar neutrino
experiments because it reaches to large energies and the detection
cross section typically scales with $E^2$, yet it is the one with the
least simple spectrum.  The decay ${}^8{\rm B}\to{}^8{\rm
  Be}+e^++\nu_e$ has no sharp cutoff because the final-state ${}^8{\rm
  Be}$ is unstable against $2\alpha$ decay. The $\nu_e$ spectrum
can be inferred from the measured $\alpha$ and $\beta^+$ spectra. The
$\nu_e$ spectrum provided by \textcite{Bahcall:1996qv} is
shown in Fig.~\ref{fig:B8-spectrum} as a solid line.  As a dashed line
we show the determination of \textcite{Winter:2004kf},
based on a new measurement of the $\alpha$ spectrum.

For comparison with keV thermal neutrinos
(Sec.~\ref{sec:Thermal-Solar}) it is useful to
consider an explicit expression for the solar flux at low energies
where the pp flux strongly dominates.  Using the observed
total pp flux from Table~\ref{table:sun}, we find that
an excellent
approximation for the flux at Earth is
\begin{equation}\label{eq:low-pp-flux}
  \frac{d\Phi_{\rm pp}}{dE}=\frac{832.7\times10^{10}}{{\rm cm}^{2}~{\rm s}~{\rm MeV}}\,
  \left(\frac{E}{{\rm MeV}}\right)^2
  \left(1-2.5\,\frac{E}{{\rm MeV}}\right).
\end{equation}
To achieve sub-percent precision, the purely quadratic term can be
used for $E$ up to a few keV. With the next correction, the
expression can be used up to 100~keV.

\subsection{Standard solar models}

The neutrino flux predictions, such as those shown in Table~\ref{table:sun},
depend on a detailed solar model that provides the variation of temperature,
density, and chemical composition with radius. While the neutrino
flux from the dominant pp reaction is largely determined by the
overall luminosity, the small but experimentally dominant
higher-energy fluxes depend on the branching between different terminations
of the pp chains and the relative importance of the CNO cycle, all of
 which depends sensitively on chemical composition and
temperature. For example, the ${}^8{\rm B}$ flux scales approximately
as $T_{\rm c}^{24}$ with solar core temperature \cite{Bahcall:1996vj} ---
the neutrino fluxes are sensitive solar thermometers.

The flux predictions are usually based on a Standard Solar Model (SSM)
\cite{Serenelli:2016dgz}, although the acronym might be more appropriately interpreted as Simplified Solar Model.
One assumes spherical symmetry and hydrostatic equilibrium,
neglecting dynamical effects, rotation, and magnetic fields. The zero-age
model is taken to be chemically homogeneous without further mass loss or gain.
Energy is transported by radiation (photons) and convection. The latter is relevant only in
the outer region (2\% by mass or 30\% by radius) and is treated phenomenologically
with the adjustable parameter $\alpha_{\rm MLT}$ to express
the mixing length in terms of the pressure scale height.

\begin{figure}[b!]
\includegraphics[width=0.88\columnwidth]{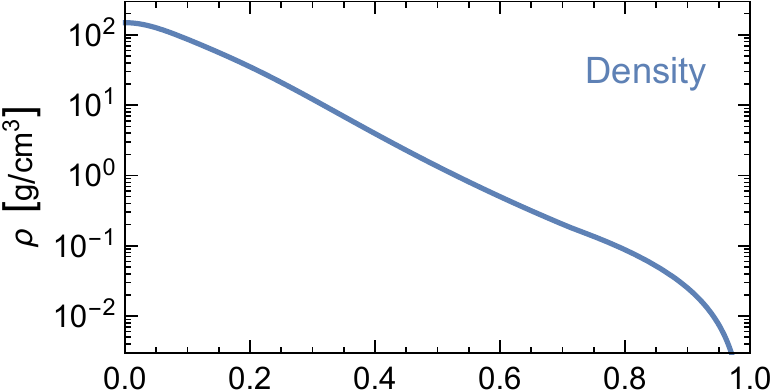}
\includegraphics[width=0.88\columnwidth]{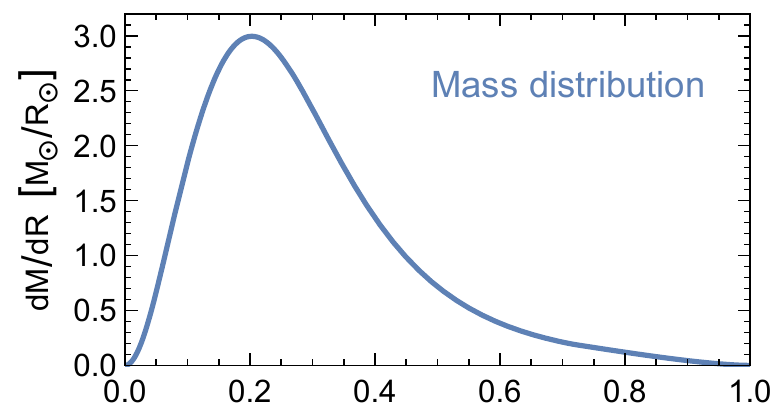}
\includegraphics[width=0.88\columnwidth]{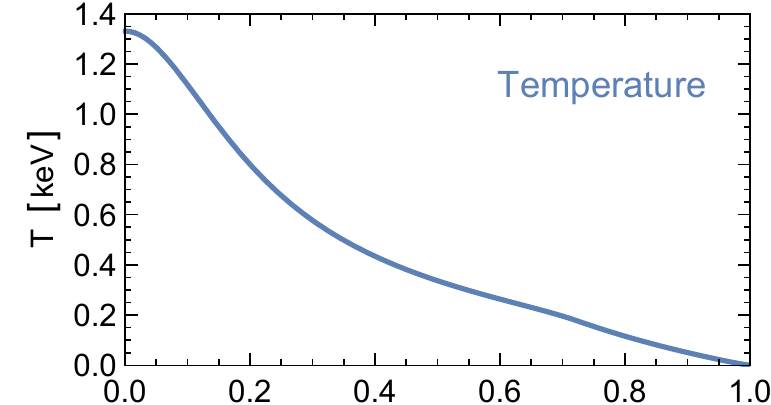}
\includegraphics[width=0.88\columnwidth]{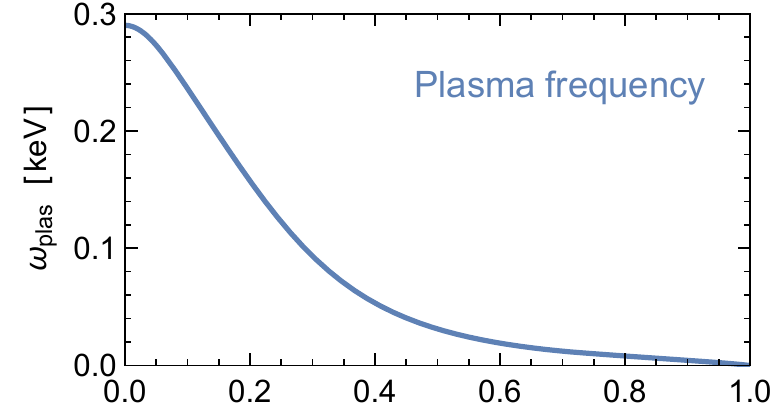}
\includegraphics[width=0.88\columnwidth]{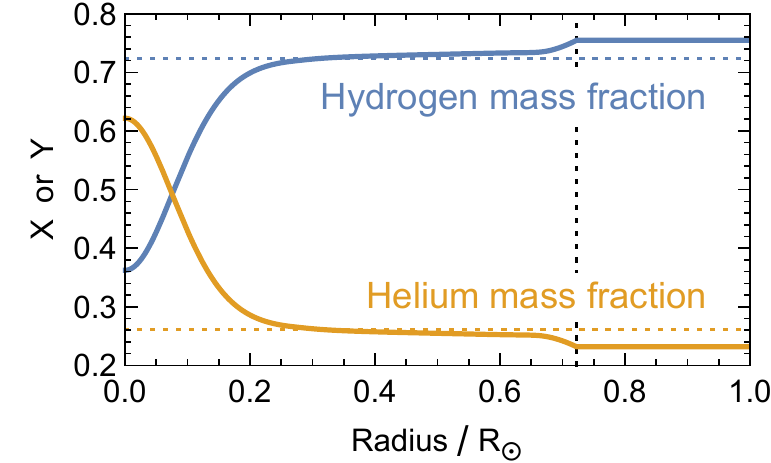}
\caption{Standard Solar Model of the Barcelona group
  \cite{Vinyoles:2016djt} with AGSS09 composition.
  {\em Bottom panel:\/} The vertical dotted line shows $R_{\rm CZ}$, the depth
  of the convection zone, whereas the horizontal lines show the initial H
  and He mass fractions.}\label{fig:SSM}
\end{figure}

Further adjustable parameters are the initial mass fractions of
hydrogen, $X_{\rm ini}$, helium, $Y_{\rm ini}$, and ``metals''
(denoting anything heavier than helium),
$Z_{\rm ini}$, with the constraint $X_{\rm ini}+Y_{\rm ini}+Z_{\rm ini}=1$.
These parameters must be adjusted such that the
evolution to the present age of
$\tau_\odot=4.57\times10^9~{\rm years}$ reproduces the measured luminosity
$L_\odot=3.8418\times10^{33}~{\rm erg}~{\rm s}^{-1}$, the radius
$R_\odot=6.9598\times10^{10}~{\rm cm}$,
and the spectroscopically observed metal abundance at the surface, $Z_{\rm S}$,
relative to that of hydrogen, $X_{\rm S}$. These surface abundances
differ from the initial ones because of
gravitational settling of heavier elements relative to lighter ones.
As an example we show in Fig.~\ref{fig:SSM}
the radial variation of several solar parameters for a SMM of the
Barcelona group \cite{Vinyoles:2016djt}.

The relative surface abundances of different elements are determined
by spectroscopic measurements which agree well, for non-volatile
elements, with those found in meteorites.  The older standard
abundances (GS98) of \textcite{Grevesse:1998bj} were
superseded in 2009 by the AGSS09 composition of Asplund, Grevesse,
Sauval and Scott and updated in 2015 \cite{Asplund:2009fu,
  Scott:2014lka, Scott:2014mka, Grevesse:2014nka}. The AGSS09
composition shows significantly smaller abundances of volatile
elements. According to \textcite{Vinyoles:2016djt}, the surface
abundances are $Z_{\rm S}=0.0170\pm0.0012$ (GS98) and
$0.0134\pm0.0008$ (AGSS09), the difference being almost entirely due
to CNO elements.

The CNO abundances not only affect CNO neutrino fluxes directly,
but determine the solar model through the photon opacities that regulate
radiative energy transfer. Theoretical opacity calculations include
OPAL \cite{Iglesias:1996}, the Opacity Project (OP) \cite{Badnell:2004rz},
OPAS \cite{Blancard:2012, Mondet:2015}, STAR \cite{Krief:2016znd},
and OPLIB \cite{Colgan:2016}, which for solar conditions agree within 5\%,
but strongly depend on input abundances.

A given SSM can be tested with helioseismology that determines the
sound-speed profile, the depth of the convective zone, $R_{\rm CZ}$,
and the surface helium abundance, $Y_{\rm S}$. The new spectroscopic surface abundances
immediately caused a problem in that these parameters deviate significantly from the solar
values, whereas the old GS98 abundances provide much better
agreement \cite{Vinyoles:2016djt,Grevesse:1998bj,Asplund:2009fu}. (See Table~\ref{table:sun-comparison} for a comparison using recent Barcelona models.)

So while SSMs with the old GS98
abundances provide good agreement with helioseismology,
they disagree with the modern surface
abundances, whereas for the AGSS09 class of models it is the other way
around.  There is no satisfactory solution to this conundrum, which is
termed the ``solar abundance problem,'' although it is not clear if
something is wrong with the abundances, the opacity calculations,
other input physics, or any of the assumptions entering the SSM
framework.

\begin{table}
  \caption{Main characteristics of two Barcelona SSMs of \textcite{Vinyoles:2016djt} with
  GS98 and AGSS09 abundances. $R_{\rm CZ}$ is the depth of the convection zone and
  $\langle \delta c/c\rangle$ the average deviation of the sound-speed profile
  relative to helioseismic measurements.}\label{table:sun-comparison}
  \vskip4pt
\begin{tabular*}{\columnwidth}{@{\extracolsep{\fill}}llll}
\hline
\hline
Quantity&B16-GS98&B16-AGSS09&Solar\footnotemark[1]\\
\hline
$Y_{\rm S}$               & $0.2426\pm0.0059$            & $0.2317\pm0.0059$ & $0.2485\pm0.0035$\\
$R_{\rm CZ}/R_\odot$      & $0.7116\pm0.0048$            & $0.7223\pm0.0053$ & $0.713\pm0.001$  \\
$\langle\delta c/c\rangle$& $0.0005^{+0.0006}_{-0.0002}$ & $0.0021\pm0.001$ & \multicolumn{1}{c}{---} \\
\hline
$\alpha_{\rm MLT}$ & $2.18\pm0.05$     & $2.11\pm0.05$     & \multicolumn{1}{c}{---} \\
$Y_{\rm ini}$      & $0.2718\pm0.0056$ & $0.2613\pm0.0055$ & \multicolumn{1}{c}{---} \\
$Z_{\rm ini}$      & $0.0187\pm0.0013$ & $0.0149\pm0.0009$ & \multicolumn{1}{c}{---} \\
$Z_{\rm S}$        & $0.0170\pm0.0012$ & $0.0134\pm0.0008$ & \multicolumn{1}{c}{---} \\
\hline
\end{tabular*}
\footnotetext[1]{\textcite{Basu:1997,Basu:2004zg}}
\end{table}

The pp-chains neutrino fluxes predicted by these two classes of models
bracket the measurements (Table~\ref{table:sun}), which however do not
clearly distinguish between them. A future measurement of the CNO
fluxes might determine the solar-core CNO abundances
and thus help to solve the ``abundance problem.''  While it
is not assured that the two classes of models actually bracket the
true case, one may speculate that the CNO fluxes might lie between the
lowest AGSS09 and the largest GS98 predictions. Therefore, this range
is taken to define the flux prediction shown in
Fig.~\ref{sunnuclear}.

\subsection{Antineutrinos}

The Borexino scintillator detector has set the latest limit on the flux of
solar $\overline\nu_e$ at Earth of $760~{\rm cm}^{-2}~{\rm s}^{-1}$, assuming a
spectral shape of the undistorted $^8$B $\nu_e$ flux
and using a threshold of 1.8~MeV
\cite{Bellini:2010gn}. This corresponds to
a 90\% C.L.\ limit on a putative $\nu_e\to\overline\nu_e$ transition
probability of $1.3\times10^{-4}$ for $E_\nu>1.8~{\rm MeV}$.

In analogy to the geoneutrinos of Sec.~\ref{sec:Geo},
the Sun contains a small fraction of the long-lived isotopes
$^{40}$K, $^{232}$Th, and $^{238}$U that produce a $\overline\nu_e$ flux
\cite{1990ApJ:767M}. However, it is immediately obvious that at the
Earth's surface, this solar flux must be much smaller than that of geoneutrinos.
If the mass fraction of these isotopes were the same in the Sun and Earth and if
their distribution in the Earth were spherically symmetric, the fluxes
would have the proportions of $M_\odot/D_\odot^2$ vs.\ $M_\oplus/R_\oplus^2$,
with the solar mass $M_\odot$, its distance $D_\odot$, the Earth mass $M_\oplus$,
and its radius $R_\oplus$. So the solar flux would be smaller in the same proportion
as the solar gravitational field is smaller at Earth,
i.e., about $6 \times10^{-4}$ times smaller.

The largest $\overline\nu_e$ flux comes from $^{40}$K decay. The solar potassium mass fraction
is around $3.5\times10^{-6}$ \cite{Asplund:2009fu},
the relative abundance of the isotope $^{40}$K is
0.012\%, so the $^{40}$K solar mass fraction is
$4\times10^{-10}$, corresponding to $8\times10^{23}~{\rm g}$
of $^{40}$K in the Sun or $1.3\times10^{46}$ atoms. With a lifetime of
$1.84\times10^9$~years, the $\overline\nu_e$ luminosity is
$2\times10^{29}~{\rm s}^{-1}$ or a flux at Earth of around $100~{\rm cm}^{-2}~{\rm s}^{-1}$.
With a geo-$\overline\nu_e$ luminosity of around $2\times10^{25}~{\rm s}^{-1}$
from potassium decay (Sec.~\ref{sec:Geo}), the average geoneutrino flux is
$5\times10^{6}~{\rm cm}^{-2}~{\rm s}^{-1}$
at the Earth's surface, although with large local variations.

An additional flux of higher-energy solar $\overline\nu_e$ comes from
photo fission of heavy elements such as uranium by the 5.5~MeV photon from the
solar fusion reaction $p+d\to{}^3{\rm He}+\gamma$ \cite{1990ApJ:767M}.
One expects a $\overline\nu_e$ spectrum similar to a power reactor,
where fission is caused by neutrons. However, this tiny flux of
around $10^{-3}~{\rm cm}^{-2}~{\rm s}^{-1}$ is vastly overshadowed by reactor $\overline\nu_e$.

\subsection{Flavor conversion}
\label{sec:FlavorConversion}

While solar neutrinos are produced as $\nu_e$, the
flux at Earth shown in Fig.~\ref{sunnuclear} (top) has a different
composition because of flavor conversion on the way out of the
Sun. The long distance between Sun and Earth relative to the
vacuum oscillation length implies that the propagation
eigenstates effectively decohere, so we can picture the neutrinos
arriving at Earth to be mass eigenstates. These can be re-projected
on interaction eigenstates, notably on $\nu_e$, if the detector is
flavor sensitive.

\begin{figure}[b]
\includegraphics[width=0.90\columnwidth]{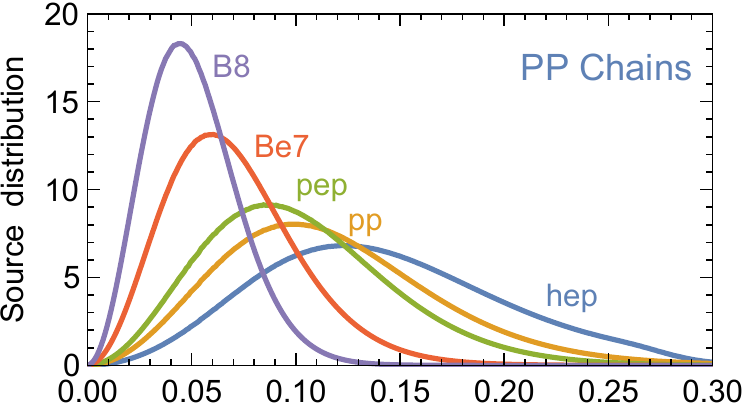}
\vskip2pt
\includegraphics[width=0.90\columnwidth]{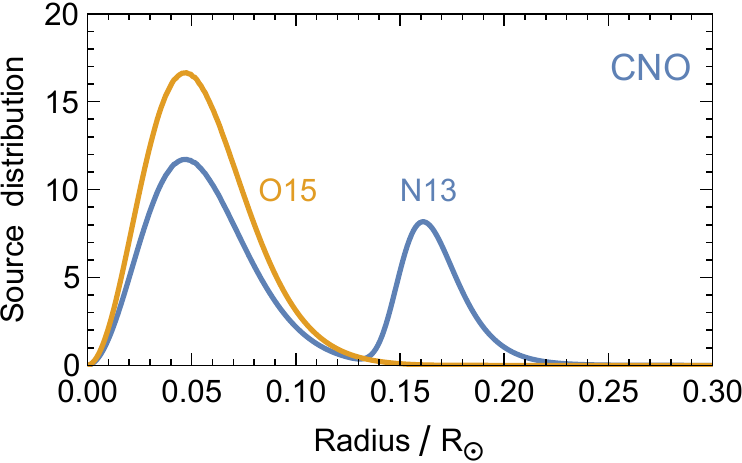}
\caption{Normalised distribution of neutrino production
  for the indicated source reactions
  according to the SSM of the Barcelona group
  \cite{Vinyoles:2016djt} with AGSS09 composition.
In the CNO cycle, the $^{17}$F distribution is very similar to that of
$^{15}$O.}\label{fig:SSM-nudis}
\end{figure}

Flavor conversion of solar neutrinos is almost perfectly adiabatic and,
because of the hierarchy of neutrino mass differences, well approximated
by an effective two-flavor treatment. The probability of a produced $\nu_e$ to
emerge at Earth in any of the three mass eigenstates is given by
Eq.~\eqref{eq:two-flavor-probabilities} and the probability to be measured
as a $\nu_e$, the survival probability, by Eq.~\eqref{eq:survival-probability}.
For the limiting case of very small $E_\nu$, the matter effect is irrelevant
and
\begin{equation}
  P_{ee}^{\rm vac}=\frac{1+\cos^22\theta_{12}}{2}\,
 \cos^4\theta_{13}+\sin^4\theta_{13}
 =0.533,
\end{equation}
corresponding to the best-fit mixing parameters in normal ordering. In the other extreme
of large energy or large matter density, one finds
\begin{equation}
  P_{ee}^{\infty}=\frac{1-\cos2\theta_{12}}{2}\,
 \cos^4\theta_{13}+\sin^4\theta_{13}
 =0.292.
\end{equation}
These extreme cases are shown as horizontal dashed lines in
the lower panel of Fig.~\ref{sunnuclear}. Otherwise,
$P_{ee}$ depends on the weak potential at the point of production,
so $P_{ee}$ for a given $E_\nu$
depends on the radial source distributions in the Sun. These are shown in
Fig.~\ref{fig:SSM-nudis} according to an AGSS09 model of the Barcelona group,
using the best-fit mixing parameters in normal ordering.
Notice that such distributions for the EC-CNO reactions have
not been provided, but would be different from the continuum
processes. The survival probabilities for the different source processes are shown in
the lower panel of Fig.~\ref{sunnuclear}. As we see from the radial distributions
of $^8{\rm B}$ and hep, the corresponding curves in
Fig.~\ref{sunnuclear} essentially bracket the range of survival probabilities
for all processes.

While neutrinos arriving at Earth have decohered into mass eigenstates, propagation
through the Earth causes flavor oscillations, producing
coherent superpositions at the far end. So if the solar flux is observed
through the Earth (``at night''), this small effect needs to be included.
This day-night asymmetry
for the $^8$B flux was measured by the Super-Kamiokande
detector to be \cite{Renshaw:2013dzu,Abe:2016nxk}
\begin{eqnarray}
  A_{\rm DN}&=&\frac{\Phi_{\rm day}-\Phi_{\rm night}}{(\Phi_{\rm day}+\Phi_{\rm night})/2}
  \nonumber\\[1ex]
  &=&\bigl(-3.3\pm1.0_{\rm stat}\pm0.5_{\rm syst}\bigr)\%\,,
\end{eqnarray}
corresponding to a $2.9\,\sigma$ significance. As measured in $\nu_e$,
the Sun shines brighter at night!

The energy-dependent $\nu_e$ survival probability for $^8$B neutrinos
shown in the lower panel of Fig.~\ref{sunnuclear} implies a spectral
deformation of the measured flux relative to the $^8$B source
spectrum. The latest Super-Kamiokande analysis \cite{Abe:2016nxk} is
consistent with this effect, but also consistent with no distortion at
all.

\subsection{Observations and detection perspectives}

Solar neutrino observations have a 50-year history, beginning in 1968
with the Homestake experiment \cite{Davis:1968cp, Cleveland:1998nv}, a
pioneering achievement that earned Raymond Davis the Physics Nobel
Prize (2002). Homestake was based on the radiochemical technique of
${}^{37}{\rm Cl}(\nu_e,e^-){}^{37}{\rm Ar}$ and subsequent argon
detection, registering approximately 800 solar $\nu_e$ in its roughly
25 years of data taking that ended in 1994. Since those early days,
many experiments have measured solar neutrinos \cite{Wurm:2017cmm},
and in particular Super-Kamiokande \cite{Abe:2016nxk}, based on
elastic scattering on electrons measured by Cherenkov radiation in
water, has registered around 80,000 events since 1996 and has thus
become sensitive to percent-level effects. The chlorine experiment was
mainly sensitive to $^8$B and $^7$Be neutrinos, whereas the lowest
threshold achieved for the water Cherenkov technique is around 4~MeV
and thus registers only $^8$B neutrinos.

Historically, the second experiment to measure solar neutrinos
(1987--1995) was Kamiokande~II and III in Japan \cite{Hirata:1989zj,
  Fukuda:1996sz}, a 2140~ton water Cherenkov detector.
Originally Kamiokande~I had been built to search for proton decay.
Before measuring solar neutrinos, however, Kamiokande
registered the neutrino burst from SN~1987A on 23 February
1987, feats which earned Masatoshi Koshiba the Physics Nobel Prize
(2002).

The lower-energy fluxes, and notably the dominant pp neutrinos, became
accessible with gallium radiochemical detectors using
$^{71}$Ga$(\nu_e, e^-){}^{71}$Ge.  GALLEX (1991--1997) and
subsequently GNO (1998--2003), using 30~tons of gallium, were mounted
in the Italian Gran Sasso National Laboratory \cite{Hampel:1998xg,
  Kaether:2010ag, Altmann:2005ix}.  The SAGE experiment in the Russian
Baksan laboratory, using 50~tons of metallic gallium, has taken data
since 1989 with results until 2007
\cite{Abdurashitov:2009tn}. However, the experiment keeps running
\cite{Shikhin:2017cim}, mainly to investigate the Gallium Anomaly, a
deficit of registered $\nu_e$ using laboratory sources
\cite{Giunti:2010zu}, with a new source experiment BEST
\cite{Barinov:2017ymq}.

A breakthrough was achieved by the Sudbury Neutrino Observatory (SNO)
in Canada \cite{Chen:1985na, Aharmim:2009gd} that took data in two
phases in the period 1999--2006. It used 1000~tons of heavy water
(D$_2$O) and was sensitive to three detection channels: (i)~Electron
scattering $\nu+e\to e+\nu$, which is dominated by $\nu_e$, but has a
small contribution from all flavors and is analogous to normal water
Cherenkov detectors. (ii)~Neutral-current dissociation of deuterons
$\nu+d\to p+n+\nu$, which is sensitive to the total flux. (iii)~Charged-current dissociation $\nu_e+d\to p+p+e$, which is sensitive to
$\nu_e$. Directly comparing the total $\nu$ flux with the
$\nu_e$ one confirmed flavor conversion, an achievement honored with
the Physics Nobel Prize (2015) for Arthur MacDonald.

Another class of experiments uses mineral oil, aug\-men\-ted with a
scintillating substance, to detect the scintillation light emitted by recoiling electrons in $\nu+e\to
e+\nu$, analogous to the detection of Cherenkov light in water.
While the scintillation light gain tends to be larger, one obtains no
significant directional information. One instrument is KamLAND, using 1000~tons of liquid scintillator, that
has taken data since 2002. It was installed in the cave of the
decommissioned Kamiokande water Cherenkov detector. Its main
achievement was to measure the $\overline\nu_e$ flux from distant power
reactors to establish flavor oscillations, it has also measured the
geoneutrino flux, and today searches for neutrinoless double beta
decay. In the solar neutrino context, it has measured the $^7$Be and
$^8$B fluxes \cite{Abe:2011em, Gando:2014wjd}.

After the question of flavor conversion has been largely settled, the focus in solar
neutrino research is precision spectroscopy, where the 300~ton liquid
scintillator detector Borexino in the
Gran Sasso Laboratory, which has taken data since 2007, plays
a leading role because of its extreme radio\-purity. It has
spectroscopically measured the pp, $^7$Be, pep and $^8$B fluxes and
has set the most restrictive constraints on the hep and CNO fluxes
\cite{Agostini:2018uly}. The detection of the latter remains
one of the main challenges in the field and might help to solve
the solar opacity problem~\cite{Cerdeno:2017xxl}.

While our paper was under review, at the
\href{https://conferences.fnal.gov/nu2020/}{Neutrino 2020} virtual conference
Borexino announced the first measurement of solar CNO neutrinos. The
flux at Earth is found to be \cite{Agostini:2020mfq}
\begin{equation}\label{eq:CNO-Borexino}
  \Phi_{\nu}=7.0^{+3.0}_{-2.0}\times 10^8~{\rm cm}^{-2}~{\rm s}^{-1}\,.
\end{equation}
This result refers to the full Sun-produced flux after including the effect of adiabatic flavor Mikheyev-Smirnov-Wolfenstein (MSW) conversion.
Comparing with the predictions shown in Table~\ref{table:sun}, after adding the C and N components there is agreement within the large experimental uncertainties. One can not yet discriminate between the different opacity cases.

Future scintillator detectors with significant solar neutrino
capabilities include the 1000~ton SNO+ \cite{Andringa:2015tza} that uses
the vessel and infrastructure of the decommissioned SNO detector,
JUNO in China \cite{An:2015jdp}, a shallow 20~kton medium-baseline
precision reactor neutrino experiment that is under construction and is meant
to measure the neutrino mass ordering, and
the proposed 4~kton Jinping neutrino experiment
\cite{JinpingNeutrinoExperimentgroup:2016nol} that would be located
deep underground in the China JinPing Laboratory (CJPL) \cite{Cheng:2018lcf}. Very
recently, the SNO+ experiment has measured the $^8$B flux during its
water commissioning phase \cite{Anderson:2018ukb}.

The largest neutrino observatory will be the approved
Hyper-Kamiokande experiment~\cite{Abe:2018uyc}, a 258~kton water Cherenkov
detector (187~kton fiducial volume), that will register
$^8$B neutrinos, threshold 4.5~MeV visible energy, with a rate of
130 solar neutrinos/day.

Other proposed experiments include THEIA~\cite{Askins:2019oqj}, which would be the
realization of the Advanced Scintillator Detector Concept
\cite{Alonso:2014fwf}. The latter would take advantage of new
developments in water based liquid scintillators and other
technological advancements, the physics case ranging from
neutrinoless double beta decay and supernova neutrinos, to beyond
standard model physics \cite{Gann:2015fba}. The liquid argon
scintillator project DUNE, to be built for long-baseline neutrino
oscillations, could also have solar neutrino
capabilities~\cite{Capozzi:2018dat}.

A remarkable new idea is to use dark-matter experiments to detect
solar neutrinos, taking advantage of coherent
neutrino scattering on large nuclei
\cite{Dutta:2019oaj}. For example, liquid argon based
WIMP direct detection experiments could be competitive in the
detection of CNO neutrinos \cite{Cerdeno:2017xxl}.

\section{Thermal Neutrinos from the Sun}
\label{sec:Thermal-Solar}

In the keV-range, the Sun produces neutrino pairs of all flavors by thermal
processes, notably plasmon decay, the Compton process, and electron
bremsstrahlung. This contribution to the GUNS has never been shown,
perhaps because no realistic detection opportunities exist at
present. Still, this is the dominant $\nu$ and $\overline\nu$ flux at
Earth for $E_\nu\alt 4$~keV. A future measurement would
carry information about the solar chemical composition.

\subsection{Emission processes}

Hydrogen-burning stars produce neutrinos effectively by
$4p+2e\to{}^4{\rm He}+2\nu_e$.
These traditional solar neutrinos were discussed in
Sec.~\ref{sec:Solar-Nuclear}, where we also discussed
details about standard solar models.
Moreover, all stars produce neutrino
pairs by thermal processes, providing an energy-loss channel that dominates in
advanced phases of stellar evolution
\cite{Clayton:1983, Kippenhahn:2012, Bahcall:1989ks, Raffelt:1996wa},
whereas for the Sun it is a minor effect.
The main processes are plasmon decay ($\gamma\to\overline\nu+\nu$), the
Compton process ($\gamma+e\to e+\overline\nu+\nu$), bremsstrahlung
($e+Ze\to Ze+e+\overline\nu+\nu$), and atomic free-bound and bound-bound
transitions.  Numerical routines exist to couple neutrino energy
losses with stellar evolution codes \cite{Itoh:1996}.
A detailed evaluation of these processes for the Sun, including spectral
information, was recently provided
\cite{Haxton:2000xb,Vitagliano:2017odj}. While
traditional solar neutrinos have MeV energies as behooves
nuclear processes, thermal neutrinos have keV energies, corresponding
to the temperature in the solar core.

Low-energy neutrino pairs are emitted by electrons, where we can use
an effective neutral-current interaction of the form
\begin{equation}
  {\cal L}_{\rm int}=\frac{G_{\rm F}}{\sqrt{2}}\,
  \bar\psi_e\gamma^\mu(C_{\rm V}-C_{\rm A}\gamma_5)\psi_e\,
  \bar\psi_\nu\gamma_\mu(1-\gamma_5)\psi_\nu\,.
\end{equation}
Here $G_{\rm F}$ is Fermi's constant and the vector and axial-vector
coupling constants are \smash{$C_{\rm
    V}=\frac{1}{2}(4\sin^2\Theta_{\rm W}\pm1)$} and \smash{$C_{\rm
    A}=\pm\frac{1}{2}$ for $\nu_e$} (upper sign) and $\nu_{\mu,\tau}$
(lower sign). The flavor dependence
derives from $W^\pm$ exchange in the effective $e$--$\nu_e$ interaction.

In the nonrelativistic limit, the emission rates for all
processes are proportional to $(a C_{\rm V}^2+bC_{\rm A}^2)G_{\rm
  F}^2$, where the coefficients $a$ and $b$ depend on the process, but
always without a mixed term proportional to $C_{\rm V}C_{\rm A}$. This
is a consequence of the nonrelativistic limit and implies that the
flux and spectrum of $\nu$ and $\overline\nu$ are the same.  Moreover,
while $C_{\rm A}^2$ is the same for all flavors, the peculiar value
$4\sin^2\Theta_{\rm W}=0.92488$ of the weak mixing angle implies that
$C_{\rm V}^2=0.0014$ for $\nu_{\mu,\tau}$. So the vector-current
interaction produces almost exclusively $\nu_e\overline\nu_e$
pairs and the thermal flux shows a strong flavor
dependence.

The emission rates involve complications caused by in-medium effects
such as screening in bremsstrahlung or electron-electron correlations
in the Compton process. One can take advantage of the solar
opacity calculations because the structure functions relevant for
photon absorption carry over to neutrino processes
\cite{Vitagliano:2017odj}.  The overall precision of the thermal
fluxes is probably on the 10\% level, but a precise error budget
is not available. Notice also that the solar opacity problem discussed
in Sec.~\ref{sec:Solar-Nuclear} shows that on the precision level
there remain open issues in our understanding of the Sun, probably
related to the opacities or metal abundances, that may also affect
thermal neutrino emission.

\subsection{Solar flux at Earth}

\begin{figure}
\vskip4pt
\includegraphics[width=0.88\columnwidth]{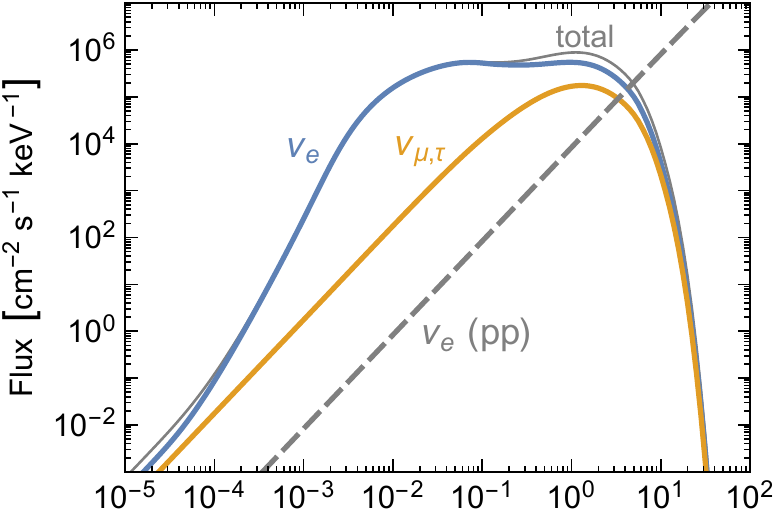}
\vskip2pt
\includegraphics[width=0.88\columnwidth]{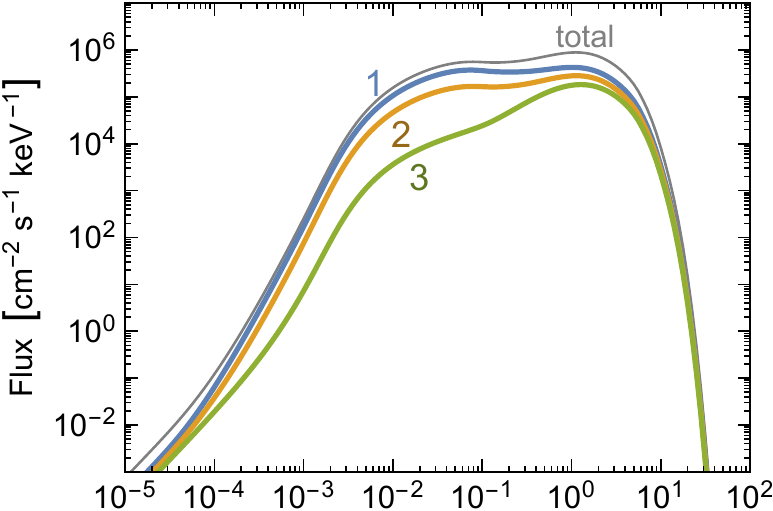}
\includegraphics[width=0.88\columnwidth]{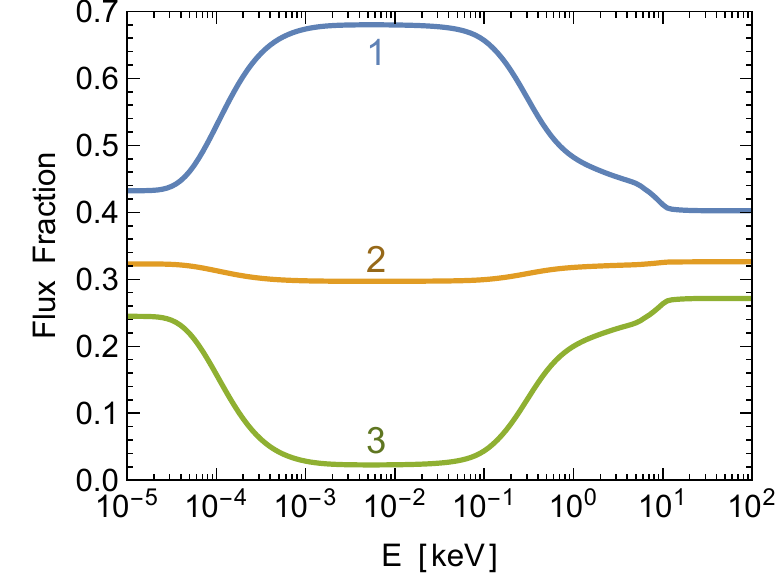}
\caption{Solar neutrino flux at Earth from thermal processes \cite{Vitagliano:2017odj}.
  The antineutrino flux is the same.
  {\em Top:\/} Flavor-eigenstate fluxes in the
  absence of oscillations.  For comparison, we also show the
  low-energy tail from the nuclear pp reaction (dashed) which produces
  only $\nu_e$. {\em Middle:\/} Mass-eigenstate
  fluxes for $\nu_i$ with $i=1$, 2 or 3 as indicated.
  This is the relevant representation for the fluxes arriving at Earth.
  {\em Bottom:\/} Fractional mass-eigenstate fluxes.}
\label{fig:Thermal-Solar}
\end{figure}

Integrating the emission rates over the Sun provides the flux at Earth
shown in Fig.~\ref{fig:Thermal-Solar}, where the exact choice of solar
model is not important in view of the overall uncertainties. In the
top panel, we show the flavor fluxes for unmixed neutrinos.
The contribution from individual processes was discussed in detail
by \textcite{Vitagliano:2017odj}. The
non-electron flavors are produced primarily by bremsstrahlung,
although Compton dominates at the highest energies. For
$\nu_e\overline\nu_e$, plasmon decay dominates, especially at lower
energies. An additional source of $\nu_e$ derives from the nuclear pp
process discussed in Sec.~\ref{sec:Solar-Nuclear} which we show as a
dashed line given by Eq.~\eqref{eq:low-pp-flux}. For $E_\nu\alt
4$~keV, thermal neutrinos vastly dominate, and they always dominate
for antineutrinos, overshadowing other astrophysical sources,
e.g.\ primordial black holes decaying via Hawking
radiation~\cite{Lunardini:2019zob}.

Solar neutrinos arriving at Earth have decohered into mass
eigenstates. They are produced in the solar interior, where according
to Eq.~\eqref{eq:Ve} the weak potential caused by electrons
is $V_e=\sqrt{2}G_{\rm F}n_e\sim 7.8\times10^{-12}~{\rm eV}$ near the solar center.
Comparing with $\delta m^2/2E=3.7\times10^{-8}~{\rm eV}~({\rm keV}/E)$
reveals that the matter effect is negligible for sub-keV neutrinos, in agreement with the discussion in
Appendix~\ref{sec:MatterMixing}. Therefore, we can use the
vacuum probabilities for a given flavor neutrino to be found in any of
the mass eigenstates.

Specifically, from the top row of Eq.~\eqref{eq:probability-matrix}, we use the best-fit probabilities for a
$\nu_e$ or $\overline\nu_e$ to show up in a given mass eigenstates to be
$P_{e1}=0.681$, $P_{e2}=0.297$, and $P_{e3}=0.022$, which add up to
unity.  These probabilities apply to vector-current processes
which produce almost pure $\nu_e\overline\nu_e$,
whereas the axial-current processes, with equal $C_{\rm A}^2$ for all
flavors, can be thought of as producing an equal mixture
of pairs of mass eigenstates. In this way we find the mass-eigenstate fluxes at
Earth shown in the middle panel of Fig.~\ref{fig:Thermal-Solar} and
the corresponding fractional fluxes in the bottom panel.

Integrating over energies implies a total flux, number density, and
energy density at Earth of neutrinos plus antineutrinos of all flavors
\begin{subequations}
  \begin{eqnarray}
    \Phi_{\nu\overline\nu} &=& 6.2\times10^{6}~{\rm cm}^{-2}~{\rm s}^{-1},\\
    n_{\nu\overline\nu}    &=& 2.1\times10^{-4}~{\rm cm}^{-3},\\
    \rho_{\nu\overline\nu} &=& 507~{\rm meV}~{\rm cm}^{-3},
  \end{eqnarray}
\end{subequations}
implying $\langle E_\nu\rangle=\langle E_{\overline\nu}\rangle=2.46~{\rm keV}$.
In analogy to traditional solar neutrinos,
the flux and density changes by $\pm3.4\%$ over the year due to the
ellipticity of the Earth orbit. The local energy density in thermal
solar neutrinos is comparable to the energy density of the CMB for
massless cosmic neutrinos.

\subsection{Very low energies}
\label{sec:SolarVeryLow}

Thermal solar neutrinos appear to be the dominant
flux at Earth for sub-keV energies all the way down to the CNB and the
BBN neutrinos discussed in Secs.~\ref{sec:CNB}
and~\ref{sec:BBN}. Therefore, it is useful to consider the asymptotic
behavior at very low energies. For $E\alt 100$~meV, bremsstrahlung
emission dominates which generically scales as $E^2$ at low energies
\cite{Vitagliano:2017odj}. A numerical integration over the Sun
provides the low-energy flux at Earth from bremsstrahlung for either $\nu$ or
$\overline\nu$
\begin{equation}\label{eq:sol-lowE-flux}
  \frac{d\Phi_{\nu}}{dE}=\frac{d\Phi_{\overline\nu}}{dE}=
  \frac{7.4\times10^{6}}{{\rm cm}^{2}~{\rm s}~{\rm keV}}\,
  \left(\frac{E}{{\rm keV}}\right)^2.
\end{equation}%
The fractions in the mass eigenstates 1, 2, and 3 are 0.432, 0.323, and 0.245,
corresponding to the low-energy plateau in the bottom panel of Fig.~\ref{fig:Thermal-Solar} and that were already shown in the BBN-context in Eq.~\eqref{eq:Sun-prob}.
As explained by \textcite{Vitagliano:2017odj}, bremsstrahlung produces an almost pure $\nu_e\bar\nu_e$ flux by the vector-current interaction which breaks down into mass eigenstates according to the vacuum-mixing probabilities given in the top row of Eq.~\eqref{eq:probability-matrix}. Moreover, bremsstrahlung produces all flavors equally by the axial-vector interaction.  Adding the vector (28.4\%) and axial-vector (71.6\%) contributions provides these numbers.

One consequence of the relatively small brems\-strah\-lung flux is that there is indeed a window of energies between the CNB and very-low-energy solar neutrinos where the BBN flux of Sec.~\ref{sec:BBN} dominates.

Of course, for energies so low that the emitted
neutrinos are not relativistic, this result needs to be modified. For
bremsstrahlung, the emission spectrum is determined
by phase space, so $d\dot n=A\,p^2 dp=A\, p E\, dE$ with $A$ some constant.
For the flux of emitted neutrinos, we need a velocity factor $v=p/E$,
so overall $d\Phi_{\nu}/dE\propto (E^2-m_\nu^2)$ for $E\geq m_\nu$ and
zero otherwise. The local density at Earth, on the other hand,
does not involve $p/E$ and so is $dn_\nu/dE\propto E\sqrt{E^2-m_\nu^2}$.

\section{Geoneutrinos}
\label{sec:Geo}

The decay of long-lived natural radioactive isotopes in the Earth,
notably $^{238}$U, $^{232}$Th and $^{40}$K, produce an MeV-range
$\overline\nu_e$ flux exceeding $10^{25}~{\rm s}^{-1}$
\cite{Marx1960, Marx1969, 1966NucPh657E, Krauss:1983zn, Fiorentini:2007te, Ludhova:2013hna, Bellini:2013wsa, Dye:2011mc, Smirnov:2019}.
As these ``geoneutrinos'' are actually antineutrinos they can be
detected despite the large solar neutrino flux in the same energy
range.  The associated radiogenic heat production is what drives much
of geological activity such as plate tectonics or vulcanism. The
abundance of radioactive elements depends on location, in principle
allowing one to study the Earth's interior with neutrinos,\footnote{See, for example,
  a dedicated conference series on Neutrino Geoscience
  \href{http://www.ipgp.fr/en/evenements/neutrino-geoscience-2015-conference}
       {http://www.ipgp.fr/en/evenements/neutrino-geoscience-2015-conference}
or Neutrino Research and Thermal Evolution of the Earth,
Tohoku University, Sendai, October 25--27, 2016
\href{https://www.tfc.tohoku.ac.jp/event/4131.html}{https://www.tfc.tohoku.ac.jp/event/4131.html}.}
although existing measurements by KamLAND and Borexino remain sparse.

\subsection{Production mechanisms}

Geoneutrinos are primarily $\bar\nu_e$
produced in decays of radioactive elements with lifetime comparable to the age of the Earth, the so-called heat producing elements (HPEs)~\cite{Ludhova:2013hna,Smirnov:2019}. Geoneutrinos carry information on the HPE abundance and distribution
and constrain the fraction of radiogenic heat contributing to the total surface heat flux
of 50~TW.
In this way, they provide indirect information on plate tectonics, mantle convection, magnetic-field generation, as well as the processes that led to the Earth formation~\cite{Ludhova:2013hna,Bellini:2013wsa}.

Around 99\% of radiogenic heat comes from the decay chains of $^{232}$Th, $^{238}$U and $^{40}$K. The main reactions are~\cite{Fiorentini:2007te}
\begin{subequations}\label{eq:geo-reactions}
\begin{eqnarray}
\kern-2em
^{238}\mathrm{U} &\rightarrow& ^{206}\mathrm{Pb} + 8\alpha + 8e + 6\overline\nu_e + 51.7\ \mathrm{MeV},
\\
\kern-2em
^{232}\mathrm{Th} &\rightarrow& ^{208}\mathrm{Pb} + 6\alpha + 4e + 4\overline\nu_e + 42.7\ \mathrm{MeV},
\\
\kern-2em
^{40}\mathrm{K} &\rightarrow& ^{40}\mathrm{Ca} + e + \overline\nu_e + 1.31\ \mathrm{MeV}\ (89.3\%),
\\
\kern-2em
e+{}^{40}\mathrm{K}  &\rightarrow& ^{40}\mathrm{Ar}^* +\nu_e + 0.044\ \mathrm{MeV}\ (10.7\%).
\end{eqnarray}
\end{subequations}
The contribution from $^{235}$U is not shown because of its small
isotopic abundance.
Electron capture on potassium is the only notable $\nu_e$ component, producing a monochromatic 44~keV line.
Notice that it is followed by the emission of a 1441~keV $\gamma$-ray to the ground state
of $^{40}{\rm Ar}$. For the
other reactions in Eq.~\eqref{eq:geo-reactions} the average energy release in neutrinos
is 3.96, 2.23 and 0.724~MeV per decay respectively~\cite{Enomoto:2005}, while the remainder of the
reaction energy shows up as heat.
An additional 1\% of the radiogenic heat comes from decays of $^{87}$Rb, $^{138}$La and $^{176}$Lu~\cite{Ludhova:2013hna}.

\begin{figure}
\includegraphics[width=0.90\columnwidth]{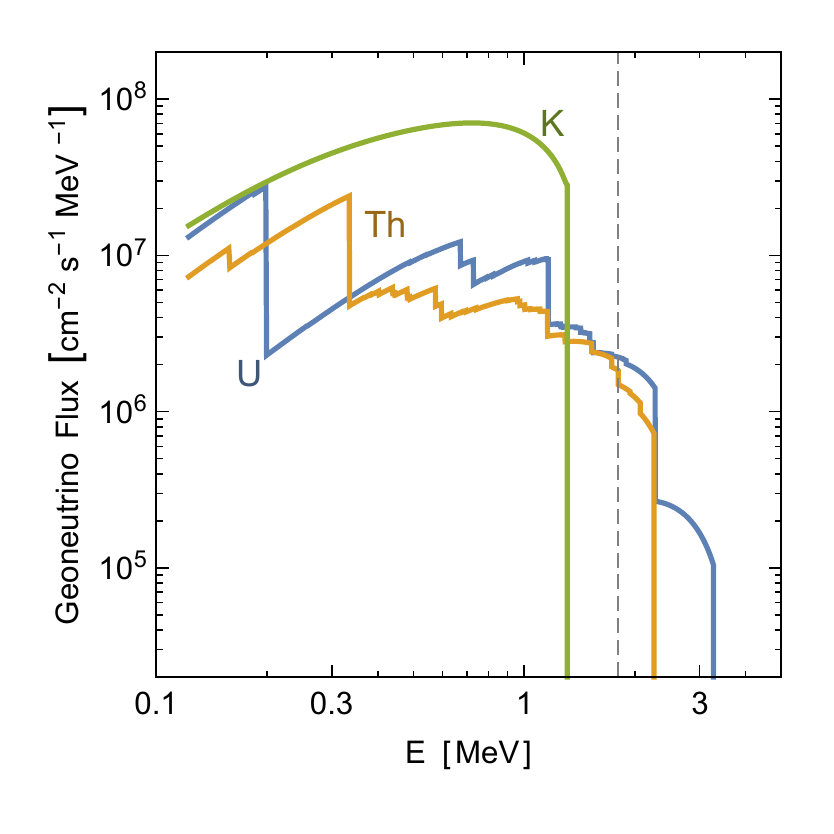}
\caption{Average geoneutrino flux ($\overline\nu_e$ before flavor conversion) from the main production chains, assuming
the BSE model of the Earth \cite{Enomoto:2005}. The vertical dashed line marks
the threshold of 1.806~MeV for inverse beta decay.}\label{geonus}
\end{figure}

The geoneutrino spectra produced in these reactions, extending up to 3.26~MeV~\cite{Ludhova:2013hna}, depend on the possible decay branches
and are shown in Fig.~\ref{geonus}. The main detection channel is inverse beta decay
$\overline\nu_e+p\to n+e^+$ with a kinematical threshold of 1.806~MeV (vertical dashed line in Fig.~\ref{geonus}), implying that
the large flux from $^{40}$K is not detectable~\cite{Bellini:2013wsa}.
On the other hand, a large fraction of the heat arises from the uranium and thorium
decay chains. The
resulting average flux is $\Phi_{\overline\nu_e} \simeq 2 \times 10^6~{\rm cm}^{-2}~{\rm s}^{-1}$,
comparable with the solar $\nu_e$ flux
from $^8$B decay. However, detecting geoneutrinos is more challenging
because of their smaller energies.

The differential $\overline\nu_e$ geoneutrino flux at position $\vec{r}$ on Earth is given by the isotope abundances $a_i(\rprime)$ for any isotope $i$ at the position $\rprime$ and integrating over the entire Earth
provides the expression~\cite{Fiorentini:2007te,Smirnov:2019}
\begin{equation}
\label{eq:geonu}
\Phi_{\overline\nu_e}(E, \vec{r}) = \sum_i A_i \frac{dn_i}{dE}\!\int_\oplus\!\! d^3 \rprime
\frac{a_i(\rprime) \rho(\rprime) P_{ee}(E, |\vec{r}-\rprime|)}{4 \pi |\vec{r}-\rprime|^2}.
\end{equation}
Here $dn_i/dE$ is the $\overline\nu_e$ energy spectrum for each decay mode, $A_i$ the decay rate per unit mass,
$\rho(\vec{r})$ the rock density, and $P_{ee}$ the $\overline\nu_e$ survival probability, where we
have neglected matter effects so that $P_{ee}$ depends only on the distance between production
and detection points.

To evaluate this expression one needs to know the absolute amount and distribution of HPEs. Although the crust composition is relatively well known, the mantle composition is quite uncertain~\cite{Fiorentini:2007te,Bellini:2013wsa}. Usually, the signal from HPEs in the crust is computed on the basis of the total amount of HPEs coming from the bulk silicate Earth (BSE) model, i.e., the model describing the  Earth region outside its metallic core~\cite{MCDONOUGH1995223,2003TrGeo...2....1P}; then the corresponding amount of elements in the mantle is extrapolated. The content of elements in the Earth mantle can be estimated on the basis of cosmochemical arguments, implying that abundances in the deep layers are expected to be larger than the ones in the upper samples.

\begin{figure}[htb]
\vskip4pt
\includegraphics[width=1.0\columnwidth]{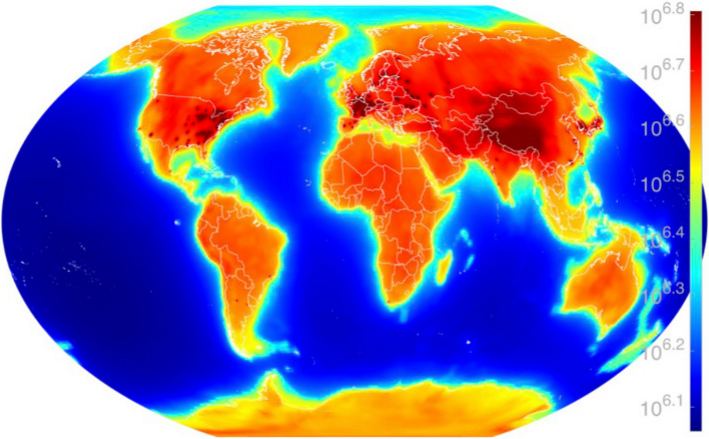}
\caption{Global map of the expected $\overline\nu_e$ flux
(in ${\rm cm}^{-2}~{\rm s}^{-1}$) of all energies
from U and Th decays in the Earth and from reactors.
The hot spots in the Eastern US, Europe and Japan are caused
by nuclear power reactors. Credits: Antineutrino Global Map 2015
\cite{Usman:2015yda} licensed under Creative Commons Attribution 4.0 International License.}\label{geonumap}
\end{figure}

Given their chemical affinity, the majority of HPEs are in the continental crust. This is useful as most of the detectors sensitive to geoneutrinos are on the continents and the corresponding event rate is dominated by the Earth contribution. Usually the continental crust is further divided in upper, lower and middle continental crust. Among existing detectors, Borexino is placed on the continental crust in Italy~\cite{doi:10.1142/S0217751X18430091, Agostini:2019dbs}, while KamLAND is in a complex geological structure around the subduction zone~\cite{Gando:2013nba,AG7388}. An example
for a global map of the expected $\overline\nu_e$ flux is shown in Fig.~\ref{geonumap}.

\subsection{Earth modeling}

The Earth was created by accretion from undifferentiated material. Chondritic meteorites seem to resemble this picture as for composition and structure. The Earth can be divided in five regions according to seismic data: core, mantle, crust (continental and oceanic), and sediment. The mantle is solid, but is affected by convection that causes plate tectonics and earthquakes~\cite{Ludhova:2013hna}.

Seismology has shown that the Earth is divided into several layers that can be distinguished by sound-speed discontinuities. Although seismology allows us to reconstruct the density profile, it cannot determine the composition. The basic structure of the Earth's interior is defined by the one-dimensional seismological profile dubbed Preliminary Reference Earth Model (PREM), which is the basis for the estimation of geoneutrino production in the mantle~\cite{DZIEWONSKI1981297}. Meanwhile, thanks to seismic tomography, a three-dimensional view of the mantle structure has become available, for example \textcite{2012EGUGA..14.3743L} and \textcite{doi:10.1002/2013JB010626}, but differences with respect to the 1D PREM are negligible for geoneutrino estimation \cite{Fiorentini:2007te}.

As discussed in the previous section, uranium and thorium are the main HPEs producing detectable geoneutrinos. After the metallic core of the Earth separated, the rest of the Earth consisted of a homogeneous primitive mantle mainly composed of silicate rocks that then led to the formation of the present mantle and crust.

The outer layer is a thin crust which accounts for 70\% of geoneutrino production~\cite{Fiorentini:2007te,articlesramek}. The crust probably hosts about half of the total uranium. The lithophile elements (uranium and thorium) tend to cluster in liquid phase and therefore concentrate in the crust, which is either oceanic or continental~\cite{Enomoto:2005}. The former is young and less than 10~km thick. The latter is thicker, more heterogeneous, and older that the oceanic counterpart. The crust is vertically stratified in terms of its chemical composition and is heterogeneous. The HPEs are distributed both in the crust and mantle. The geoneutrino flux strongly depends on location. In particular, the continental crust is about one order of magnitude richer in HPEs than the oceanic one.
The continental crust is 0.34\% of the Earth's mass, but contains $40\%$ of the U and Th budget~\cite{Bellini:2013wsa,doi:10.1002/ggge.20129}.

The mantle, which consists of pressurized rocks at high temperature,
can be divided in upper and lower mantle \cite{Fiorentini:2007te}. However, seismic discontinuities between the two parts do not divide the mantle into layers. We do not know whether the mantle moves as single or multiple layers, its convection dynamics, and whether its composition is homogeneous or heterogeneous. The available data are scarce and are restricted to the uppermost part.

Two models have been proposed~\cite{articlehofmann}. One is a two-layer model with a demarcation surface and a complete insulation between the upper mantle (poor in HPEs) and the lower layer. Another one is a fully mixed model, which is favored by seismic tomography. Concerning the estimation of the related geoneutrino flux, both models foresee the same amount of HPEs, but with different geometrical distributions~\cite{Mantovani:2003yd,2005hep.ph....8049E}. In the following, we assume a homogeneous distribution of U and Th in the mantle. Geophysicists have proposed models of mantle convection predicting that 70\% of the total surface heat flux is radiogenic. Geochemists estimate this figure to be 25\%; so the spread is large~\cite{Bellini:2013wsa,Meroni:2016kmz}.

The Earth's innermost part is the core, which accounts for 32\% of the Earth's mass, and is made of iron with small amounts of nickel~\cite{Fiorentini:2007te}. Because of their chemical affinity, U and Th are believed to be absent in the core.

BSE models adopted to estimate the geoneutrino flux fall into three classes: geochemical, geodynamical, and cosmochemical \cite{Ludhova:2013hna}. Geochemical models are based on the fact that the composition of carbonaceous chondrites matches the solar photospheric abundances of refractory lithophile and siderophile elements. A typical bulk-mass Th/U ratio is 3.9. Geodynamical models look at the amount of HPEs needed to sustain mantle convection. Cosmochemical models are similar to geochemical ones, but assume a mantle composition based on enstatite chondrites and yield a lower radiogenic abundance.

A reference BSE model to estimate the geoneutrino production is the starting point for studying the expectations and potential of various neutrino detectors. It should incorporate the best available geochemical and geophysical information. The geoneutrino flux strongly depends on location, so the global map shown in Fig.~\ref{geonumap} is only representative.
It includes the geoneutrino flux from the U and Th decay chains as well as the reactor neutrino flux~\cite{Usman:2015yda}.

A measurement of the geoneutrino flux could be used to estimate our planet's radiogenic heat production and to constrain the composition of the BSE model. A leading BSE model~\cite{MCDONOUGH1995223} predicts a radiogenic heat production of 8~TW from $^{238}\mathrm{U}$, 8~TW from $^{232}\mathrm{Th}$, and 4~TW from $^{40}\mathrm{K}$, together about half the heat dissipation rate from the Earth's surface.
According to measurements in chondritic meteorites, the concentration mass ratio Th/U is~3.9. Currently, the uncertainties on the neutrino fluxes are as large as the predicted values.

The neutrino event rate is often expressed in Terrestrial Neutrino Units (TNUs), i.e., the number of interactions detected in a target of $10^{32}$ protons (roughly correspondent to 1~kton of liquid scintillator) in one year with maximum efficiency~\cite{Ludhova:2013hna}. So the neutrino event rates can be expressed as
\begin{subequations}
\begin{eqnarray}
 {}^{232}\mathrm{Th} &\quad& S= 4.07~\mathrm{TNU}\times \Phi_{\overline\nu_e}\big/10^6\,{\rm cm}^{-2}\,{\rm s}^{-1}
 \\
 {}^{238}\mathrm{U~} &\quad& S= 12.8~\mathrm{TNU}\times \Phi_{\overline\nu_e}\big/10^6\,{\rm cm}^{-2}\,{\rm s}^{-1}
\end{eqnarray}
\end{subequations}
for the thorium and uranium decay chains.

\subsection{Detection opportunities}

Geoneutrinos were first considered in 1953 to explain a puzzling background in the Hanford reactor neutrino experiment of Reines and Cowan, but even Reines' generous estimate of $10^8~{\rm cm}^{-2}~{\rm s}^{-1}$ fell far short (the real explanation turned out to be cosmic radiation).\footnote{See \textcite{Fiorentini:2007te} for a reproduction of the private exchange between G.~Gamow and F.~Reines.} First realistic geoneutrino estimates appeared in the 1960s by \textcite{Marx1960} and \textcite{Marx1969} and independently by \textcite{1966NucPh657E}, to be followed in the 1980s by \textcite{Krauss:1983zn}.

The first experiment to report geoneutrino detection was KamLAND in 2005, a 1000~t liquid scintillator detector in the Kamioka mine \cite{Araki:2005qa}. The detection channel is inverse beta decay (IBD), $\bar\nu_e+p\to n+e^+$, using delayed coincidence between the prompt positron and a delayed $\gamma$ from neutron capture. The results were based on 749.1 live days, corresponding to an exposure of $(0.709\pm0.035)\times10^{32}~{\rm protons}\times{\rm year}$, providing 152 IBD candidates, of which
$25^{+19}_{-18}$ were attributed to \hbox{geo-$\bar\nu_e$}. This signal corresponds to about one geoneutrino per month to be distinguished from a background that is five times larger. About $80.4\pm7.2$ of the background events were attributed to the $\bar\nu_e$ flux from nearby nuclear reactors --- KamLAND was originally devised to detect flavor oscillations of reactor neutrinos (Sec.~\ref{sec:Reactors}).

Over the years, the detector was improved, notably by background reduction through liquid-scintillator purification. A dramatic change was the shut-down of the Japanese nuclear power reactors in 2011 following the Fukushima Daiichi nuclear disaster (March 2011). For KamLAND this implied a reactor-off measurement of backgrounds and geoneutrinos that was included in the latest published results, based on data taken between March 9, 2002 and November 20, 2012 \cite{Gando:2013nba}. Preliminary results from data taken up to 2016, including 3.5 years of a low-reactor period (and of this 2.0 years reactor-off), were shown at a conference in October 2016 \cite{Watanabe:2016} and also reported in a recent review article~\cite{Smirnov:2019}. We summarize these latest available measurements in Table~\ref{table:geonu}.

\begin{table}
\vskip-4pt
\caption{Geoneutrino observations. The Th/U abundance ratio is assumed to be the chondritic value.}
\vskip4pt
\begin{tabular*}{\columnwidth}{@{\extracolsep{\fill}}lll}
\hline
\hline
&KamLAND\footnotemark[1]&Borexino\footnotemark[2]\\
\hline
Period   &2002--2016&2007--2019\\
Live days&3900.9    &3262.74\\
Exposure &$6.39$&$1.29\pm0.05$\\
\quad($10^{32}$~protons$\times$year)\\
IBD Candidates&1130&154\\
Reactor $\bar\nu_e$&$618.9\pm33.8$&$92.5^{+12.2}_{-9.9}$\\
Geoneutrinos (68\% C.L.)\\
\quad Number of $\bar\nu_e$&139--192&43.6--62.2\\
\quad Signal (TNU)&29.5--40.9&38.9--55.6\\
\quad Flux ($10^6~{\rm cm}^{-2}~{\rm s}^{-1}$)&3.3--4.6&4.8--6.2\\
\hline
\hline
\end{tabular*}\label{table:geonu}
\footnotetext[1]{\textcite{Watanabe:2016}}
\footnotetext[2]{\textcite{Agostini:2019dbs}}
\end{table}

A second experiment that has detected geoneutrinos is Borexino, a 300~t liquid scintillator detector in the Gran Sasso National Laboratory in Italy, reporting first results in 2010 \cite{doi:10.1142/S0217751X18430091}. Despite its smaller size, Borexino is competitive because of its scintillator purity, large underground depth, and large distance from nuclear power plants, effects that all help to reduce backgrounds. Comprehensive results for the data-taking period December 2007--April 2019 were recently published \cite{Agostini:2019dbs} and are summarized in Table~\ref{table:geonu}.

Reactor neutrinos (Sec.~\ref{sec:Reactors}) are the main background for geoneutrino detection, whereas atmospheric neutrinos and the diffuse supernova neutrino background (Sec.~\ref{sec:DSNB}) are negligible. Other spurious signals include intrinsic detector contamination, cosmogenic sources, and random coincidences of non-correlated events. While the reactor flux at Borexino is usually much smaller than at KamLAND, the shut-down of the Japanese reactors has changed this picture for around 1/3 of the KamLAND live period. From Table~\ref{table:geonu} we conclude that at Borexino, the reactor signal was around 1.7 times the geoneutrino signal, whereas at KamLAND this factor was on average around~3.7. Any of these $\bar\nu_e$ measurements refer to the respective detector sites and include the effect of flavor conversion on the way between source and detector.

The overall results in Table~\ref{table:geonu} assume a Th/U abundance ratio fixed at the chondritic value of 3.9, but both Borexino and KamLAND provide analyses with independent contributions. For example, KamLAND finds for the ratio $4.1^{+5.5}_{-3.3}$, consistent with the simplest assumption but with large uncertainties \cite{Watanabe:2016}. KamLAND is also beginning to discriminate between the geodynamical, geochemical and cosmochemical BSE models, somewhat disfavoring the latter. The radiogenic heat production is allowed to be roughly in the range 8--30~TW.
Borexino finds a measured mantle signal of $21.2^{+9.5}_{-9.0}({\rm stat})^{+1.1}_{-0.9}({\rm sys})$~TNU, corresponding to the production of a radiogenic heat of
$24.6^{+11.1}_{-10.4}$~TW (68\% interval) from $^{238}$U and $^{232}$Th in the mantle.
Assuming 18\% contribution of $^{40}$K in the mantle and $8.1^{+1.9}_{-1.4}$~TW of the total radiogenic heat of the lithosphere, Borexino estimates the total radiogenic heat of the Earth to be
$38.2^{+13.6}_{-12.7}$~TW \cite{Agostini:2019dbs}. Overall, while the observation of geoneutrinos in these two detectors is highly significant, detailed geophysical conclusions ultimately require better statistics.

Several experiments, in different stages of development, will improve our knowledge. For example, SNO+ in Canada expects a geoneutrino rate of 20/year \cite{Arushanova:2017jbk}. The site is in the old continental crust containing felsic rocks which are rich in U and Th. The crust at the SNO+ location is especially thick, about 40\% more than at Gran Sasso and Kamioka. JUNO in China also plans to measure geoneutrinos. Finally,  the  Hanohano project has been proposed in Hawaii, a 5~kton detector on the oceanic crust \cite{Cicenas:2012cta}. Because the oceanic crust is thin, 75\% of the signal would come from the mantle.

\section{Reactor Neutrinos}
\label{sec:Reactors}

Nuclear power plants release a few percent of their energy production in the form of MeV-range $\overline\nu_e$ arising from the decay of fission products. In contrast to other human-made neutrino sources such as accelerators, reactors produce a diffuse flux that can dominate over geoneutrinos in entire geographic regions and are therefore a legitimate GUNS component. Historically, reactor neutrinos were fundamental to the study of neutrino properties, including their very first detection by Cowan and Reines in the 1950s, and they remain topical for measuring neutrino mixing parameters and possible sterile states \cite{Giunti:2007ry,Tanabashi:2018,Qian:2018wid}, as well as for many applied fields~\cite{Bergevin:2019tcg}.

\subsection{Production and detection of reactor neutrinos}

Nuclear power plants produce $\overline\nu_e$'s through $\beta$-decays of neutron-rich nuclei. The main contributions come from the fission of $^{235}\rm U$ (56\%), $^{239}\rm Pu$ (30\%), $^{238}\rm U$ (8\%) and $^{241}\rm Pu$ (6\%), where the percentages vary over time and we reported typical values of fission fractions during operation \cite{Giunti:2007ry,baldonciniat}. In addition, below the detection threshold of inverse beta decay $E_{\rm min}=1.8~\rm MeV$ there is another $\overline\nu_e$ source due to neutron captures. The most important is the decay of $^{239}\rm U$ produced by the neutron capture on $^{238}\mathrm{U}$, which is usually written as $^{238}\mathrm{U}(n,\gamma)^{239}\rm U$. To obtain a basic estimate of the $\overline\nu_e$ flux from a reactor we note that on average a fission event releases about 6~neutrinos and a total energy of about $200~\rm MeV$. A nuclear power plant producing $1~\rm GW$ of thermal power will then produce a $\overline\nu_e$ flux of $2\times 10^{20}~\rm s^{-1}$~\cite{Giunti:2007ry}.

The globally installed nuclear power corresponds to around 0.4~TW electric\footnote{International Atomic Energy Agency (IAEA), Power Reactor Information System (PRIS),   \href{https://pris.iaea.org/PRIS/}{https://pris.iaea.org/PRIS/}} or, with a typical efficiency of 33\%,\footnote{Thermal Efficiency of Nuclear Power Plants, see for example
  \href{https://www.nuclear-power.net/nuclear-engineering /thermodynamics/laws-of-thermodynamics/thermal-efficiency/thermal-efficiency-of-nuclear-power-plants/}
  {https://www.nuclear-power.net/nuclear-engineering/thermodyn amics/laws-of-thermodynamics/thermal-efficiency/thermal-effici ency-of-nuclear-power-plants/}}
to 1.2~TW thermal. This is a few percent of the natural radiogenic heat production in the Earth, so the overall reactor neutrino flux is only a few percent of the geoneutrino flux, yet the former can dominate in some geographic regions and has a different spectrum.

Predicting the spectrum is a much more complicated task as many different decay branches must be included. In the last 50~years, two main approaches were used. One method predicts the time dependent total flux by summing over all possible $\beta$-decay branches, but the spectrum is very uncertain because the fission yields and endpoint energies are often not well known, one needs a good model for the Coulomb corrections entering the Fermi function, and so forth. The alternative is to use the measured electron spectrum for different decay chains, which can be inverted taking advantage of the relation
\begin{equation}
E_\nu=E_e+T_n+m_n-m_p\simeq E_e+1.293~\rm MeV\,,
\end{equation}
where $T_n$ is the small recoil kinetic energy of the neutron and $E_e$ is the energy of the outgoing positron.

\begin{figure}[htb]
\includegraphics[width=0.90\columnwidth]{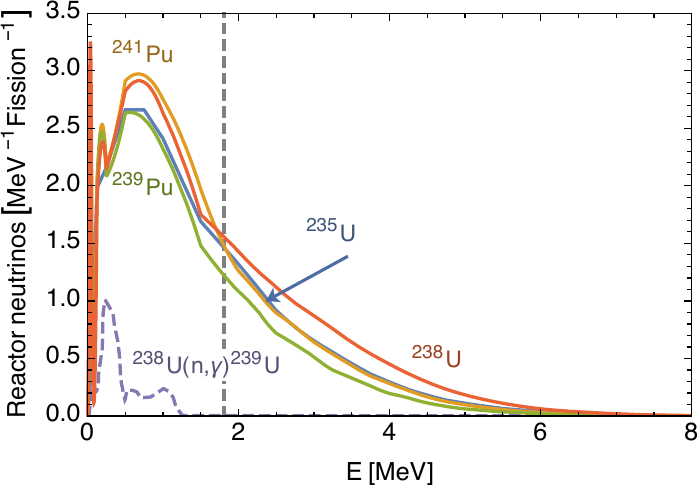}
\caption{The $\overline\nu_e$ energy spectra for $^{235}\rm U$, $^{238}\rm U$, $^{239}\rm Pu$, and $^{241}\rm Pu$ fissions. The inverse beta decay threshold is marked by a vertical dashed line. At low energies, the dominant contribution is due to neutron capture processes $^{238}\mathrm{U}(n,\gamma)^{239}\rm U$, here rescaled by a factor $1/20$.}
\label{fig:reac}
\end{figure}

Until 2011, the standard results were the ones obtained by \textcite{Vogel:1989iv}, but then two papers recalculated the spectrum at energies larger than 2~MeV
with different methods, one by \textcite{Huber:2011wv} and one by \textcite{Mueller:2011nm}. Figure~\ref{fig:reac} shows the spectrum due to the dominant processes reported in Table~II of \textcite{Vogel:1989iv} for energies smaller than 2~MeV, while for larger energies we use Tables~VII--IX of \textcite{Huber:2011wv} for $^{235}\rm U$, $^{239}\rm Pu$ and $^{241}\rm Pu$ and Table~III of \textcite{Mueller:2011nm} for $^{238}\rm U$; finally, the low-energy spectrum of neutron capture $^{238}\mathrm{U}(n,\gamma)^{239}\rm U$ are directly extracted from \textcite{Qian:2018wid}. Notice that fits to the tables are reported in these references.

The detection of reactor neutrinos typically relies on inverse beta decay on protons, $\overline{\nu}_e+p\to n+e^+$. The cross section is usually expressed in terms of well measured quantities such as the neutron lifetime $\tau_n$ and the electron mass $m_e$~\cite{Vogel:1999zy},
\begin{equation}
\sigma_{\mathrm{CC}}^{\overline{\nu}_e p}=\frac{2\pi^2}{\tau_n m_e^5 f}E_e \sqrt{E_e^2-m_e^2}\,.
\end{equation}
Here $f$ is the dimensionless phase-space integral
\begin{equation}
f=\int_{m_e}^{m_n-m_p}\!\!dE_e \frac{(m_n-m_p-E_e)^2E_e \sqrt{E_e^2-m_e^2}}{m_e^5}\,,
\end{equation}
where we neglect the small neutron recoil energy. The detection signature features a prompt signal due to the positron, followed by neutron capture. Alternatives to this process include charged and neutral current deuteron break-up using heavy water, $\overline\nu$-$e$ elastic scattering, and coherent $\overline\nu$-nucleus interactions \cite{Giunti:2007ry}.

\begin{figure}
\includegraphics[width=0.90\columnwidth]{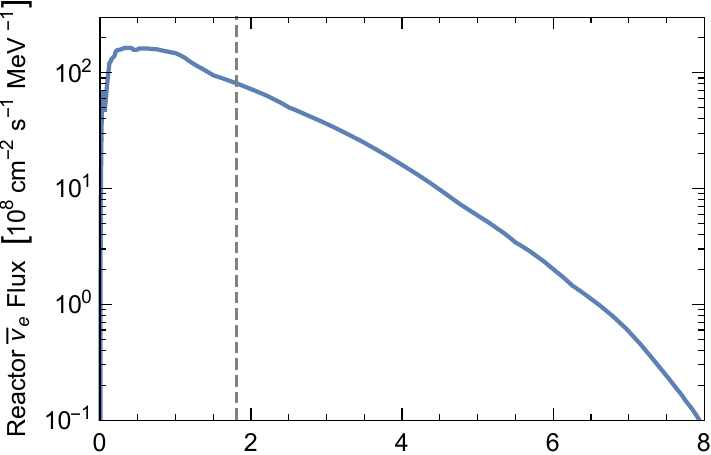}
\includegraphics[width=0.90\columnwidth]{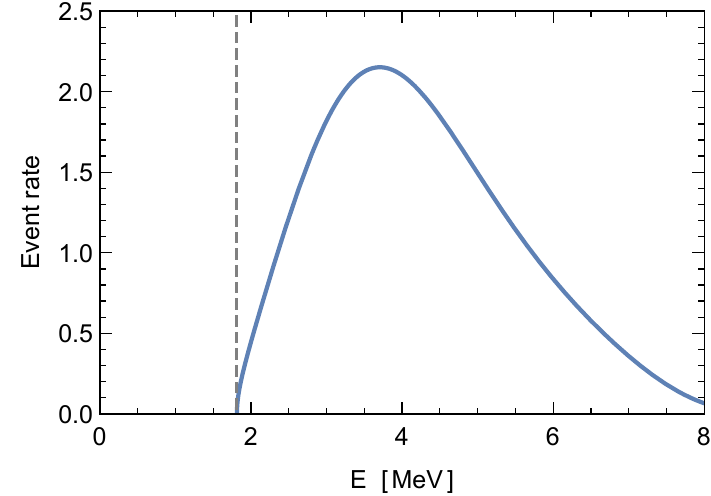}
\caption{Typical reactor neutrino spectrum, here from the Japanese experimental fast reactor JOYO \cite{Furuta:2011iu}. {\em Top:\/} Flux at a
  distance of 24.3~m. {\em Bottom:\/} Event rate (arbitrary units) as
  a function of neutrino energy, i.e., the neutrino energy
  distribution folded with the interaction cross section for inverse
  $\beta$ decay. Only $\overline\nu_e$ above the threshold of 1.8~MeV
  (vertical dashed line) are detectable. The event rate peaks at
  $E_\nu\sim4~{\rm MeV}$.}
\label{fig:reacflux}
\end{figure}

As a typical example we show in Fig.~\ref{fig:reacflux}
(top) the flux from the Japanese experimental fast
reactor JOYO \cite{Furuta:2011iu}, which has 140~MW thermal power and
a detector at the close distance of 24.3~m. Convolution with the
inverse $\beta$ cross section (bottom panel) shows that the
interactions peak for $E_{\overline{\nu}_e}\sim4~{\rm MeV}$. While
the quantitative details strongly depend on the reactor and detector,
several general features can be pointed out
\cite{Giunti:2007ry}. First, the large threshold implies that only
reactions with large \hbox{Q-value} can be observed, so only one fourth
of the total flux can be detected. Another important
point is that reactor shutdowns can be used to measure background and
that the intensity of the flux is proportional to the thermal power,
which is accurately monitored. Moreover, the flux is very large, so
the detectors do not need large shielding against cosmic rays. All of
these advantages make reactor neutrinos a fundamental tool for
measuring intrinsic neutrino properties such as mixing angles and mass
differences.

\begin{figure}
\includegraphics[width=0.80\columnwidth]{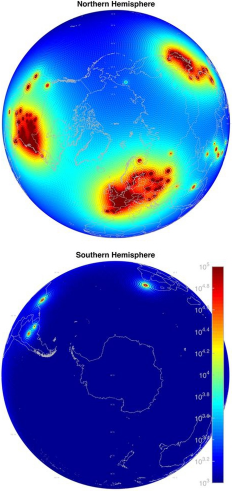}
\caption{Global map of the expected $\overline\nu_e$ flux (in $1/100\,{\rm
    cm}^{2}/{\rm s}$) in the narrow energy range 3.00--3.01~MeV from all
  power reactors. Flavor oscillations driven by the ``solar'' mixing
  parameters are visible on the 100~km scale.  Figure credit:
  Antineutrino Global Map 2015 \cite{Usman:2015yda} licensed under Creative Commons Attribution 4.0 International License.}\label{fig:reac-map}
  \vskip-6pt
\end{figure}

The global flux is produced by around 500 reactors worldwide with a
very uneven geographic distribution. In Fig.~\ref{fig:reac-map} we
show a global map, restricted to the very narrow energy range
3.00--3.01~MeV, allowing one to discern flavor oscillations over
100~km distances. The three main centers of production are the Eastern
US, Europe, and East Asia, notably Japan.

\subsection{Measurements}

While we have included reactor neutrinos in our discussion of the
diffuse neutrino background at Earth, of course reactors, or clusters
of reactors, are typically used as  nearby sources to study
neutrino properties.  Reactor experiments were the first successful
attempt to detect the elusive neutrinos. The proposal of using inverse
beta decay dates back to Bethe and Peierls~\cite{Bethe:1934qn}, but it was only in 1953
that Reines and Cowan started their experiments at Hanford and Savanah
River which eventually detected neutrinos for the first time~\cite{Cowan:1953mw}. Nuclear
power plants have been employed in the following decades to study
neutrino properties many times.

Concerning flavor conversion, a mile-stone
discovery was the detection
of $\overline\nu_e$ disappearance by KamLAND in 2002
over an approximate distance of 180~km \cite{Eguchi:2002dm}, driven by
the ``solar'' mixing parameters $\theta_{12}$ and $\delta m^2$.  The
subsequent measurement of a spectral distortion \cite{Araki:2004mb}
revealed first direct evidence for the phenomenon of flavor
oscillations with the usual energy dependence. Notice that matter
effects are here subdominant, so essentially one is testing vacuum
oscillations.

The earlier search over much shorter distances for oscillations
driven by the mixing angle $\theta_{13}$ and the ``atmospheric'' mass
difference $\Delta m^2$ by the CHOOZ \cite{Apollonio:1999ae} and
Palo Verde \cite{Boehm:2001ik} experiments proved elusive. However,
since 2012, a new generation of reactor experiments (Double Chooz
\cite{Abe:2011fz}, Daya Bay \cite{An:2012eh}, and RENO
\cite{Ahn:2012nd}) has succeeded to measure a non-zero value for
$\theta_{13}$ with high precision.

The frontier of reactor neutrino measurements will be advanced by the
JUNO detector \cite{An:2015jdp} that is currently under construction
in China. One of the prime goals is to detect subtle three-flavor
interference effects at an approximate distance of 60~km to establish
the neutrino mass ordering.

Meanwhile, the ``reactor antineutrino anomaly'' \cite{Mention:2011rk} remains unsettled, i.e., a few-percent deficit of the measured $\overline\nu_e$ flux at very close distances from reactors. Other anomalies include $\overline\nu_e$ appearance and $\nu_e$ disappearance,  see \textcite{Boser:2019rta}~and references therein. One interpretation are oscillations to sterile
neutrinos driven by an eV-scale mass difference~\cite{Conrad:2016sve, Giunti:2019aiy}. Recent antineutrino flux predictions~\cite{Estienne:2019ujo,Hayen:2019eop} have been used to reevaluate the significance of the reactor anomaly; in the ratios of measured antineutrino spectra an anomaly may still persist~\cite{Berryman:2019hme}.

\section{Supernova neutrinos}
\label{sec:snnu}

The core collapse of a massive star within a few seconds releases the
gravitational binding energy of a neutron star (NS), $E_{\rm
  b}\sim3\times10^{53}~{\rm erg}$, in the form of neutrinos in what is
known as a supernova (SN) explosion. This energy release is roughly
comparable to that of all stars in the Universe within the same
period.  While the neutrino burst from the next nearby SN is one of
the most cherished targets of neutrino astronomy, it is a transient
signal and thus not part of the GUNS. We here summarize the main
features of core-collapse neutrino emission primarily as an ingredient for the
diffuse SN neutrino background (DSNB) presented in
Sec.~\ref{sec:DSNB}. For reviews of SN neutrinos see
\textcite{Janka:2012wk}, \textcite{Scholberg:2012id}, \textcite{Mirizzi:2015eza},
and \textcite{Janka:2016fox}.

\begin{figure*}
\hbox to\textwidth{\includegraphics[scale=0.495]{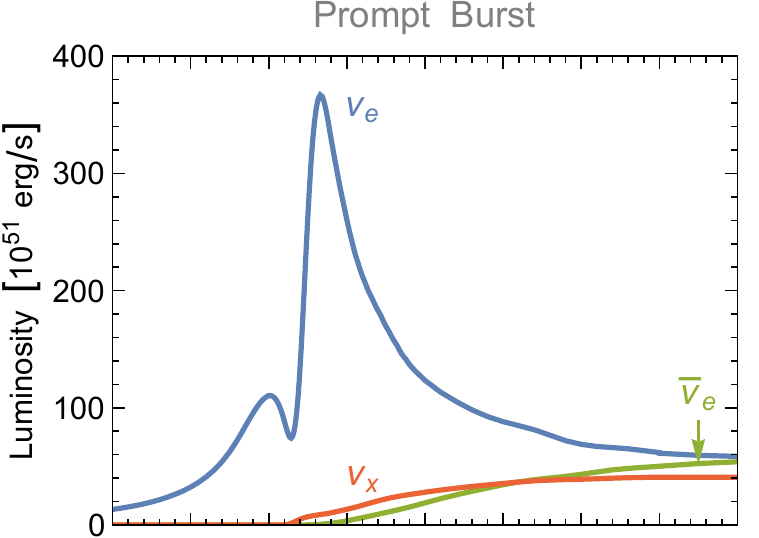}\hfil
\includegraphics[scale=0.495]{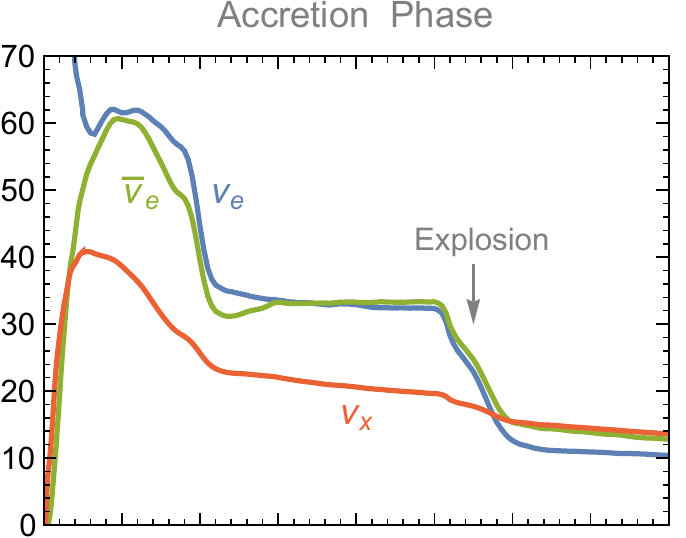}\hfil
\includegraphics[scale=0.495]{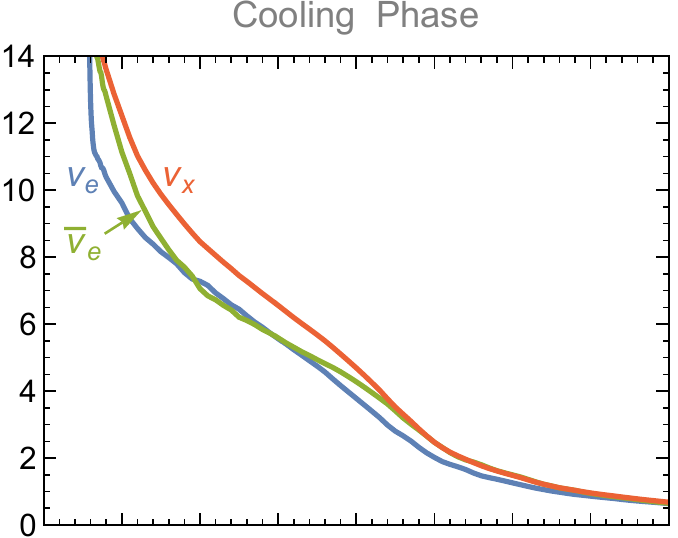}}
\vskip4pt
\hbox to\textwidth{\includegraphics[scale=0.495]{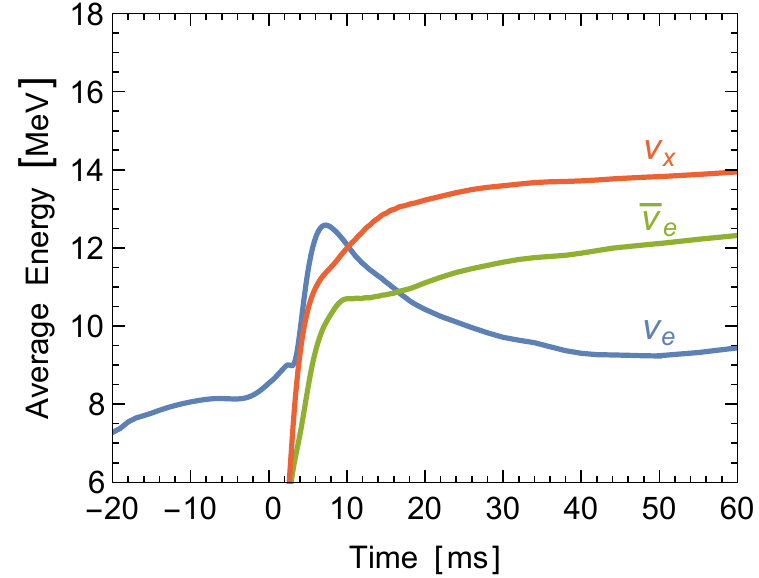}\hfil
\includegraphics[scale=0.495]{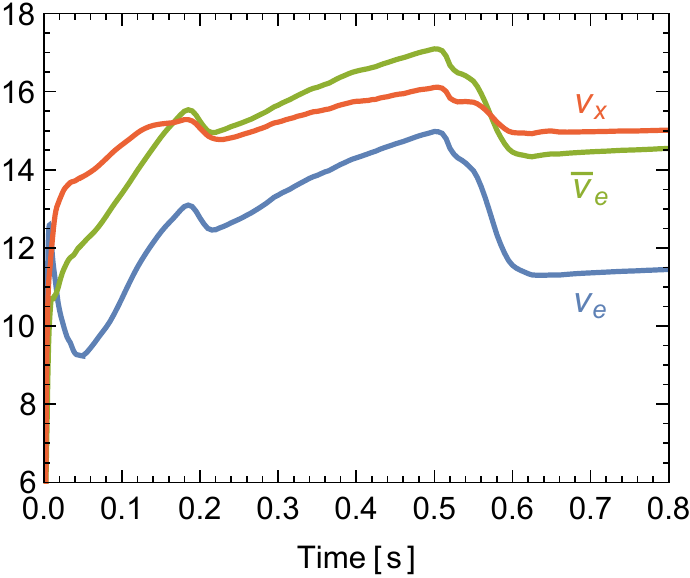}\hfil
\includegraphics[scale=0.495]{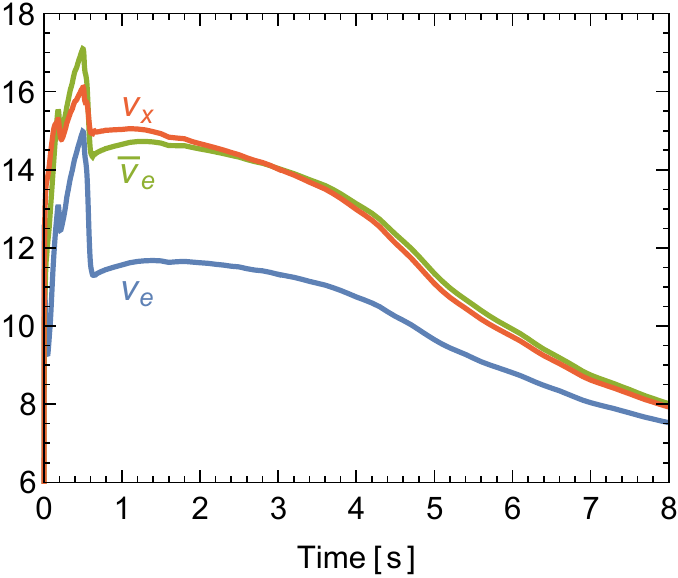}}
\caption{Luminosities and average energies for $\nu_e$, $\overline\nu_e$
  and $\nu_x$ (representing any of $\nu_{\mu}$, $\overline\nu_{\mu}$,
  $\nu_{\tau}$ or $\overline\nu_{\tau}$) for the main phases of neutrino
  emission from a core-collapse SN. From left to right: (1)~Infall,
  bounce and initial shock-wave propagation, including prompt $\nu_e$
  burst. (2)~Accretion phase with significant flavor differences of
  fluxes and spectra. (3)~Cooling of the proto neutron star (PNS),
  only small flavor differences between fluxes and spectra.  Based on
  a spherically symmetric $27\,M_\odot$ Garching model with explosion
  triggered by hand during 0.5--0.6~s. It uses the nuclear equation
  of state of Lattimer and Swesty with nuclear incompressibility
  modulus $K=220$~MeV and includes a mixing-length treatment of PNS
  convection. The final gravitational mass is $1.592\,M_\odot$ (or
  89.6\% of the baryonic mass of $1.776\,M_\odot$), so the mass
  deficit is $E_{\rm b}=0.184\,M_\odot=3.31\times10^{53}~{\rm erg}$
  that was lost in neutrinos. See the Garching Core-Collapse Supernova
  Data Archive at
  \hbox{\href{https://wwwmpa.mpa-garching.mpg.de/ccsnarchive/}{https://wwwmpa.mpa-garching.mpg.de/ccsnarchive/}}
  for several suites of SN models.  }
  \label{fig:SN-Burst}
\end{figure*}

\subsection{Generic features of supernova neutrinos}

At the end of its life, the compact core of an evolved star becomes
unstable and collapses to nuclear density, where the equation of state
stiffens~\cite{Janka:2012wk,Burrows:2012ew,Janka:2017vcp}. At this core bounce,
a shock wave forms, moves outward, and ejects most of the mass in the form of a SN explosion,
leaving behind a compact remnant that cools to become a
NS. Typical masses are around $M_{\rm{NS}} \simeq 1.5\,M_\odot$, with $2\,M_\odot$ the
largest observed case. The radius is $R_{\rm{NS}} \simeq 12$--$14$~km, the exact
value and NS structure depending on the nuclear equation of
state. Within these uncertainties one expects the
release of  the following binding energy
\begin{equation}
E_{\rm b} \simeq \frac{3}{5} \frac{G_{\rm N} M_{\rm{NS}}^2}{R_{\rm{NS}}} \simeq 3\times10^{53}~{\rm erg} \simeq 2\times10^{59}~{\rm MeV}\ ,
\end{equation}
with $G_{\rm N}$ being Newton's gravitational constant.

This huge amount of energy appears in the form of neutrinos
because the interaction rate of $\gamma$ and $e^\pm$ is so large
in dense matter that they contribute little to energy transfer, whereas
gravitons interact far too weakly to be effective. Moreover, in hot
nuclear matter the neutrino mean free path is short compared to the
geometric dimension of the collapsed object, so $\nu$ and $\overline\nu$ of
all flavors thermalize, for example by nucleon-nucleon
bremsstrahlung and other processes~\cite{Bruenn:1985en}. Hence, very approximately we may think of the collapsed SN core as a blackbody source for $\nu$ and $\overline\nu$
of all flavors.

The diffusion character of neutrino transport leads to an estimated
time of a few seconds for most of the energy trapped in the SN core to
escape.  The emission temperature depends on radiative transfer in the
decoupling region (``neutrino sphere''); typically
$T_\nu \simeq 3$--$5$~MeV, i.e., average
energies $\langle E_\nu\rangle=(3/2)\,T_\nu \simeq 10$--$15$~MeV after neutrino decoupling~\cite{Janka:2012wk,Burrows:2012ew,Janka:2017vcp}.  This scale is similar to that of solar and geoneutrinos, where however it is set by nuclear physics.
Overall one expects the emission of
around $E_{\rm b}/\langle E_\nu\rangle \simeq 3\times10^{57}$ particles for each of
the six $\nu$ and $\overline\nu$ species.

Besides energy, the SN core must also radiate lepton number (deleptonization).
The final NS contains only a small proton
(or electron) fraction, whereas the collapsing material, consisting of
elements between O and Fe, initially has an electron fraction
$Y_e=0.46$--$0.5$. A baryonic mass of $1.5\,M_\odot$ corresponds to
$2\times10^{57}$ nucleons, implying that $1\times10^{57}$
units of electron lepton number must escape in the form of
$\nu_e$, ignoring for now flavor conversion. Comparison
with the estimated $6\times10^{57}$ of $\nu_e$ plus $\overline\nu_e$ to be
radiated by the required energy loss reveals a significant excess of
$\nu_e$ over $\overline\nu_e$ \hbox{emission~\cite{Mirizzi:2015eza}}.

The overall picture of neutrino energies and time scale of
emission was confirmed on 23~February 1987 by the neutrino burst from
SN~1987A in the Large Magellanic Cloud with a total of about two dozen
events in three small detectors \cite{Hirata:1987hu, Bionta:1987qt, Alekseev:1988gp}. However, the data was too sparse for detailed quantitative tests. The next nearby SN would provide high
statistics, especially in Super-Kamiokande, in IceCube, or in upcoming large detectors such as Hyper-Kamiokande or DUNE. The expected large number of neutrino events in these detectors may show detailed imprints of SN \hbox{physics~\cite{Scholberg:2017czd}}.

\subsection{Reference neutrino signal}

\label{sec:reference-SN-signal}

The standard paradigm of stellar core-collapse and
SN explosions has evolved over decades of
numerical modeling \cite{Janka:2012wk, Janka:2016fox}, first in
spherical symmetry (1D) and over the past years with ever more refined
3D models.  After the collapse has begun and when the density exceeds
some $10^{12}~{\rm g}~{\rm cm}^{-3}$, neutrinos are entrained by the
infalling matter because of coherent scattering on large nuclei.  When
nuclear density of $3\times 10^{14}~{\rm g}~{\rm cm}^{-3}$ is reached,
the core bounces and a shock wave forms within the core at an enclosed
mass of around $0.5\,M_\odot$. As the shock propagates outward, it loses
energy by dissociating iron and eventually stalls at a radius of some
150~km, while matter keeps falling in. Meanwhile the neutrino flux
streaming through this region deposits some of its energy,
rejuvenating the shock wave, which finally moves on and ejects the
outer layers. It leaves behind a hot and dense
proto neutron star (PNS), which cools
and deleptonizes within a few seconds.  This is the essence of the
neutrino-driven explosion mechanism, also called delayed explosion
mechanism or Bethe-Wilson \hbox{mechanism~\cite{Bethe:1984ux}}.

The corresponding neutrino signal falls into three main phases shown
in Fig.~\ref{fig:SN-Burst}, using a $27\,M_\odot$ spherically
symmetric model for illustration. As in most simulations, neutrino
radiative transfer is treated in a three-species approximation
consisting of $\nu_e$, $\overline\nu_e$ and $\nu_x$ which stands for any of
$\nu_\mu$, $\overline\nu_\mu$, $\nu_\tau$ and $\overline\nu_\tau$. This approach
captures the main flavor dependence caused by charged-current
interactions of the $e$-flavored states. However, heavy-flavor $\nu$
and $\overline\nu$ do not interact exactly the same because of recoil
corrections and weak magnetism \cite{Horowitz:2003yx}. Moreover, the
$\mu$ and $\tau$ flavored states differ by the presence of some muons
($m_\mu=105.7~{\rm MeV}$) in matter that reaches temperatures
of several ten~MeV~\cite{Bollig:2017lki}.

\bigskip

\noindent{\em Prompt Burst.}---Soon after bounce, the shock wave
breaks through the edge of the iron core, liberating the conspicuous
prompt $\nu_e$ burst that corresponds to a significant fraction of the
overall lepton number. It is therefore also called the deleptonization
or neutronization or breakout burst. During the
post-bounce time window $-20$~ms to 60~ms shown in the left
panels in Fig.~\ref{fig:SN-Burst}, the SN core radiates about 5\%
of the total energy that corresponds to the period shown in the
rightmost panels, whereas it radiates $0.4\times10^{57}$ units of
lepton number, i.e., around 50\% of what is emitted over the full
period. The features of the prompt-burst phase are thought to be essentially
universal~\cite{Liebendoerfer:2002ny, Kachelriess:2004ds}. The little
dip in the $\nu_e$ luminosity curve at $t=0$--4~ms, for
example, is explained by the shock first compressing matter to
opaque conditions before the post-shock layer re-expands to
\hbox{become transparent}.

\bigskip

\noindent{\em Accretion Phase.}---As the shock wave stalls, neutrino
emission is powered by the accretion flow of matter onto the SN core,
emitting $\nu_e$ and $\overline\nu_e$ with almost equal luminosities, but
somewhat different average energies, so the $\nu_e$ particle flux is
some 20\% larger than the $\overline\nu_e$ one. The production and
interaction is mostly by $\beta$ processes on protons and
neutrons. Heavy-flavor $\nu$ and $\overline\nu$, on the other hand, are
produced in pairs and emerge from somewhat deeper layers, with a
smaller radiating region and therefore smaller fluxes. Their average
energies, however, are very similar to that of $\overline\nu_e$. The large
hierarchy of flavor-dependent average energies that was seen, for
example, in the often-cited Livermore model \cite{Totani:1997vj} is
not physical \cite{Raffelt:2001kv} and is not borne out by present-day
simulations.  The luminosity drop at around 200~ms represents the
infall of the Si/O interface, after which the accretion rate and
luminosity become smaller. Over the entire accretion phase,
the mass gain and concomitant contraction
of the SN core show up in the increasing neutrino energies.

\bigskip

\noindent{\em Explosion.}---Spherically symmetric numerical models do
not explode except for the smallest-mass progenitors, such as electron-capture supernovae~\cite{Kitaura:2005bt}, so the duration of the accretion phase, and if an explosion occurs at all, can not be inferred from these models. For example, the explosion in the case of
Fig.~\ref{fig:SN-Burst} was triggered by hand during the 500--600~ms
period. Three-dimensional simulations  successfully explode and suggest that the explosion time
strongly depends on the SN model and may vary up to few hundreds of ms relative to
what is shown in Fig.~\ref{fig:SN-Burst}, see e.g.\ \textcite{Bollig:2017lki}, \textcite{Summa:2017wxq}, \textcite{Burrows:2019zce}, and \textcite{Vartanyan:2018iah}. The quenching of accretion strongly reduces the $\nu_e$ and $\overline\nu_e$ luminosities which drop to the component provided by core emission.

\bigskip

\noindent{\em Cooling.}---The remaining evolution consists of cooling
and deleptonization of the PNS. The luminosity is essentially
equipartitioned among the six species, whereas $\langle E_{\nu_e}\rangle$
is smaller than the others, i.e., there is a net lepton number flux.
The quantitative details depend strongly on the PNS mass and the nuclear equation of state, see e.g.\ \textcite{Oertel:2016bki}, \textcite{Nakazato:2019ojk}, and \textcite{Nakazato:2020ogl}.
Note that we show the neutrino signal until $8$~s in Fig.~\ref{fig:SN-Burst}, but this is not a hard cut-off. We refer the reader to \textcite{Nakazato:2012qf} and \textcite{Nakazato:2020ogl}
for discussions of late SN neutrino emission.
The neutronization burst and the accretion phase release about $50\%$ of the total
energy; the other half is emitted during the cooling phase.

\bigskip

The instantaneous neutrino spectra are quasi thermal, but do not
follow exactly a Fermi-Dirac distribution. Rather they are ``pinched,'' i.e.,
the spread of energies around the mean is less than in the
thermal case. Phenomenologically, the numerical spectra are well described
by a Gamma distribution of the form~\cite{Keil:2002in,Tamborra:2012ac}
\begin{equation}\label{eq:gamma-dist}
f(E)\propto\left(\frac{E}{E_{\rm av}}\right)^{\alpha}
\exp\left[-\frac{(\alpha+1) E}{E_{\rm av}}\right]\,,
\end{equation}
where $\alpha$ is the ``pinching parameter'' with $\alpha=2$
corresponding to a Maxwell-Boltzmann distribution. For any $\alpha$,
the parameter $E_{\rm av}$ matches $\langle E\rangle$, whereas
$\alpha$ is fixed to match, for example, $\langle E^2\rangle$ of the
numerical spectrum by $\langle E^2\rangle/\langle E\rangle^2 =
(2+\alpha)/(1+\alpha)$. In addition, the overall normalization is
fixed to match the numerical case. The pinching is largest for $\nu_e$,
especially during the prompt burst, and smallest for $\nu_x$.

\begin{figure}[b]
\includegraphics[width=0.90\columnwidth]{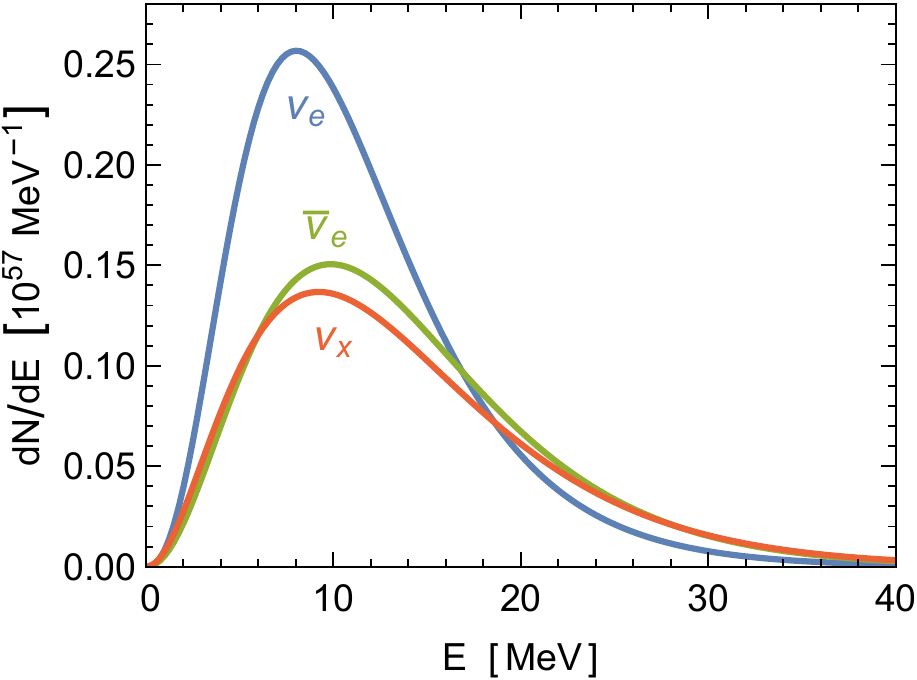}
  \caption{Time-integrated spectra of the reference model of
    Fig.~\ref{fig:SN-Burst}. The total particle emission
    is $3.2\times10^{57}~\nu_e$, $2.4\times10^{57}~\overline\nu_e$,
    and $2.3\times10^{57}$ for each of $\nu_{\mu}$, $\overline\nu_{\mu}$,
    $\nu_{\tau}$ or $\overline\nu_{\tau}$.}
  \label{fig:SN-spec}
\end{figure}

The time-integrated flux spectra of our reference model are shown in
Fig.~\ref{fig:SN-spec}. They are superpositions of pinched spectra
with different $E_{\rm av}$, which broadens their spectral shape,
and therefore do not need to be pinched
themselves. We find the average energies $\langle
E_{\nu_e}\rangle=11.3~{\rm MeV}$, $\langle
E_{\overline\nu_e}\rangle=13.9~{\rm MeV}$, and $\langle
E_{\nu_x}\rangle=13.8~{\rm MeV}$, as well as the pinching parameters
$\alpha_{\nu_e}=2.5$, $\alpha_{\overline\nu_e}=2.4$, and
$\alpha_{\nu_x}=2.0$. So in this example, the integrated
spectra follow nearly a Maxwell-Boltzmann distribution.

\subsection{Electron-capture supernovae}

Assuming this scenario to capture the main features of a SN neutrino
signal, one still expects large case-by-case variations depending on
progenitor properties. The lowest-mass SN progenitors (about
8--$10\,M_\odot$) become unstable due to electron capture before
nuclear burning of their O-Ne-Mg core can be ignited, so they never
reach an iron core. These ``electron capture SNe'' or ``O-Ne-Mg-core
SNe'' could represent 30\% of all cases because the initial mass
function decreases rapidly with increasing mass. Spherically
symmetric models of these low-mass progenitors
explode after a very short accretion phase,
but otherwise resemble what was shown earlier
\cite{Fischer:2009af,Huedepohl:2009wh}.

\subsection{Failed explosions}

For higher-mass progenitors, numerical models do not explode.  It
remains open if this question depends, for example, on quantitative
details of neutrino energy transfer and 3D effects, on details of the
progenitor models, or if a crucial piece of input physics is
missing. Moreover, probably not all collapsing stellar cores lead to
successful explosions --- the class of failed SNe, leaving a black
hole (BH) instead of a NS as a compact remnant. Using the
``compactness parameter'' as a criterion, recent theoretical work
hints that up to 40\% of all collapsing cores may lead to BH
formation~\cite{OConnor:2010moj, Ertl:2015rga, Sukhbold:2015wba}.

The cosmic star-formation rate predicts perhaps twice the observed SNe
rate at high redshifts, suggesting a significant fraction of failed
explosions \cite{Hopkins:2006bw, Horiuchi:2011zz}.  Likewise, the
``red supergiant problem'' suggests a cutoff of around $18\,M_\odot$
in the mass range of identified SN progenitors \cite{Smartt:2008zd,
  Jennings:2014mam}.  A significant fraction of failed SNe would also
naturally explain the compact-object mass distribution
\cite{Kochanek:2014mwa}.  Motivated by these hints, a survey looks for
disappearing red supergiants in 27 galaxies within 10~Mpc with the
Large Binocular Telescope~\cite{Adams:2016hit}. Over the first seven
years, ending in early 2016, this survey found six core-collapse SNe
and one candidate for a failed SN, providing $0.14_{-0.10}^{+0.33}$
for the fraction of failed SNe.

In the neutrino signal of a failed SN,
the cooling phase would be missing, whereas the accretion phase would abruptly
end. The average neutrino energies would increase until this point and the $\nu_e$ and $\overline\nu_e$ fluxes dominate \cite{Sumiyoshi:2006id,Nakazato:2008vj,Walk:2019miz}. The overall emitted neutrino energy could exceed that of an exploding~SN.
The crucial point is that BH formation is delayed, not prompt, so
the core bounce and shock scenario is crucial for
the expected neutrino burst of both exploding and failed cases.

An intermediate class between exploding and failed progenitors are fallback SNe, where
BH formation is delayed, if the explosion energy is not sufficient to unbind the star~\cite{Fryer:2007cf,2014arXiv1401.3032W}. Hence, a fraction of the stellar mantle may fall back and push the NS beyond the BH limit.

\subsection{Broken spherical symmetry in the stellar explosion}

Observations of SN remnants as well as large NS kick velocities
reveal that core-collapse SNe are not spherically symmetric. Recently,
numerical simulations without global symmetries (3D
simulations) with sophisticated neutrino transport have become
available. They show that large-scale convective overturns develop
during the accretion phase~\cite{Bethe:1990mw}. Moreover, the neutrino
emission properties are also affected by large-scale instabilities,
notably the Standing Accretion Phase Instability (SASI)
\cite{Blondin:2002sm,Tamborra:2013laa}, a global sloshing or spiral
hydrodynamical oscillation, and the Lepton Emission Self-Sustained
Asymmetry (LESA) \cite{Tamborra:2014aua}, whose effect is that
deleptonization mostly occurs into one hemisphere.

For the neutrino signal, these phenomena imply that during the
accretion phase the detailed signal properties depend on the observer
direction.  Moreover, the SASI mode would imprint periodic signal
modulations that probably could be picked up with large detectors such
as IceCube or Super-Kamiokande.  The neutrino signal of the next
nearby SN may provide details about the hydrodynamical behavior.

\subsection{Flavor conversion}

\label{sec:SN-flavor-conversion}

Numerical SN models treat neutrino transport usually in a
three-species formalism consisting of $\nu_e$, $\overline\nu_e$ and
$\nu_x$, representing any of $\nu_\mu$, $\overline\nu_\mu$, $\nu_\tau$, or
$\overline\nu_\tau$, and completely ignore flavor conversion. From a
numerical perspective, including flavor conversion is completely out
of the question. From a theoretical perspective, many questions remain
open because the matter effect of neutrinos on each other leads to
collective flavor conversion phenomena that are not yet fully
understood \cite{Mirizzi:2015eza,Chakraborty:2016yeg}.

The flavor evolution of the prompt $\nu_e$ burst is probably similar
to MSW conversion of solar neutrinos, except that the starting point
is at far larger densities, requiring a three-flavor
treatment. Moreover, neutrino-neutrino refraction would cause
synchronized oscillations and, depending on the matter profile, cause
a spectral split, i.e., a discontinuity in the conversion probability.

During the accretion phase, the $\nu_e\overline\nu_e$ flux is larger than
the $\nu_\mu\overline\nu_\mu$ or $\nu_\tau\overline\nu_\tau$ one. Collective
effects can lead to pair conversion of the type
$\nu_e\overline\nu_e\leftrightarrow\nu_\mu\overline\nu_\mu$ or
$\nu_e\overline\nu_e\leftrightarrow\nu_\tau\overline\nu_\tau$, i.e., pair
annihilation on the level of forward scattering with a rate much
faster than the usual non-forward scattering process. Conceivably it
could lead to flavor equilibration not far from the neutrino
decoupling region~\cite{Sawyer:2015dsa,Izaguirre:2016gsx}. In addition
to collective effects, one expects MSW conversion by the ordinary
matter profile~\cite{Dighe:1999bi}, although the matter effect could
be modified by density variations caused, e.g., by turbulence in the
convective regions.  Far away from the SN, neutrinos would decohere
into mass eigenstates. However, unlike for solar neutrinos, one cannot
easily predict the energy-dependent probability for the various
$\nu_i$ and $\overline\nu_i$ components.

\subsection{Detection perspectives}

The neutrino signal of SN~1987A on 23~February 1987 in three small
detectors was a historical achievement, but the event statistics was
sparse~\cite{Loredo:2001rx,Lunardini:2004bj,Pagliaroli:2008ur}. The next nearby (probably galactic) SN will be observed in a large number of detectors of different size, ranging from a few events to thousands (Super-Kamiokande) or even millions
(IceCube), although in the latter case without event-by-event
recognition \cite{Scholberg:2012id,Mirizzi:2015eza,Scholberg:2017czd}.
The various detectors will provide complementary information. What
exactly one will learn depends, of course, on the exact type of
core-collapse event that could range from an electron-capture SN to a
failed explosion with BH formation. It will also depend on
concomitant electromagnetic and possibly gravitational-wave
observations.

While the next nearby SN is perhaps the most cherished target of
low-energy neutrino astronomy and will provide a bonanza of
astrophysical and particle-physics information, its transient
nature sets it apart from the general neutrino background.
Therefore, a detailed discussion of the detection
opportunities is beyond our remit and we refer to several pertinent
reviews~\cite{Mirizzi:2015eza,Scholberg:2017czd}.

\section{Diffuse Supernova Neutrinos}
\label{sec:DSNB}

All collapsing stars in the visible universe, a few per second,
provide the diffuse supernova neutrino background (DSNB).  It
dominates at Earth for 10--25~MeV and in future could be measured by
the JUNO and Gd-enriched Super-Kamiokande detectors, providing hints
on the SN redshift distribution, the fraction of electromagnetically
dim progenitors, and average SN energetics.

\subsection{Basic estimate}

\label{sec:DSNB-Basic}

The idea that the accumulated neutrinos from all collapsed stars in
the universe form an interesting cosmic background goes back to the
early 1980s \cite{Bisnovatyi-Kogan:1982,Krauss:1983zn,Domogatskii:1984},
while modern reviews are by \textcite{Ando:2004hc}, \textcite{Beacom:2010kk},
\textcite{Lunardini:2010ab}, and \textcite{Mirizzi:2015eza}. The DSNB flux and spectrum depend
on the overall core-collapse rate that is uncertain within perhaps
a factor of two and on the average neutrino emission spectrum.
Our baseline case (see below) predicts for the sum of all species
\begin{subequations}\label{eq:Mix1-results}
  \begin{eqnarray}
    \Phi_{\Sigma\,\nu\overline\nu} &=& 126~{\rm cm}^{-2}~{\rm s}^{-1},\\
    n_{\Sigma\,\nu\overline\nu}    &=& 4.2\times10^{-9}~{\rm cm}^{-3},\\
    \rho_{\Sigma\,\nu\overline\nu} &=& 25~{\rm meV}~{\rm cm}^{-3},
    \label{eq:rho-DSNB}
  \end{eqnarray}
\end{subequations}
with an average energy of 6.0~MeV, corresponding to an
emission energy, averaged over all species, of 12.8~MeV.
The DSNB energy density is almost the same as the CNB energy density
of massless neutrinos that was given, for a single flavor,
in Eq.~\eqref{eq:CNB-density-b}. If the lightest neutrino
mass is so large that all CNB neutrinos
are dark matter today, the DSNB is the dominant neutrino
radiation density in the present-day universe.

We can compare the DSNB with the accumulated photons from
all stars, the extra-galactic background
light (EBL), that provides a radiation density of around
$50~{\rm nW}~{\rm m}^{-2}~{\rm sr}^{-1}$ \cite{Dole:2006de}.
Integrating over directions yields a flux of
$400~{\rm MeV}~{\rm cm}^{-2}~{\rm s}^{-1}$ and thus
an energy density of $13~{\rm meV}~{\rm cm}^{-3}$.
Photons and neutrinos are redshifted in the same way,
so the stars of the universe have emitted about
twice as much energy in the form of core-collapse neutrinos
as in the form of light.

We can express the time-averaged neutrino luminosity
$L_\nu$ of a given stellar population in units of the number of core-collapse
events per unit time, assuming one event releases
$2.5\times10^{53}~{\rm erg}$. Moreover, we can express the
photon luminosity $L_\gamma$ in units of the solar luminosity
of $L_\odot=4\times10^{33}~{\rm erg}/{\rm s}$, so
a ratio $L_\nu/L_\gamma=2$ corresponds to
$1/100~{\rm years}/10^{10}\,L_\odot$ core-collapse events.
This rate corresponds approximately to the usual SN unit
that is defined as
$1~{\rm SNu}=
1~{\rm SN}/10^{10}\,L_{\odot{\rm B}}/100\,{\rm yr}$ with $L_{\odot{\rm B}}$
the solar luminosity in the blue spectral band.
While the SN rate depends
strongly on galaxy type, e.g.\ no core-collapse SNe in elliptical
galaxies where no star formation takes place, averaged over all galaxies
it is around 1~SNu \cite{Cappellaro:1999qy, Cappellaro:2000ez}.
Very roughly, 1~SNu corresponds to one SN per century per galaxy. In other words,
$L_\nu/L_\gamma\sim 2$ of an average stellar population corresponds
to the usual astronomical measure of the SN rate.
Within uncertainties, the DSNB density of Eq.~\eqref{eq:rho-DSNB}
follows from expressing 1~SNu as a neutrino-to-photon
luminosity ratio.

For DSNB detection, the $\overline\nu_e$ component is of particular interest.
For energies below 10~MeV it is hidden under the reactor $\overline\nu_e$ background,
so the higher-energy part of the DSNB spectrum is particularly
important. It requires a more detailed discussion than a simple prediction
of the overall DSNB density.

\subsection{Redshift integral}

The DSNB depends on the core-collapse rate $R_{\rm cc}(z)$ at cosmic
redshift $z$ and the average spectrum $F_{\nu}(E)=dN_\nu/dE$ emitted
per such event, where $\nu$ can be any of the six species of neutrinos
or antineutrinos. The long propagation distance implies loss of flavor
coherence, so each $\nu$ represents a mass eigenstate. Each neutrino
burst lasts for a few seconds, but this time structure plays no
practical role because one will need to integrate for several years to
detect even a small number of DSNB neutrinos. Moreover, the bursts
sweeping through the detector somewhat overlap. Therefore,
$F_{\nu}(E)$ is the average time-integrated number of neutrinos per
energy interval emitted by a collapsing star.

The neutrino density spectrum accumulated from all cosmic epochs is
given by the redshift integral
\begin{equation}\label{eq:DSNB-1}
  \frac{dn_\nu}{dE}=\int_{0}^{\infty}dz\,(z+1)\,F_\nu(E_z)\,n'_{\rm cc}(z)\,,
\end{equation}
to be multiplied with the speed of light to obtain the diffuse flux
(see Appendix~\ref{sec:Units}).  Here $E_z=(1+z)E$ is the blue-shifted
energy at emission of the detected energy $E$. The first factor
$(1+z)$ arises as a Jacobian $dE_z/dE=(1+z)$ between emitted and
detected energy interval. It is assumed that the average neutrino flux
spectrum $F_\nu(E)$ is the same at all cosmic epochs.

Finally $n'_{\rm cc}(z)=dn_{\rm cc}/dz$ is the core-collapse number
per comoving volume per redshift interval. It is usually expressed in
the form
\begin{equation}\label{eq:RSN-1}
  n'_{\rm cc}(z)=
  \frac{R_{\rm cc}(z)}{H_0\,(1+z)\sqrt{\Omega_{\rm M}(1+z)^3+\Omega_\Lambda}}\,,
\end{equation}
where $H_0$ is the Hubble expansion parameter, while $\Omega_{\rm M}$
and $\Omega_\Lambda$ are the present-day cosmic matter and dark-energy
fractions.  In the literature one usually finds $R_{\rm cc}(z)$, the
number of core-collapse events per comoving volume per unit time
(units ${\rm Mpc}^{-3}~{\rm yr}^{-1}$). However, $R_{\rm cc}(z)$ is
derived in terms of an assumed cosmological model because observations
for a given redshift interval need to be translated to intervals of
cosmic time, i.e., only $n'_{\rm cc}(z)$ has direct meaning. So a given
$R_{\rm cc}(z)$ makes sense only in conjunction with the
assumed underlying cosmological model.

We may further express $n'_{\rm cc}(z)=n_{\rm cc} f_{\rm cc}(z)$ in
terms of the comoving density $n_{\rm cc}$ of all past core-collapse
events, times its normalised redshift distribution with
$\int_{0}^{\infty}dz\,f_{\rm cc}(z)=1$.  Likewise, the neutrino
emission spectrum is expressed as $F_\nu(E)=N_\nu\,f_\nu(E)$ with
$N_\nu$ the total number of species $\nu$ emitted by an average core
collapse times its normalised spectrum with
$\int_{0}^{\infty}dE\,f_{\nu}(E)=1$.  With these definitions,
Eq.~\eqref{eq:DSNB-1} is
\begin{equation}\label{eq:DSNB-2}
  \frac{dn_\nu}{dE}=N_\nu\,n_{\rm cc}\,g_\nu(E)
\end{equation}
with the energy spectrum of the accumulated neutrinos
\begin{equation}\label{eq:DSNB-3}
 g_\nu(E)=\int_{0}^{\infty}dz\,(z+1)\,f_\nu[(z+1)E]\,f_{\rm cc}(z)\,.
\end{equation}
It fulfills the normalization $\int_{0}^{\infty}dE\,g_{\nu}(E)=1$ if
$f_\nu(E)$ and $f_{\rm cc}(z)$ are normalised.

\subsection{Cosmic core-collapse rate}

\begin{figure}[!b]
\includegraphics[width=0.90\columnwidth]{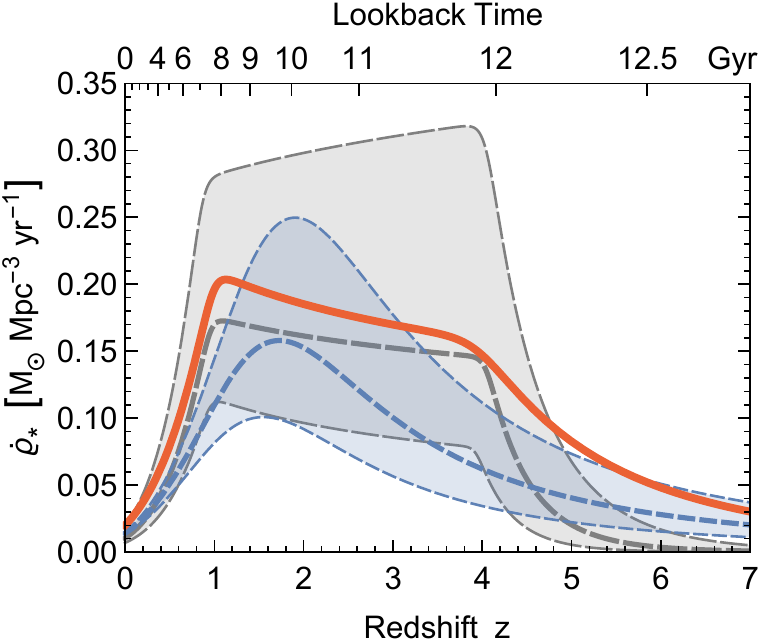}
\caption{Cosmic star-formation rate.
  {\em Red solid line:\/} Best fit of \textcite{Yuksel:2008cu} that we use as our reference case.
  {\em Gray (long dashed) lines:\/} Best fit and allowed range of \textcite{Mathews:2014qba}. {\em Blue (short dashed) lines:\/} Best fit and allowed range of \textcite{Robertson:2015uda}. All of these authors provide analytic fit functions that were used here.}
\label{fig:SFR}
\end{figure}

The core-collapse rate as a function of redshift can be determined by
direct SN observations. However, this approach may be significantly
incomplete because core-collapse SNe can be electromagnetically dim
or, for non-exploding cases, completely invisible. Therefore, usually
one estimates the rate from the star-formation activity, essentially
translating ultraviolet and infrared astronomical observations into a
neutrino emission rate.  The star-formation rate (mass processed into
stars per comoving volume per time interval) as a function of redshift
determined by different authors is shown in Fig.~\ref{fig:SFR}.
In keeping with previous DSNB studies we use the star-formation rate of
\textcite{Yuksel:2008cu} as our reference case (red solid line). A similar representation, including an explicit allowed range, was provided by
\textcite{Mathews:2014qba}, shown as a gray (long dashed) line and shaded region.
These rates considerably increase from the present to $z\sim 1$, then level
off to form a plateau, and decrease at larger redshift.
Following these authors we use the somewhat schematic
cosmological parameters $H_0=70~{\rm km}~{\rm s}^{-1}~{\rm Mpc}^{-1}=(13.9~{\rm Gyr})^{-1}$,
$\Omega_{\rm M}=0.3$, and $\Omega_\Lambda=0.7$.
A different form and allowed range was provided by \textcite{Robertson:2015uda}, here shown as a blue (short dashed) line and shaded region. These authors have used a somewhat different
cosmological model which we have transformed to our reference parameters.
These results are similar to those provided by \textcite{Madau:2014bja} that we do not show
in this figure.

To convert the star-formation rate into a core-collapse rate $R_{\rm
  cc}=k_{\rm cc}\,\dot\varrho_*$ we need the factor
\begin{equation}\label{eq:cc-conversion}
  k_{\rm cc}=\frac{\int_{M_{\rm min}}^{M_{\rm max}}dM\,\psi(M)}{\int_{M_{\rm l}}^{M_{\rm u}}dM\,M\,\psi(M)}
  =(135\,M_\odot)^{-1}\,,
\end{equation}
where $\psi(M)\propto M^{-2.35}$ is the \textcite{Salpeter:1955it} initial mass function
and $(M_{\rm l},M_{\rm u})=(0.1,125)\,M_\odot$ the overall stellar mass range.
For stars that develop collapsing cores we use $(M_{\rm min},M_{\rm max})=(8,125)\,M_\odot$,
including those cases that do not explode as a SN but rather form a
black hole (BH) because these non-exploding cases are also powerful
neutrino sources.

With this conversion factor we find for the integrated core-collapse
density $n_{\rm cc}$ of the past cosmic history for the best-fit star-formation
rates of the mentioned authors
\begin{equation}\label{eq:cc-rate}
\begin{matrix}
               1.05\times10^{7}~{\rm Mpc}^{-3}& \hbox to 128pt{\textcite{Yuksel:2008cu},\hfil}\\
               0.84\times10^{7}~{\rm Mpc}^{-3}& \hbox to 128pt{\textcite{Mathews:2014qba},\hfil}\\
               0.69\times10^{7}~{\rm Mpc}^{-3}& \hbox to 128pt{\textcite{Robertson:2015uda},\hfil}\\
               0.58\times10^{7}~{\rm Mpc}^{-3}& \hbox to 128pt{\textcite{Madau:2014bja}.\hfil}
             \end{matrix}
\end{equation}
If every core collapse emits on average $N_\nu\sim 2\times10^{57}$
neutrinos of each species, $n_{\rm cc}\sim 10^7~{\rm Mpc}^{-3}$ yields
a DSNB density in one species of $n_\nu\sim 2\times 10^{64}~{\rm
  Mpc}^{-3}=0.7\times10^{-9}~{\rm cm}^{-3}$ or, after multiplying with
the speed of light, an isotropic flux of $20~{\rm cm}^{-2}~{\rm s}^{-1}$ in one species.

\begin{figure}
\includegraphics[width=0.90\columnwidth]{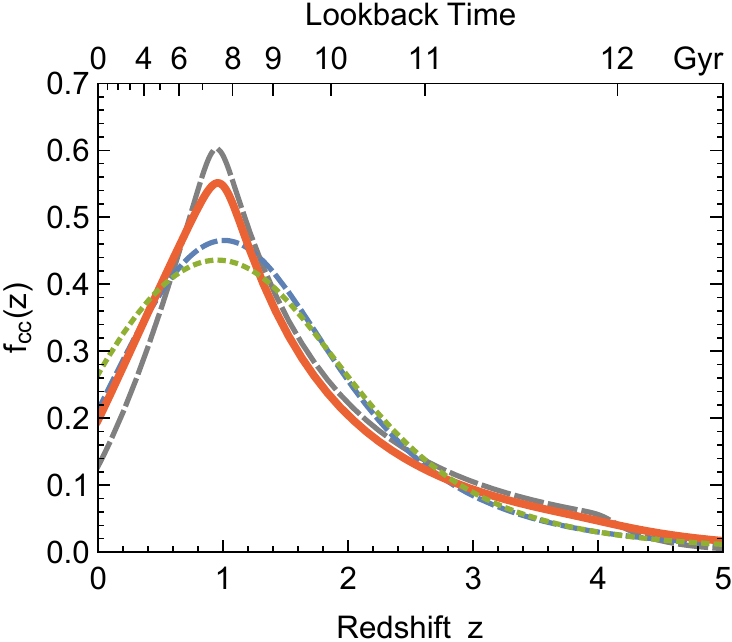}
\caption{Normalised core-collapse distribution as a function of redshift
for the best-fit cases of Fig.~\ref{fig:SFR} and in addition  that of \textcite{Madau:2014bja} in green
(dotted).}
\label{fig:fcc}
\end{figure}

\begin{figure}
\includegraphics[width=0.90\columnwidth]{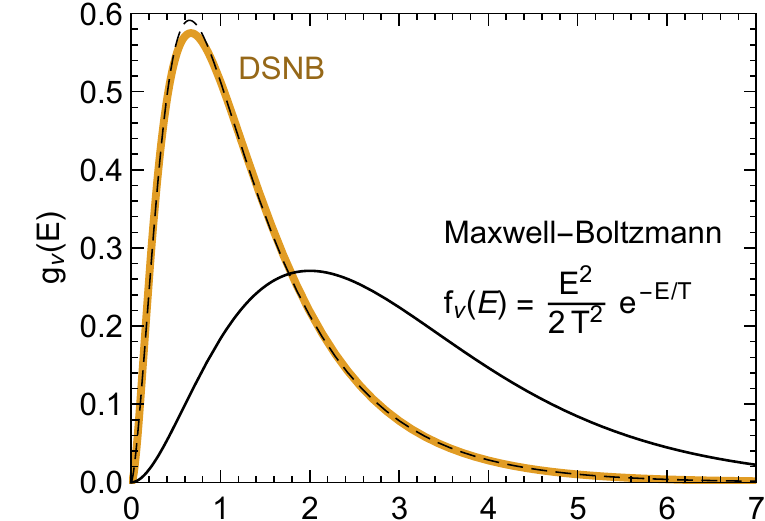}
\includegraphics[width=0.90\columnwidth]{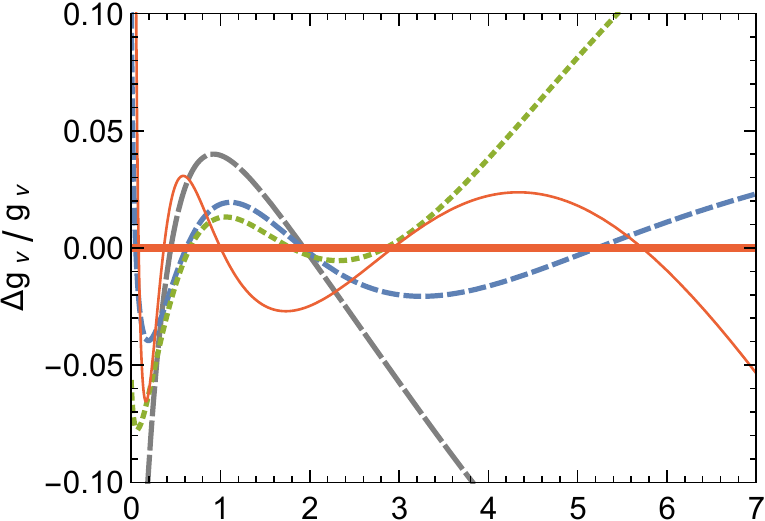}
\includegraphics[width=0.90\columnwidth]{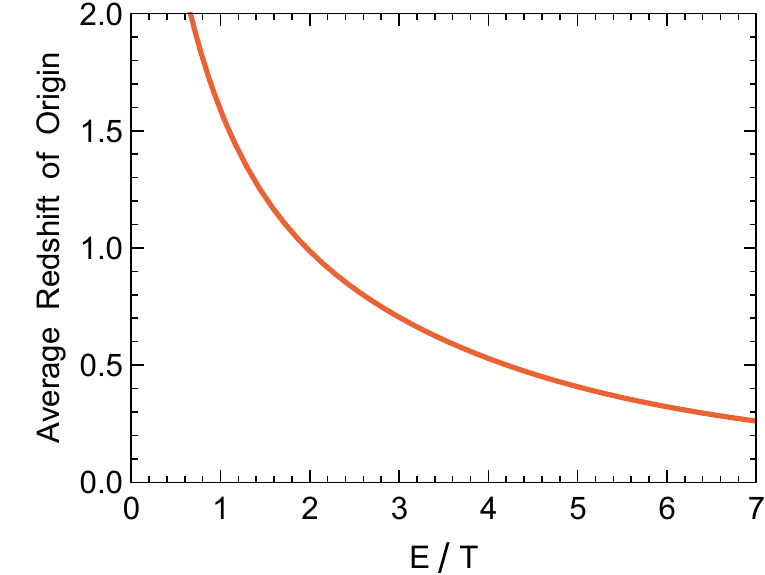}
\caption{Spectral properties of the DSNB. {\em Top:\/} Maxwell-Boltzmann source spectrum
with temperature $T$ (black) and corresponding DSNB spectrum (orange) for our fiducial redshift distribution of \textcite{Yuksel:2008cu}. The dashed line is the fit function of Eq.~\eqref{eq:DSNB-fit-1};
the fractional deviation is the red (thin solid) line the next panel.
 {\em Middle:\/} Fractional difference of the \textcite{Mathews:2014qba} case (gray, long dashed), the \textcite{Robertson:2015uda} case
  (blue, short dashed), and the \textcite{Madau:2014bja} case (green, dotted)
  to the fiducial spectrum (red, thick solid).
  {\em Bottom:\/} Average of the source redshift for the fiducial case.}
\label{fig:Maxwell}
\end{figure}

We show the normalised redshift distributions $f_{\rm cc}(z)$ in
Fig.~\ref{fig:fcc}. After convolution with the SN emission spectrum they
yield very similar neutrino distributions. To illustrate this point we
assume a Maxwell-Boltzmann distribution $f_\nu(E)=(E^2/2T^2)\,e^{-E/T}$
for the time-integrated SN emission spectrum (see
Sec.~\ref{sec:reference-SN-signal}). In this case the
fiducial redshift distribution produces the DSNB spectrum shown in the
top panel of Fig.~\ref{fig:Maxwell}. The other redshift distributions
produce similar spectra, so we show the fractional difference to the
reference case (middle panel).  The detection interval is 10--25~MeV,
so for $T\sim4~{\rm MeV}$ this corresponds to roughly 2--6 on the
horizontal axis of Fig.~\ref{fig:Maxwell}. At the lower end of this
interval, where the detectable flux is largest, the differences are
very small, but up to 30\% at the upper end for the \textcite{Mathews:2014qba}
 case. Even though the star-formation history looks quite
different for the red and blue cases, the final DSNB spectrum is
nearly the same.

Overall the DSNB spectrum is fairly insensitive to the exact redshift
distribution $f_{\rm cc}(z)$.  In keeping with previous studies we use
the distribution provided by \textcite{Yuksel:2008cu} as a fiducial case that
was shown as a red line in the figures of this section.

In the bottom panel of Fig.~\ref{fig:Maxwell} we show the average
redshift contributing to the DSNB spectrum (only for our fiducial case)
at a given energy. Even at the lower end of the detection interval, $\langle z\rangle$ is less than
1, and significantly smaller at higher energies. Therefore, the main
contribution comes from relatively low redshifts.

The DSNB derived from a Maxwell-Boltz\-mann source spectrum
is strongly anti-pinched (average energy for our fiducial case
$\langle E\rangle=1.41\,T$ and pinching parameter 0.84) and not well
represented by a Gamma distribution of the form of
Eq.~\eqref{eq:gamma-dist}. However, one finds that the decreasing part
of the spectrum is very close to an exponential $e^{-E/T}$ and a good
overall fit to the fiducial case is
\begin{equation}\label{eq:DSNB-fit-1}
  g_\nu(E/T)=1.15\,{\rm arctan}\bigl[3\,(E/T)^{3/2}\bigr]\,e^{-1.03\,E/T}.
\end{equation}
The deviation of this fit from our fiducial spectrum is shown in the
middle panel of Fig.~\ref{fig:Maxwell} as a thin red line. The
deviation is smaller than the spread of different cases of
star-formation histories.

The main uncertainty of the DSNB prediction is the total number of
core-collapse events shown in Eq.~\eqref{eq:cc-rate}.  Moreover, these
predictions involve an overall uncertainty in converting the
star-formation rate into a core-collapse rate (the factor $k_{\rm
  cc}$).  A mismatch of about a factor of 2 between direct SN
observations and the core-collapse rate estimated from star formation
was found, the so-called SN-rate problem~\cite{Horiuchi:2011zz}. The
most likely explanation is dust extinction, especially at higher
redshift, or a relatively large fraction of dim SNe, and in particular
of non-exploding, BH forming cases~\cite{Horiuchi:2011zz,Horiuchi:2014ska,Adams:2013ana,Kochanek:2013yca}.

\subsection{Average emission spectrum}

\begin{figure}[b]
\includegraphics[scale=0.55]{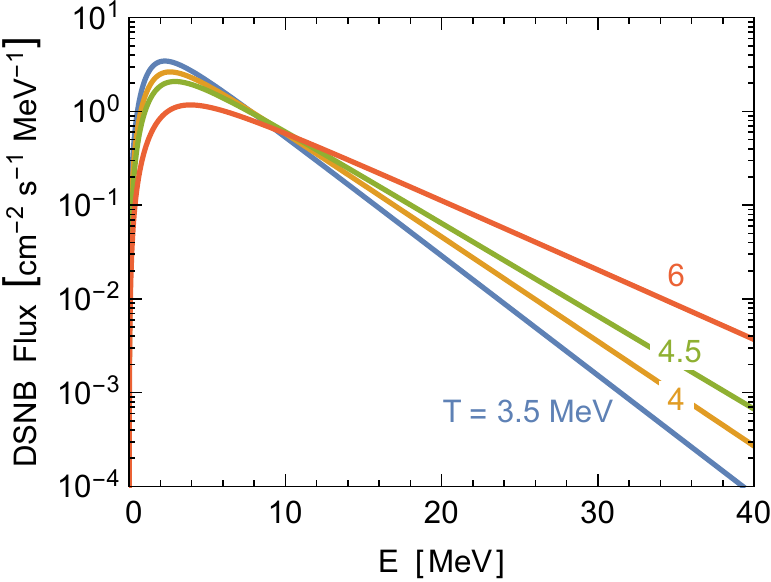}
\caption{DSNB flux in one species according to
  Eq.~\eqref{eq:DSNB-MB-1}. The emission spectrum is taken to be
  Maxwell-Boltzmann with the indicated temperatures.}
\label{fig:DSNB-simple}
\end{figure}

The sparse data of SN~1987A are not detailed enough to give a good
estimate of the neutrino spectrum and also need not be representative
of the average case. Therefore, DSNB predictions depend on numerical
SN models. To get a first impression we assume that the
time-integrated spectrum is of Maxwell-Boltzmann type.  With
Eq.~\eqref{eq:DSNB-2} the DSNB flux for a given species $\nu$ is
\begin{eqnarray}\label{eq:DSNB-MB-1}
  \frac{d\Phi_\nu}{dE}&=&4.45~{\rm cm}^{-2}~{\rm s}^{-1}~{\rm MeV}^{-1}
  ~\frac{n_{\rm cc}}{10^7~{\rm Mpc}^{-3}}
  \nonumber\\
  &\times&\frac{6\,E^{\rm tot}_\nu}{2\times10^{53}~{\rm erg}}\,
  \left(\frac{4~{\rm MeV}}{T}\right)^2\,g_\nu(E/T)\,,
\end{eqnarray}
where $E^{\rm tot}_\nu$ is the total emitted energy in the
considered species $\nu$, and $g_\nu(E/T)$ is the
normalised spectrum of Eq.~\eqref{eq:DSNB-fit-1} that includes our
fiducial redshift distribution.  We show this spectrum for
$T=3.5$, 4, 4.5 and 6~MeV in Fig.~\ref{fig:DSNB-simple}, where values
around 4~MeV would be typical for a core-collapse SN, whereas 6~MeV
could represent a BH forming event with larger spectral
energies.

\begin{table*}[p]
  \caption{Characteristics of the time-integrated neutrino emission of
  the core-collapse models used to synthesize our
  illustrative DSNB example; for details
  see \textcite{Moller:2018kpn}.
  For each species $\nu$ we give the total number of emitted particles $N_\nu$, the
  emitted energy $E_\nu^{\rm tot}$, the average energy $E_{\rm av}$, the pinching
  parameter $\alpha$, and $E_{\rm av}^{\rm DSNB}$
  after convolution with our fiducial redshift distribution.
  The remaining parameters determine the fit function
  of Eq.~\eqref{eq:DSNB-fit-2} for the normalised DSNB spectrum.}\label{tab:DSNB-models}
\begin{tabular*}{\textwidth}{@{\extracolsep{\fill}}llcccccccccc}
\hline
\hline
&& $N_\nu$   & $E_\nu^{\rm tot}$   & $E_{\rm av}$ & $\alpha$& $E_{\rm av}^{\rm DSNB}$ &$a$ & $b$ & $q$ & $p$ & $T$\\
&& $10^{57}$ & $10^{52}~{\rm erg}$ & MeV          &         & MeV                     &    &     &     &     & MeV\\
\hline
$9.6\,M_\odot$ (SN)&
$\nu_e$       &2.01&3.17&~9.8&2.81&4.59&1.347&1.837&1.837&0.990&2.793\\
&$\overline\nu_e$  &1.47&2.93&12.4&2.51&5.83&1.313&1.770&1.703&0.969&3.483\\
&$\nu_x$      &1.61&3.09&12.0&2.10&5.62&1.173&2.350&1.672&0.953&3.432\\
&$\overline\nu_x$  &1.61&3.27&12.7&1.96&5.95&1.145&2.401&1.620&0.944&3.617\\
\hline
$27\,M_\odot$ (SN)&
$\nu_e$       &3.33&5.87&11.0&2.17&5.16&1.575&0.489&1.775&0.794&1.824\\
&$\overline\nu_e$  &2.61&5.72&13.7&2.25&6.41&1.260&1.791&1.667&0.942&3.700\\
&$\nu_x$      &2.56&5.21&12.7&1.88&5.95&1.153&2.106&1.615&0.916&3.400\\
&$\overline\nu_x$  &2.56&5.53&13.5&1.76&6.32&1.111&2.337&1.569&0.916&3.690\\
\hline
$40\,M_\odot$ (BH)&
$\nu_e$               &3.62&9.25&16.0&1.66&7.47&1.065&2.340&1.809&0.866&3.904\\
&$\overline\nu_e$          &2.88&8.61&18.7&1.99&8.75&1.089&3.199&1.801&0.951&5.486\\
&$\nu_x$, $\overline\nu_x$ &1.72&4.83&17.5&1.46&8.21&1.227&1.090&1.314&0.822&3.707\\
 \hline
Mix 1&
$\langle \nu\rangle$    &2.14&4.14&12.1&1.74&5.66&1.471&0.356&1.755&0.724&1.592\\
(59,32,9)&
$\langle\overline\nu\rangle$ &1.94&4.20&13.5&1.80&6.34&1.308&0.940&1.614&0.822&2.687\\
\hline
Mix 2&
$\langle \nu\rangle$    &2.09&4.25&12.7&1.52&5.95&1.362&0.318&1.768&0.690&1.486\\
(59,20,21)&
$\langle\overline\nu\rangle$ &1.88&4.26&14.2&1.64&6.63&1.301&0.734&1.617&0.777&2.439\\
\hline
\end{tabular*}
\vskip12pt
\end{table*}

\begin{figure*}
\hbox to\textwidth{\hfil
  \includegraphics[scale=0.5]{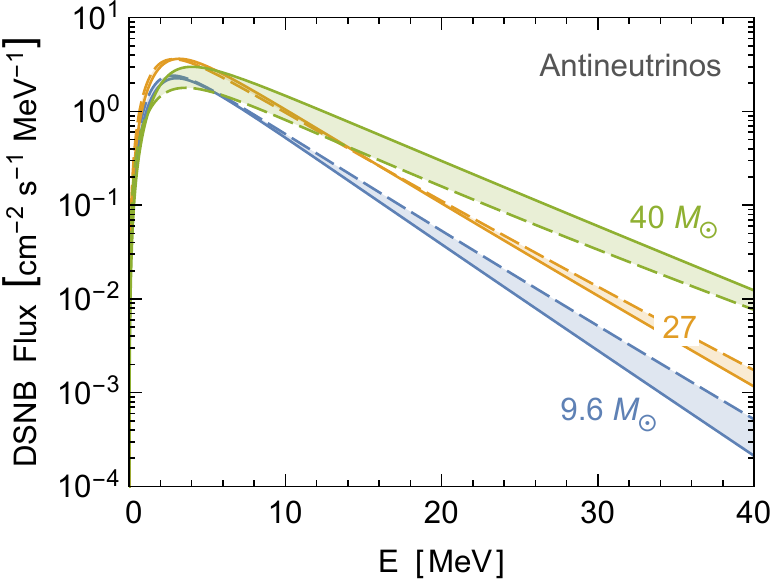}\hskip24pt
  \includegraphics[scale=0.5]{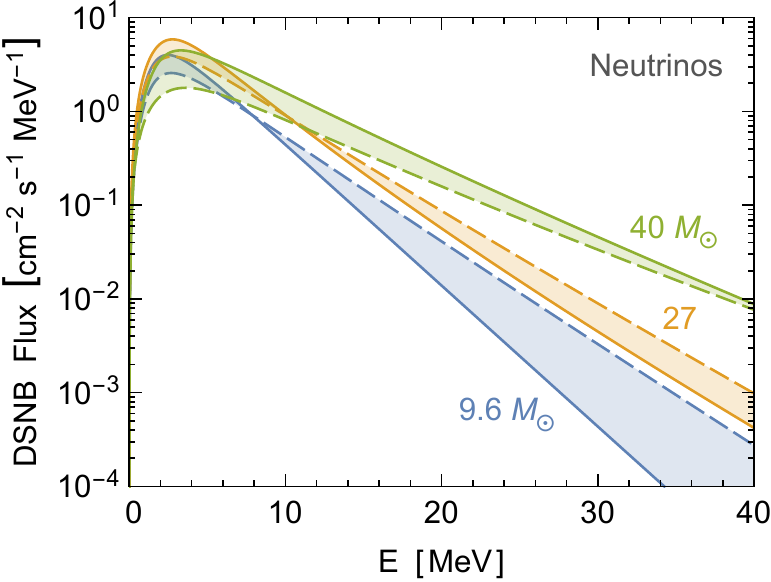}
  \hfil}
\caption{DSNB flux for the 9.6, 27 and $40\,M_\odot$ core-collapse
  models described in the text, where the actual flux would be a
  superposition of these reference cases.  For each neutrino species,
  the flux is given by Eq.~\eqref {eq:DSNB-fit-3} with the parameters
  of Table~\ref{tab:DSNB-models}. Solid lines are for $\nu_e$ or
  $\overline\nu_e$, dashed lines for $\nu_x$ or $\overline\nu_x$, so the flux
  for each mass eigenstate is a superposition of solid and dashed
  spectra, i.e., in the shaded bands, depending on flavor evolution
  upon leaving the source.
  \label{fig:DSNB-models}}
\end{figure*}

\begin{figure*}
\hbox to\textwidth{\hfil
  \includegraphics[scale=0.5]{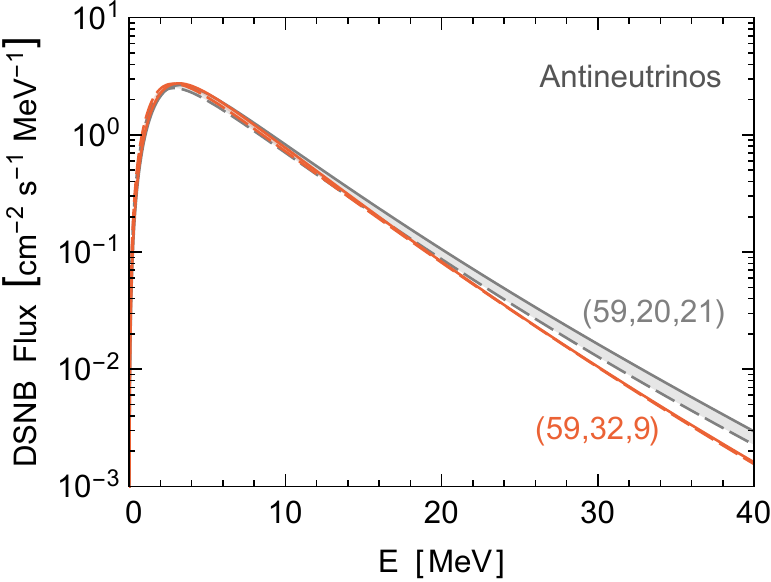}\hskip24pt
  \includegraphics[scale=0.5]{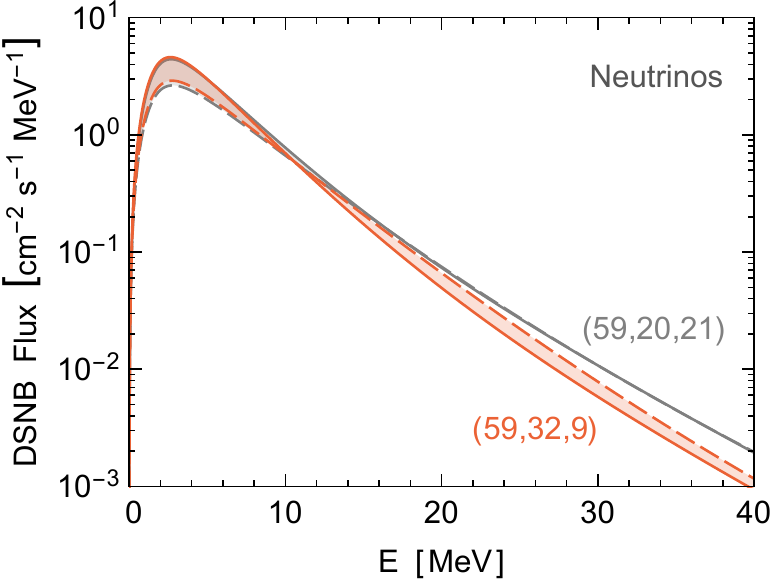}
  \hfil}
\caption{DSNB flux for the two indicated mixes of the 9.6, 27 and $40\,M_\odot$ source models.
  \label{fig:DSNB-mix}}
\end{figure*}

We conclude that around the detection threshold of 10~MeV, the flux
predictions are very similar and depend primarily on the overall
normalization, i.e., the cosmic core-collapse rate and the average
energy release. For larger $E$, the spectra scale essentially as
$e^{-E/T}$ and thus depend strongly on the effective emission
temperature. Therefore, the flux at the upper end of the detection
window ($\sim 20$~MeV) is particularly sensitive to the fraction of
BH forming events~\cite{Nakazato:2008vj,Lunardini:2009ya,Priya:2017bmm,Moller:2018kpn,Schilbach:2018bsg}.

More recent and sophisticated predictions synthesize the average emission
spectrum from a suite of numerical SN models
\cite{Nakazato:2015rya,Horiuchi:2017qja,Moller:2018kpn}. Note that the high-energy tail
of the DSNB spectrum is higher in these papers than what was adopted in the Super-Kamiokande analysis~\cite{Bays:2011si} that was based on
simplified modeling of the SN population and relied on older and more approximate SN models.
For illustration, we here follow \textcite{Moller:2018kpn} and consider
three components, a $9.6\,M_\odot$ progenitor model, representing the
range 8--$15\,M_\odot$ including electron-capture SNe, a $27\,M_\odot$ model, representing the
higher-mass exploding cases, and a $40\,M_\odot$ non-exploding case
called ``slow BH formation'' in \textcite{Moller:2018kpn}.
Using a Salpeter initial mass function, the 8--$15\,M_\odot$ range
encompasses 59\% of all collapsing stars.  The minimal prediction
further assumes that the 15--$40\,M_\odot$ progenitors (32\%)
explode, and all progenitors with larger masses (9\%) follow the
slow-BH case. The fiducial case of \textcite{Moller:2018kpn}
assumes a larger fraction of 21\% of BH formation, whereas an extreme case would be
with 41\% such cases, leaving only the 8--$15\,M_\odot$ range to explode.

We show the main characteristics of these spectral components in
Table~\ref{tab:DSNB-models}.  The exploding models use four-species
neutrino transport and thus provide separate emission spectra for
$\nu_e$, $\overline\nu_e$, $\nu_x$ and $\overline\nu_x$. For each
model and each
species we show the total number of emitted particles $N_\nu$, the
emitted energy $E^{\rm tot}_\nu$, and the average neutrino energy
$E_{\rm av}$ and pinching parameter $\alpha$ of the time-integrated
spectrum. The overall emitted energy is $E_{\rm tot}=E^{\rm
  tot}_{\nu_e}+E^{\rm tot}_{\overline\nu_e}+ 2E^{\rm tot}_{\nu_x}+2E^{\rm
  tot}_{\overline\nu_x}$. We convolve these emission spectra with our
fiducial redshift distribution, which does not change $N_\nu$ but only
the spectral shape.  Inspired by Eq.~\eqref{eq:DSNB-fit-1}, we
approximate the spectra by fit functions of the form
\begin{equation}\label{eq:DSNB-fit-2}
  g_\nu(E)= \frac{a}{T}\,{\rm arctan}\left[b \left(\frac{E}{T}\right)^q\right]\,
  \exp\left[-\left(\frac{E}{T}\right)^p\right]\,,
\end{equation}
where the global factor $a$ is constrained by normalization
$\int_{0}^{\infty}dE\,g_\nu(E)=1$. In the measurement region,
$E\gtrsim10~{\rm MeV}$, the spectrum scales as $\exp[-(E/T)^p]$ with
$p\sim1$. The fit parameter $T$ sets the energy scale and is one way
of defining an effective temperature for the non-thermal emission
spectrum. For a given species $\nu$, the DSNB flux is
\begin{equation}\label{eq:DSNB-fit-3}
  \frac{d\Phi_\nu}{dE}=\frac{10.3}{{\rm cm}^{2}~{\rm s}~{\rm MeV}}\,
  \frac{n_{\rm cc}}{10^7~{\rm Mpc}^{-3}}\,\frac{N_\nu}{10^{57}}\,g_\nu(E)\,,
\end{equation}
where the parameter $T$ is assumed to be in units of MeV.  These fits
represent the numerical spectra to better than a few percent,
especially in the detection region. The fractional deviation
always looks similar to the dashed red line in the middle
panel of Fig.~\ref{fig:Maxwell}.

We show the DSNB fluxes for each species in Fig.~\ref{fig:DSNB-models}
for each core-collapse model as if the entire DSNB was caused by only
one of them. The higher-mass models, with a longer period of accretion,
have hotter spectra, especially the BH forming case.
For the exploding cases, the $x$ spectra provide larger fluxes at
high energies than the $e$-flavored
ones, whereas for the BH case, it is opposite because
the long accretion period produces larger fluxes of $\nu_e$ and
$\overline\nu_e$ than of the other species.

In Fig.~\ref{fig:DSNB-mix} we show DSNB fluxes based on the minimal case
of \textcite{Moller:2018kpn} with a mixture
of 59, 32 and 9\% of the 9.6, 27 and $40\,M_\odot$ models (red) and
their fiducial case with 59, 20 and 21\% (gray).
It is intriguing that the mixed cases show a much smaller spread
between the flavor-dependent spectra (solid vs.\ dashed lines).
This effect partly owes to the inverted flavor dependence between
the exploding and non-exploding models.
Moreover, near the detection threshold of around 10~MeV, the
exact mix leaves the DSNB prediction nearly unchanged.

\subsection{Flavor Conversion}

Neutrinos are produced with flavor-dependent fluxes and spectra
so that flavor conversion on the way from the decoupling region
modifies the escaping flavor composition, or rather, the
final mix of mass eigenstates. Moreover, in which way this
effect is relevant depends on the detection method.
As argued in Sec.~\ref{sec:SN-flavor-conversion}, SN neutrino
flavor conversion is not yet fully understood, so by the
current state of the art, there is no reliable prediction.
On the other hand, flavor conversion would be a small effect
on the overall DSNB prediction as seen in
Fig.~\ref{fig:DSNB-mix}, where the difference between the dashed
and solid lines is quite small, especially near
the detection threshold of 10~MeV where most events would
be measured.

Therefore, as a baseline prediction we use fluxes
that are flavor averages of the form
$\Phi_{\langle\nu\rangle}=(\Phi_{\nu_e}+2\Phi_{\nu_x})/3$.
We show the spectral characteristics for our two illustrative
mixtures in Table~\ref{tab:DSNB-models}. For our minimal
Mix~1, we find a total DSNB flux, number density, and
energy density at Earth of neutrinos plus antineutrinos of all flavors
that was shown in Eq.~\eqref{eq:Mix1-results}.
Concerning normalization,
the main uncertainty is the overall core-collapse rate.
Concerning the spectral shape, the main uncertainty
is the fraction of BH-forming cases.

\begin{figure}[b]
\includegraphics[width=0.90\columnwidth]{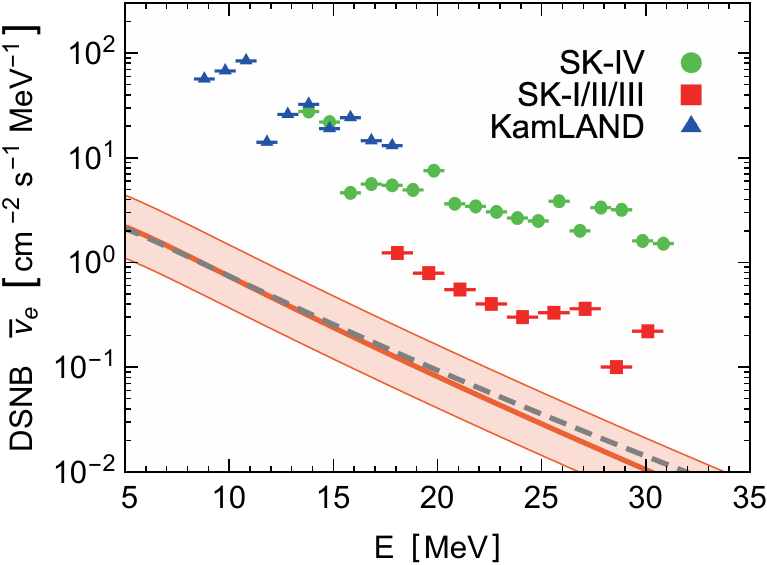}
\caption{Experimental limits on the $\overline\nu_e$ component of the DSNB
by Super-Kamiokande (SK) I/II/III \cite{Bays:2011si}, SK~IV \cite{Zhang:2013tua},
and KamLAND \cite{Gando:2011jza}. [Plot of limits adapted from \textcite{Nakazato:2015rya} with permission.]
The fiducial predicted flux is our Mix~1 (red solid) with an adopted normalization uncertainty of a factor of 2 in either direction (shaded band). We also show Mix~2 (gray dashed)
that includes a larger fraction of BH-forming cases.}
\label{fig:DSNB-limits}
\end{figure}

\subsection{Detection perspectives}

The DSNB has not yet been detected, but the experimental limits
shown in Fig.~\ref{fig:DSNB-limits} have been obtained by the Super-Kamiokande
(SK) I/II/III water Cherenkov detector \cite{Bays:2011si}, SK~IV with neutron tagging
\cite{Zhang:2013tua}, and the KamLAND liquid scintillator detector \cite{Gando:2011jza}.
All of these limits are based on the inverse-beta decay reaction
$\overline\nu_e+p\to n+e^+$.
These limits do not yet reach predictions, but keeping in mind that the
cosmic core-collapse rate and its BH forming component could be larger
than assumed here means that any significant experimental improvement can lead to
a detection.

DSNB detection is not only a question of event rate,
but of identification and rejection of several backgrounds
that can mimic DSNB events. A first detection should become
possible over the next decade
with the upcoming Gd-enhanced Super-Kamiokande
water Cherenkov detector \cite{Beacom:2003nk,Labarga:2018fgu}
and later with a possible Gd-enhanced version
of the upcoming Hyper-Kamio\-kande \cite{Abe:2018uyc}.
Another promising contender is the upcoming
JUNO 20~kt scintillator detector \cite{An:2015jdp}.
A complementary detection channel, using the $\nu_e$ flux, may be
offered by the upcoming liquid argon detector DUNE
at the LBNF facility in the US \cite{Cocco:2004ac,Acciarri:2016ooe}.
A detailed forecast of these opportunities is beyond the scope
of our discussion.
\section{Atmospheric Neutrinos}
\label{sec:ATM}

Atmospheric neutrinos are produced by cosmic rays interacting with the Earth or Sun atmosphere~\cite{Seckel:1991ffa,Ingelman:1996mj,Gaisser:2016uoy,Ng:2017aur,Edsjo:2017kjk,Arguelles:2017eao,Ng:2017aur}. Historically, they were the first
``natural neutrinos'' to be detected \cite{Achar:1965ova,Reines:1965qk} and later
played a fundamental role in establishing flavor oscillations
by the Super-Kamiokande water Cherenkov detector~\cite{Fukuda:1998mi}.
Nowadays, atmospheric neutrinos are employed to measure the neutrino mass and mixing parameters with high precision, while on the other hand
they are a background to the detection of astrophysical neutrinos.

\subsection{Cosmic rays}

Charged particles like electrons, protons and heavier nuclei are accelerated within cosmic reservoirs or on their way to Earth in the presence of astrophysical shocks and magnetic turbulence. These particles constitute the cosmic-ray flux. It further interacts with the Earth atmosphere, producing a secondary particle flux that includes neutrinos. The origin of cosmic rays as well as their composition (the fraction of heavy nuclei and protons) remains subject of vivid debate. The correspondent neutrino flux depends on the cosmic-ray composition, the scattering cross section
with the atmosphere as well as radiative losses, and the branching ratios of the by-products.

Comparing the cosmic-ray composition with the chemical composition of the solar system reveals
interesting differences \cite{Gaisser:2016uoy}.
One is that the relative contribution of heavy nuclei with respect to hydrogen is larger in cosmic rays~\cite{dartois,Wang_2002,DENOLFO20061558,George_2009,Lodder:2003zy}. This could be due to the relative greater ionization energy of hydrogen compared to heavy elements; in fact only ionized or charged particles can be accelerated. An additional, straightforward reason could be a difference in the source composition itself \cite{Casse:1975tg}. Finally, for volatile elements, it is possible that this could be due to a mass-to-charge dependence of the acceleration efficiency, with heavier ions being more favorably accelerated~\cite{Meyer:1997vz}.
Another striking difference is that two groups of elements (Li, Be and B is one; Sc, Ti, V, Cr, and Mn the other) are more abundant in cosmic rays. This is because they are produced in spallation processes (scattering of cosmic rays in the interstellar medium)
instead of stellar nucleosynthesis~\cite{Tanabashi:2018}.

Turning to the energy distribution, above 10~GeV a good approximation to the differential spectrum per nucleon is given by an inverse power law of the form
\begin{equation}
\frac{dN_N}{dE}\propto E^{-(\gamma+1)} \ ,
\end{equation}
where $\gamma\approx 1.7$ up to around $3\times10^6$ GeV, e.g., $\gamma_{\rm{proton}}= 1.71\pm0.05$~\cite{Gaisser:2016uoy}, and $\gamma\approx2.0$ at larger energies. This spectral break is known as the knee of the cosmic-ray flux. A second break, known as second knee, is near $10^{8}$~GeV. Near $3\times10^{9}$~GeV there is another break known as the ankle. Including the normalization given in \textcite{Tanabashi:2018}, the spectrum between several~GeV and 100~TeV is
\begin{equation}\label{cosmic_spectrum}
\frac{dN_N}{dE}=\frac{1.8 \times 10^4}{(\mbox{GeV/nucleon})~\mbox{m}^2~\mbox{s}~\mbox{sr}}
  \left(\frac{E}{\mbox{GeV/nucleon}}\right)^{-(\gamma+1)}\!\!.
\end{equation}
Below $10$ GeV, all cosmic-ray spectra show ``solar modulation''~\cite{Gleeson:1968zza,straussr,Maccione:2012cu,Cholis:2015gna}, a time variation
caused by the solar wind, a low-energy plasma of electrons and protons ejected by the Sun with its 11~year cycle. The shield-like effect of the solar activity translates to an anti-correlation between the latter and cosmic-ray spectra. Moreover, low-energy particles entering the atmosphere also suffer geomagnetic effects. Therefore, low-energy secondary particle fluxes, including neutrinos, depend on both location and time.

\subsection{Conventional neutrinos}\label{convatmo}

Cosmic rays entering the atmosphere scatter and produce secondary particles, especially charged or neutral pions and kaons, which in turn decay and produce the ``conventional neutrinos''
\cite{Stanev:2004ys} as a main contribution at low energies. The detailed decay chains are\footnote{More details on the decay channels and their branching ratios can be found in \textcite{Agashe:2014kda} and \textcite{Tanabashi:2018}.}
\begin{align}
\pi^{\pm}\to\, &\mu^{\pm}+\nu_\mu  (\overline{\nu}_\mu)  \nonumber\\
&\downarrow \nonumber \\
 &e^\pm +\nu_e (\overline{\nu}_e)+\overline{\nu}_\mu  (\nu_\mu) \ ,
\end{align}
\begin{equation}
K^\pm\rightarrow \mu^\pm+\nu_\mu(\overline{\nu}_\mu) \ .
\end{equation}
Three-body decays of kaons also occur, for example
\begin{equation}
K^\pm\rightarrow \pi^0+e^\pm+\nu_e(\overline{\nu}_e) \ .
\end{equation}
Some of the kaons decay purely into pions, for example in processes such as
\begin{equation}
K^\pm\rightarrow \pi^\pm+\pi^0 \ ,
\end{equation}
which in turn produce neutrinos.
On the other hand, cosmic rays also produce $\pi^0$ that decay to photons
\begin{equation}
\pi^0 \rightarrow \gamma + \gamma \ ,
\end{equation}
establishing a connection between high-energy astrophysical photons
and neutrinos (see Sec.~\ref{sec:HE}).

Up to 1~GeV, all muons decay before reaching the ground, implying a neutrino $\mu/e$ flavor ratio of
\begin{equation}
\frac{\nu_\mu+\overline{\nu}_\mu}{\nu_e+\overline{\nu}_e}\simeq 2\ .
\end{equation}
At somewhat higher energies, $\mu$ decay becomes negligible and $\pi$ and $K$ decays dominate.
The resultant $\nu_\mu$ plus $\overline\nu_\mu$ flux is given by the fit~\cite{Stanev:2004ys,Gaisser:2002jj,Gaisser:2019efm}\footnote{We thank T.~Gaisser for insightful clarifications concerning the semi-analytical fit to the atmospheric muon neutrino flux.}
\begin{align}
\frac{dN_\nu}{dE_\nu}\simeq{}&0.0096 \, \frac{1}{\rm  cm^2 \, s\, sr\, GeV } \left(\frac{E_\nu}{\rm GeV}\right)^{-2.7} \nonumber
\\
&{}\times\left(\frac{1}{1+\frac{3.7E_\nu\cos\theta}{\epsilon_\pi}}+\frac{0.38}{1+\frac{1.7E_\nu\cos\theta}{\epsilon_K}}\right)\ ,
\end{align}
where $\epsilon$ is the energy scale for the most probable process in propagation (decay vs.~interaction); for pions  $\epsilon_\pi\simeq 115$~GeV and for kaons $\epsilon_K\simeq 850$~GeV.
Moreover, $\theta$ is the zenith angle of observation.

\begin{figure}
  \includegraphics[width=0.80\columnwidth]{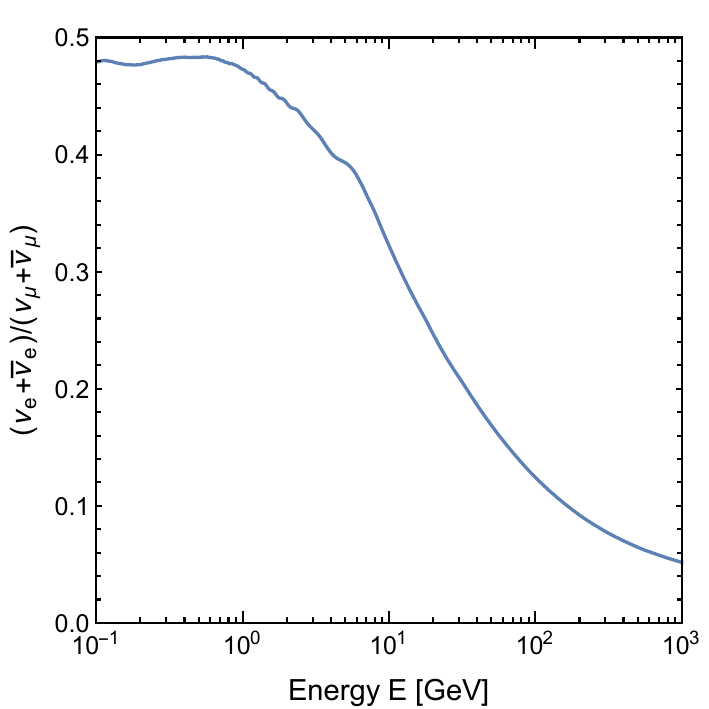}
\vskip-4pt
\caption{{Atmospheric neutrino $e/\mu$ flavor ratio, corresponding
to the source fluxes (no oscillations)
shown in Fig.~\ref{atmospheric}.}}\label{atmospheric-ratio}
\end{figure}

Asymptotically at high energies the $\pi:K$ neutrino production is $1:3$. The resultant flavor is mainly muonic, while the electronic one becomes negligible as shown in Fig.~\ref{atmospheric-ratio}. This feature can be understood observing that, at high enough energies, the muon neutrino flux scales as $E_\nu^{-3.7}$ while electron neutrinos originating from muons scale
as $E_\nu^{-4.7}$, where the extra power of $E_\nu^{-1}$ comes from the muon Lorentz factor that reduces
its decay probability, making scattering more likely. At some energy, of course, the $\nu_e$ and $\overline\nu_e$ produced directly by kaon decays take over and then their flux also scales as $E_\nu^{-3.7}$, but with a much smaller flux than the muon flavor.

In first approximation, atmospheric neutrinos seen by a detector at or below the ground are
isotropic. The atmospheric source mass intersected by a differential solid angle $d\Omega$ scales with $r^2$ which cancels
exactly the geometric $1/r^2$ flux dilution, where $r$ is the distance between detector and atmosphere in the chosen direction.

There are important corrections to this simple picture. In fact in the few-GeV range, that was crucial for establishing flavor oscillations, the flux is essentially up-down symmetric, but it is enhanced in the horizontal direction because there is a longer decay path before muons reach the ground~\cite{Gaisser:2002jj,Honda:2004yz,Stanev:2004ys}. A similar effect
pertains at high energies for kaon decays.

At energies beyond a few TeV, the Earth is no longer completely transparent to neutrinos so that the flux from below is diminished~\cite{Nicolaidis:1987fe,Nicolaidis:1990jm,GonzalezGarcia:2007gg,Donini:2018tsg}. For energies up to about 1~PeV, neutrinos of any flavor are more efficiently absorbed than antineutrinos
because they scatter on nuclei, while the scattering on electrons is negligible. Nuclei in the Earth matter are heavy and contain more neutrons (quark content $udd$) than protons (quark content $uud$). Taking into account that neutrinos (respectively antineutrinos) can exchange a $W$ boson with $d$ (respectively $u$), the reaction $\nu+A\rightarrow l+B$ becomes more likely than
$\overline{\nu}+A\rightarrow l+B$. With increasing energy, valence quarks become negligible relative to sea quarks, so the cross sections of $\nu$ and $\overline\nu$ become asymptotically equal. The only exception is provided by the Glashow resonance~\cite{Glashow:1960zz}
\begin{equation}\overline{\nu}_e+e^-\rightarrow W^-\rightarrow X
\end{equation}
at $E_\nu\simeq m_W^2/2m_e\simeq 6$ PeV, so there is a region in which $\overline{\nu}_e$ are more likely to be absorbed than $\nu_e$.

\subsection{Prompt neutrinos}

Atmospheric neutrinos produced by charmed mesons are called prompt neutrinos \cite{Volkova:1980sw,Gondolo:1995fq,Pasquali:1998xf,Enberg:2008te,Gaisser:2019efm}. They consist
of equal amounts $e$ and $\mu$ flavor and a very small $\tau$ component.
The prompt flux contribution was expected to be large in the TeV to PeV range ($\epsilon_{\rm charm}\simeq$ PeV), where the only other contribution comes from kaon decay. The latter is distinguishable thanks to its angular distribution
which is enhanced in the horizontal direction because of the larger kaon decay path.
Prompt neutrinos, instead, are isotropic up to high energies because of the short charmed-meson
lifetime of $10^{-12}$--$10^{-13}$~s. Moreover, the prompt flux is harder so that it will dominate
beyond a certain energy. Semi-analytical expressions for the prompt flux can be found in \textcite{Volkova:1999uz}. The uncertainty in the estimates of this flux is quite large~\cite{Garzelli:2016xmx}; while conventional atmospheric neutrino predictions are affected by a $\sim10\%$ uncertainty \cite{Honda:2006qj}, the prompt flux has large uncertainties due to poor knowledge of the charm meson production processes~\cite{Garzelli:2015psa}.

Concerning recent developments, IceCube has not found a significant prompt component~\cite{Aartsen:2014muf,Aartsen:2016xlq}. Moreover, recent calculations accounting for the latest measurements of the hadronic cross sections predict a prompt neutrino flux that is generally lower than its previous benchmark estimation~\cite{Bhattacharya:2016jce}.

\subsection{Predictions and observations}

\begin{figure}
\includegraphics[width=1.0\columnwidth]{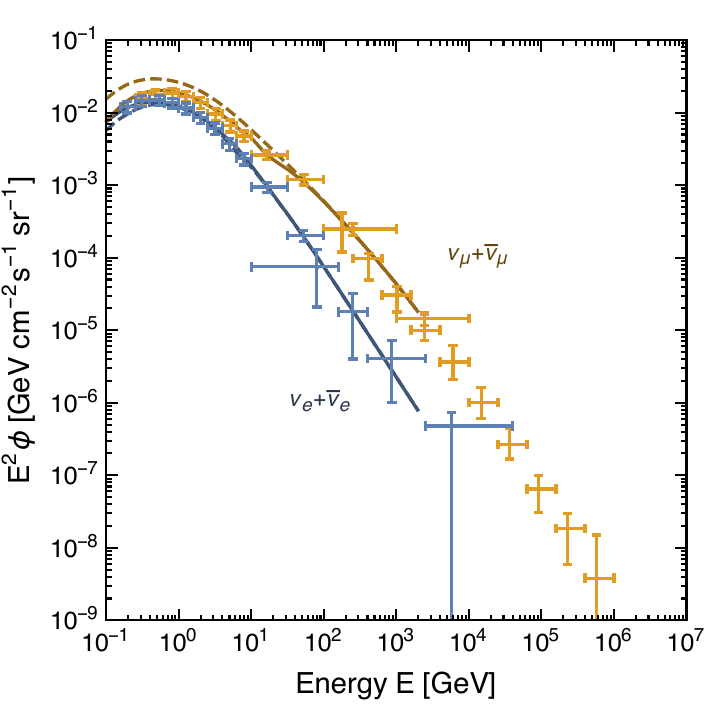}
\caption{Atmospheric neutrino flux per solid angle, averaged over directions,
as a function of energy for $\nu_\mu+\overline{\nu}_\mu$ (upper curve, orange)
and $\nu_e+\overline{\nu}_e$ (lower curve, blue). The data points
at low and medium energies represent the Super-Kamiokande observations
\cite{Richard:2015aua} and of IceCube at high energies \cite{Aartsen:2014qna,Aartsen:2015xup}. The dashed lines are theoretical predictions at the Kamioka site for average solar activity \cite{Honda:2015fha}; the solid lines are the expected fluxes including flavor oscillations.
The $\nu_\tau+\overline{\nu}_\tau$ flux appearing in this case is not shown, corresponding to
the difference between the orange dashed and solid lines.
We thank F.~Capozzi for providing the tables used to include oscillations.}\label{atmospheric}
\end{figure}

To predict the atmospheric neutrino flux one needs to solve a set of transport equations, which are coupled integro-differential equations. While semi-analytical approximations exist, numerical solutions are more reliable.
To reproduce the theoretically expected flux in Fig.~\ref{atmospheric}
(dashed lines), we use the tables publicly available on
\href{http://www.icrr.u-tokyo.ac.jp/~mhonda/}{Mitsuhito Honda's
  homepage} \cite{Honda:2015fha}. For the very low energy ($\lesssim
100\, \rm MeV$) flux, included in Fig.~\ref{fig:GUNS0}, we use FLUKA results~\cite{Battistoni:2005pd}.
We choose Kamioka as a site because the geographic dependence is more important for low-energy neutrinos which are measured mostly
by Super-Kamiokande. Because we are not aiming for a high-precision fit, we consider the flux under the mountain in Kamioka and take the Sun at average magnetic activity.
As explained earlier, the $\mu/e$ flavor ratio begins approximately at 2 at low energies and then increases.
Concerning uncertainties, the authors of \textcite{Fedynitch:2012fs} have quantified the systematic
influence caused by the choice of primary cosmic-ray flux models and the interaction model. The average errors
on the $\nu_\mu$ and $\nu_e$ fluxes at high energies were found to be
$^{+32\%}_{-22\%}$ and $^{+25\%}_{-19\%}$ respectively.

Neutrinos produced in the atmosphere can change flavor before reaching the detector. For the given mixing parameters (Appendix~\ref{sec:MassMatrix}) and for GeV energies, $\nu_e$ and $\overline\nu_e$ remain essentially unaffected, because the mean weak potential describing the Earth matter effect is large compared with $\delta m^2/2E$ and because they have only a small admixture of the third mass eigenstate. The main effect derives from two-flavor oscillations in the $\nu_\mu$-$\nu_\tau$ sector, driven by the ``atmospheric mass difference'' $\Delta m^2\sim (50~{\rm meV})^2$ with an oscillation length $L_{\rm osc}=4\pi E/\Delta m^2\sim990~{\rm km}~E/{\rm GeV}$. Therefore, neutrinos from above show the primary flavor content, whereas those from below, after traveling thousands of kilometers,
show significant $\nu_\mu$ disappearance. This up-down asymmetry was the smoking-gun signature detected by Super-Kamiokande \cite{Fukuda:1998mi} and honored with the 2015 Physics Nobel Prize for Takaaki Kajita.

The solid lines in Fig.~\ref{atmospheric} show the predicted angle-averaged fluxes when flavor oscillations are included.
Without aiming for a precision comparison in our plot, they agree well with the measured fluxes which are taken
from Super-Kamiokande at low and medium energies ($10^{-1}$--$10^3$ GeV) \cite{Richard:2015aua} and from IceCube at high energies ($10^2$--$10^6$~GeV) \cite{Aartsen:2014qna,Aartsen:2015xup}.
While the primary $\nu_\tau$ flux of prompt neutrinos is very small, there is a large $\nu_\tau$ component from flavor conversion
that we do not show and that is difficult to measure because of the short lifetime of the
$\tau$ lepton produced in charged-current interactions. It was only recently that first evidence for atmospheric $\nu_\tau$ appearance
was reported by IceCube DeepCore \cite{Aartsen:2019tjl}.

\subsection{Experimental facilities}

The main experimental facilities sensitive to atmospheric neutrinos
have been IceCube~\cite{Aartsen:2014qna,Aartsen:2015xup},
Super-Kamiokande~\cite{Richard:2015aua}, SNO~\cite{Aharmim:2009gd},
and MINOS~\cite{Adamson:2011ig}. SNO, although mainly built to detect
solar neutrinos, also detected high-energy atmospheric neutrinos. SNO
was located 2~km underground, and therefore near-horizontal
downward-going muons with typical energies of 100~GeV originated as a
result of the atmospheric neutrino interactions. Given the measured
atmospheric $\Delta m^2$, the effect of flavor conversion is
tiny for the  near-horizontal downward-going muons. Hence, these muon
data were used to calibrate the estimated atmospheric neutrino
flux. SNO is currently being replaced with its successor
SNO+~\cite{Lozza:2019ptk}, which however does not have atmospheric
neutrinos as a main goal for the next future.
MINOS was a long-baseline neutrino oscillation experiment and has been the first  magnetized tracking detector for atmospheric neutrinos.

IceCube measures atmospheric neutrinos as a background for very
high-energy astrophysical neutrinos in the range 100~GeV--400~TeV.
Neutrinos with energies up to 1~GeV will have the final-state
particle ``fully contained'' in the detector. Muon
neutrinos with higher energies may result in a muon leaving the
detector, the so-called ``partially contained'' events. To
measure the atmospheric flux accurately, it is important to pinpoint
the vertex position of the interaction and classify the neutrino event
accordingly. IceCube DeepCore~\cite{Collaboration:2011ym} is an infill
of 8~strings added to the IceCube array and is dedicated to the
detection of neutrinos with energy below 100~GeV. Similarly to IceCube, the deep-sea Cherenkov detectors, ANTARES
and its successor Km3NeT~\cite{VanElewyck:2019uch}, allow us to
exploit atmospheric neutrinos to study flavor oscillation
physics.

The largest statistics of atmospheric neutrinos for neutrino
oscillation studies is dominated by the Super-Kamiokande data.
Hyper-Kamiokande \cite{Abe:2018uyc} will provide
an even larger amount of data.

Among the detectors specifically dedicated to the observation of
atmospheric neutrinos, there is the project of the
India-based Neutrino Observatory (INO)~\cite{Indumathi:2015hfa}.
It will be located in a 1.2~km deep cave near Theni in India.
INO promises to provide a precise measurement of neutrino mixing parameters.

\subsection{Solar atmospheric neutrinos}

\begin{figure}[htb]
  \includegraphics[width=0.90\columnwidth]{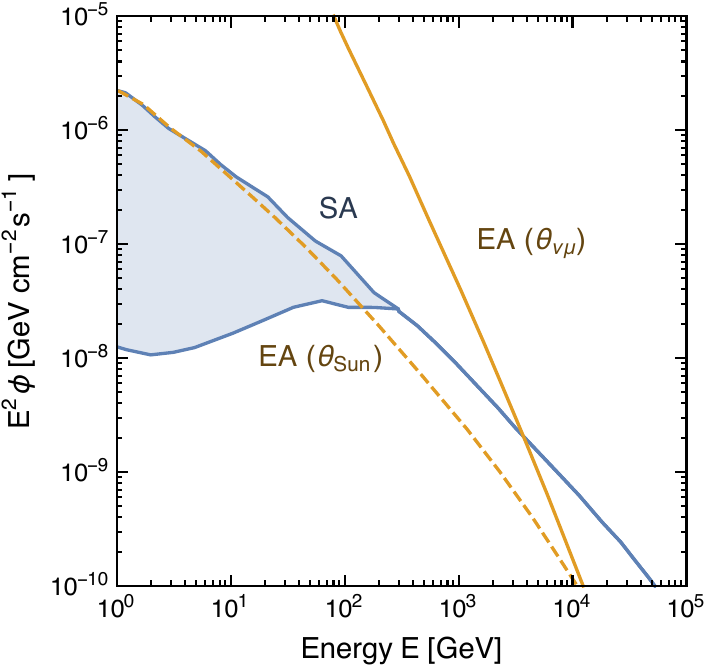}
\caption{The $\nu_\mu+\overline{\nu}_\mu$ solar atmospheric neutrino (SA) flux (blue) compared to the Earth (EA) one (orange), the latter integrated over the solar angular cone (dashed) and over the muon-neutrino angular cone (solid), see \textcite{Ng:2017aur} and references therein. The SA uncertainty at low energies (blue shaded region) is due to the modelling of the magnetic fields at the solar surface and those carried by the solar wind. We thank Kenny C.~Y.~Ng for providing the data for this figure.}\label{atmosphericbeacom}
\end{figure}

An additional contribution to the GUNS comes from the Sun in the form of ``solar atmospheric neutrinos''
that are produced by cosmic-ray interactions in the solar atmosphere \cite{Ingelman:1996mj,Ng:2017aur,Edsjo:2017kjk,Arguelles:2017eao}.
While the production processes are analogous to those in the Earth atmosphere,
the Sun atmosphere is thinner so that pions and kaons can travel much larger distances without collisions. This results in a neutrino flux that is both larger and harder at high energies as shown in Fig.~\ref{atmosphericbeacom}. The detection of low-energy ($\lesssim 1\rm \, TeV$) solar atmospheric neutrinos, while not possible with on-going experiments, would be very useful to probe the magnetic field of the solar atmosphere~\cite{Ng:2017aur}.

The fluxes shown in Fig.~\ref{atmosphericbeacom} correspond to
Fig.~1 of \textcite{Ng:2017aur}, where the authors assumed the solar magnetic field of \textcite{Seckel:1991ffa}
for the flux up to 300~GeV and the model of \textcite{Ingelman:1996mj} at larger energies.
What is shown is the primary $\nu_\mu+\overline\nu_\mu$ flux that will be diminished by
flavor oscillations before reaching Earth.

To compare the solar flux with the diffuse background of Earth atmospheric neutrinos, the latter
must be integrated over a suitable solid angle. A naive estimate is the solar angular cone $\theta_{\rm Sun}\simeq0.17^\circ$
(dashed orange line in Fig.~\ref{atmosphericbeacom}). However, muons
coming from different directions can decay producing neutrinos along a
direction lying in the solar angular cone, so the flux must be
integrated over the energy-dependent muon-neutrino separation angle $\theta_{\nu\mu}\simeq 1^\circ \sqrt{\rm 1\, TeV/E_\nu}$~\cite{Ng:2017aur} (solid orange line in Fig.~\ref{atmosphericbeacom} marked EA).

The detection of high-energy ($\gtrsim 1\, \rm TeV$) solar atmospheric neutrinos is conceivable
in ten years of data taking by IceCube and KM3NeT \cite{Ng:2017aur,In:2017qma}, and would mark an important milestone for neutrino astronomy, as well as being an important calibration source for future neutrino telescopes in different hemispheres.

\section{Extra-terrestrial high energy neutrinos}\label{sec:HE}

The era of high-energy neutrino astronomy was born with the detection
of neutrinos of astrophysical origin by the IceCube Neutrino
Observatory~\cite{Aartsen:2013bka,Aartsen:2013jdh,Aartsen:2016xlq}. These
events have energies between few TeV to few PeV. Their arrival
directions do not exhibit anisotropies, suggesting that only up to
$\sim 1\%$ of the observed flux may come from our
Galaxy~\cite{Denton:2017csz,Albert:2018vxw,Ahlers:2013xia}.  More neutrinos are instead
expected from sources distributed on cosmological scales, such as dim
or choked astrophysical jets, star-forming galaxies (SFGs), gamma-ray
bursts (GRBs), active galactic nuclei (AGNs), and galaxy clusters.
For recent dedicated reviews see for example \textcite{Ahlers:2018fkn}, \textcite{Ahlers:2018mkf}, \textcite{Meszaros:2015krr}, \textcite{Waxman:2015ues}, and \textcite{Murase:2015ndr}.

\subsection{Production mechanisms and detection prospects}

Neutrinos in the energy range of interest are produced by cosmic-ray interactions in the source, its surroundings, or during cosmic-ray propagation en route to Earth. The reactions involve proton-proton (``$pp$'') or proton-photon (``$p\gamma$'') interactions, leading to the following production channels for neutrinos and gamma rays: $\pi^0 \rightarrow \gamma + \gamma$, $\pi^{\pm} \rightarrow \mu^{\pm} + \nu_\mu(\overline\nu_\mu)$ and $\mu^{\pm} \rightarrow e^{\pm} + \overline{\nu}_\mu(\nu_\mu) + \nu_e(\overline\nu_e)$ in analogy to atmospheric neutrino production (Sec.~\ref{convatmo}).

Before absorption and reprocessing of very high-energy $\gamma$ rays, the relative fluxes of neutrinos and gammas is approximately regulated by the ratio of $\pi^{\pm}$ to $\pi^0$ production, whereas the $\nu$ flavor distribution would be $\nu_e:\nu_\mu:\nu_\tau \simeq 1:2:0$. After a long distance of propagation, the oscillation-averaged composition reaching the detector is expected to be $\nu_e:\nu_\mu:\nu_\tau \simeq 1:1:1$~\cite{Learned:1994wg,Farzan:2008eg}.

The diffuse neutrino intensity at Earth from extragalactic sources is given
by the integral of the spectral distribution for each source, $F_{\nu_\alpha}$, convolved with the source distribution (a function of redshift $z$ and source luminosity $L$) over the co-moving volume $\rho(z,L)$
\begin{eqnarray}
\phi(E_\nu) &=& \frac{1}{4\pi} \int_0^\infty \!dz \int_{L_{\rm min}}^{L_{\rm max}} dL_\nu\,\frac{1}{H(z)} \rho(z,L_\nu)
\nonumber\\
&&\kern6em{}\times\sum_\alpha F_{\nu_\alpha}\left[(1+z) E_\nu\right]
\,,
\label{eq:phinu}
\end{eqnarray}
with $H(z)$ being the Hubble factor at redshift $z$.

\begin{figure}
\includegraphics[width=1.0\columnwidth]{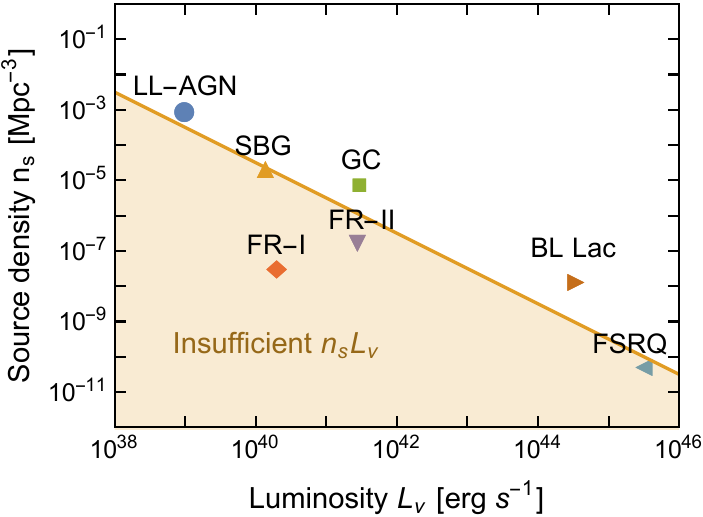}
\caption{Source density vs.\ neutrino luminosity for potential sources of high-energy neutrinos. The markers are examples of benchmark astrophysical sources:
low-luminosity active galactic nuclei (LL-AGN), starburst galaxies (SBG), galaxy clusters (GC), BL Lacertae objects (BL-Lac), Fanaroff-Riley galaxies of type I and II (FR-I and FR-II), flat-spectrum radio quasars (FSRQ). See \textcite{Mertsch:2016hcd} and \textcite{Ackermann:2019ows} for details on the estimation of $(L_\nu,n_{\rm s})$.}\label{fig:kowalski}
\end{figure}

This equation can be approximately expressed in the form~\cite{Waxman:1998yy}
\begin{equation}
\label{eq:phidiff}
\phi_\nu = \xi\,\frac{L_\nu n_{\rm s} R_H}{4\pi}\ ,
\end{equation}
where $\xi$ accounts for the redshift evolution of sources ($\xi=2$ or 3 is usually assumed for sources following the star-formation rate), $n_{\rm s}$ is the source density, and $R_H = c/H_0 \simeq 400$~Mpc is the Hubble radius. Comparing Eq.~\eqref{eq:phidiff} with the diffuse flux observed by IceCube ($2.8 \times 10^{-8}$~GeV$/$cm$^2$~s~sr), we obtain~\cite{Gaisser:2016uoy}
\begin{equation}
n_{\rm s} L_\nu = \frac{4 \times 10^{43}}{\xi} \frac{\mathrm{erg}}{\mathrm{Mpc}^3~\mathrm{yr}} \sim 10^{43} \frac{\mathrm{erg}}{\mathrm{Mpc}^3~\mathrm{yr}}\ .
\label{eq:upplim}
\end{equation}
This relation  provides the minimum power density necessary  to produce the neutrino flux observed by IceCube. Hence any viable neutrino source needs to sit above the line defined by Eq.~(\ref{eq:upplim}) in the luminosity-density plane in Fig.~\ref{fig:kowalski}; such a plot was originally shown in various forms in \textcite{Silvestri:2009xb}, \textcite{Murase:2012df}, and \textcite{Kowalski:2014zda}.

\subsection{Multi-messenger connections}

Assuming that all particles populating the high-energy sky originate from the same source classes, the cosmic energy density of high-energy neutrinos should be connected to the one of $\gamma$-rays observed by the {\it Fermi\/} Telescope~\cite{Fornasa:2015qua} and to the one of ultra-high-energy cosmic rays seen by the Auger Observatory~\cite{doi:10.1146/annurev-astro-082214-122457}, see e.g.\ \textcite{Murase:2013rfa} and \textcite{Ahlers:2018fkn} for more details.

The extragalactic $\gamma$-ray background observed by {\it Fermi\/} derives from point-like sources and an isotropic component.  Intriguingly, the current IceCube data cannot be consistently interpreted by employing the same composition of sources. This is especially true for the 10--100~TeV neutrino energy spectrum that cannot be fitted  by invoking a common origin for neutrinos and $\gamma$-rays \cite{Denton:2018tdj,Murase:2015xka}.

A direct correlation between TeV--PeV neutrinos and ultra-high-energy cosmic rays should also exist, but no clear evidence has been found yet~\cite{Aartsen:2015dml,Moharana:2015nxa}. Cosmic rays could be trapped in the source because of strong magnetic fields and hence produce neutrinos through collisions with the gas. The efficiency of this process is related to the total energy stored in the source under the assumption that it is calorimetric.

Cosmic rays above $3 \times 10^{18}$~eV are thought to be of extragalactic origin, while a mainly galactic origin is expected at smaller
energies. The observation of extragalactic cosmic rays allows one to establish an upper bound on the fluence of neutrinos of astrophysical origin produced in cosmic reservoirs; this leads to the so-called Waxman and Bahcall bound~\cite{Waxman:1998yy,Bahcall:1999yr}
\begin{equation}\label{eq:WB}
E_\nu \phi_\nu < 2 \times 10^{-8} \xi~\mathrm{GeV}/(\mathrm{cm}^2~\mathrm{s}~\mathrm{sr})\,,
\end{equation}
where $\xi$ is the same as in Eq.~\eqref{eq:phidiff}. Equation~\eqref{eq:WB}
should be considered as an upper limit on neutrino emission from the sources of ultra-high-energy cosmic rays under the assumption that the spectrum scales as $E^{-2}$, as predicted by Fermi acceleration. Notably, the Waxman and Bahcall bound was derived under the assumption that sources are optically thin to photo-meson and proton-nucleon interactions such that protons are free to escape. If optically thick sources exist, this bound does not hold.

High-energy neutrinos are emitted by a plethora of astrophysical sources, however we will focus on SFGs, GRBs and AGNs. These are the most efficient neutrino emitters. In particular, for what concerns AGNs, we will focus on a sub-class, blazars, currently considered to constitute the bulk of the extra-galactic $\gamma$-ray diffuse emission~\cite{TheFermi-LAT:2015ykq}. Notably, a dozen of the IceCube neutrino events is likely to be connected to a blazar~\cite{IceCube:2018cha,IceCube:2018dnn}, but given their energy, those neutrino events do not contribute to the diffuse IceCube flux.  The non-detection of point sources generating multiple neutrino events from astrophysical sources provides a lower limit on the local density of these sources and an upper limit on their effective neutrino luminosity~\cite{Murase:2016gly,Mertsch:2016hcd}. Finally, we will discuss the predicted flux of cosmogenic neutrinos, produced by cosmic-ray interactions en route to Earth.

\subsection{Star-forming galaxies}

SFGs are stationary sources compared with transient ones, such as GRBs and AGNs, that will be discussed later. SFGs are perfect examples of calorimetric sources \cite{Loeb:2006tw,Waxman:2015ues}. Presumably they produce high-energy neutrinos mostly through $pp$ interactions~\cite{Loeb:2006tw,Thompson:2006np,Lacki:2010vs}.

Beyond normal galaxies, such as our Milky Way, another class of SFGs consists of starburst galaxies. These are individually more luminous than SFGs as they undergo a phase of enhanced star-formation activity (up to 100 times higher than normal galaxies).

Our understanding of star formation has dramatically improved in the last decade. In particular,  the Herschel Space Observatory~\cite{Gruppioni:2013jna}  provided an unprecedented estimation of the infrared luminosity function of SFGs up to redshift 4 and made possible the distinction among different sub-classes. In fact, beyond  normal and starburst galaxies, Herschel provided for the first time information on SFGs containing low-luminosity AGNs or AGNs obscured by dust (after correcting for the contribution of AGNs~\cite{Gruppioni:2013jna}). All these classes contribute to the star-formation activity.

\begin{figure}[b]
\includegraphics[width=1.0\columnwidth]{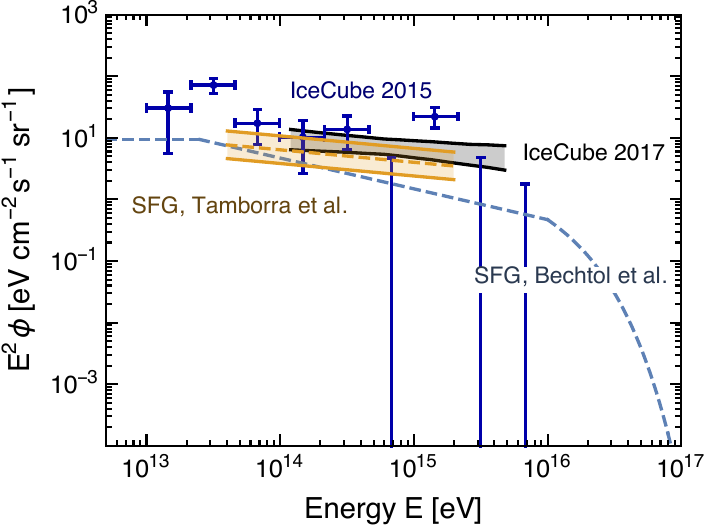}
\caption{Diffuse neutrino flux per flavor $\nu_\alpha+\overline{\nu}_\alpha$ from SFGs. The dashed line reproduces the results of \textcite{Bechtol:2015uqb}, where an upper limit to the  contribution from blazars was calculated by analyzing the $\gamma$-ray flux. The orange band (marked ``SFG, Tamborra et al.'')
reproduces the results of \textcite{Tamborra:2014xia} based on the infrared data. Notice that the spectral shape is slightly different given the different injection spectral indexes adopted in the theoretical estimations. Also shown is the  IceCube neutrino flux per flavor according to \textcite{Aartsen:2015knd} (data points) and a more recent estimation including high-energy data only~\cite{Aartsen:2017mau} (black band marked ``IceCube 2017'').}\label{sfg}
\end{figure}

Among all galaxies, about $38\%$ are normal, $7\%$ are of starburst type, and the remaining ones are SFGs containing AGNs. The abundance of each class varies as a function of redshift, with normal galaxies being more abundant at low redshifts ($z< 1.5$). The $\gamma$-ray energy distribution of normal galaxies is observed to be softer ($F_{\gamma} \propto E_\gamma^{-2.7}$)  on average  than the one of starburst galaxies ($F_{\gamma} \propto E_\gamma^{-2.2}$--$E_\gamma^{-2.3}$~\cite{Ackermann:2012vca,Bechtol:2015uqb}. Finally, star-forming galaxies containing active galactic nuclei can have an energy spectral distribution resembling normal galaxies or starburst galaxies depending on redshift~\cite{Tamborra:2014xia}.

Neutrinos are thought to be produced in SFGs through $pp$ interactions under the assumption that $\mathcal{O}(100)$~PeV cosmic rays  are produced and confined in these sources. This assumption might be optimistic given that  the Galactic cosmic-ray spectrum breaks at $3$~PeV.

As a consequence of $pp$ interactions in the source, a direct connection
between the estimated neutrino and \hbox{$\gamma$-ray} emission can be established~\cite{Thompson:2006np,Lacki:2010vs,Tamborra:2014xia,Senno:2015tra,Sudoh:2018ana}.
One can then estimate the neutrino emission following the modelling proposed in \textcite{Tamborra:2014xia} using the infrared data from the Herschel Space Observatory~\cite{Gruppioni:2013jna}. As the infrared luminosity function is connected to that of \hbox{$\gamma$-rays} \cite{TheFermi-LAT:2015ykq}, one can estimate the correspondent neutrino spectrum by applying the relation
\begin{equation}
\sum_{\nu_\alpha} \phi_{\nu_\alpha}(E_{\nu_\alpha}) \simeq 6 \phi_{\gamma}(E_{\gamma})\ ,
\end{equation}
with $E_\gamma \simeq 2 E_\nu$, and  $\phi_{\gamma}$ the $\gamma$-ray diffuse intensity.
The expected $\phi_{\nu_\alpha}$ from SFGs as a function of $E_\nu$ is shown in Fig.~\ref{sfg} (orange band).

Note that \textcite{Bechtol:2015uqb} recently provided a more conservative upper limit on the expected neutrino emission from star-forming galaxies by relying on the most recent constraints on the diffuse extra-galactic $\gamma$-ray sky from {\em Fermi\/}  \cite{Bechtol:2015uqb,TheFermi-LAT:2015ykq}; this corresponds to the dashed blue line in Fig.~\ref{sfg}.
These results are in agreement with current tomographic constraints~\cite{Ando:2015bva}. Notably, the detection of neutrinos from  stacked searches of star-forming galaxies is currently statistically disfavored~\cite{Feyereisen:2016fzb,Mertsch:2016hcd,Murase:2016gly}.

\subsection{Gamma-ray bursts}

GRBs are among the most energetic transients in our Universe. They are divided in long (${>}\,2$~s) and short (${<}\,2$~s) duration bursts according to the electromagnetic observations by BATSE~\cite{Meszaros:2006rc}. Long-duration GRBs are thought to originate from the death of massive stars. They are usually classified as low and high-luminosity GRBs according to their isotropic luminosity.

High-luminosity GRBs are routinely observed by Swift and the {\em Fermi} Gamma-ray Burst Monitor.
They are characterized by a Lorentz boost factor of $\Gamma \simeq 500$ and isotropic luminosity of about $10^{52}$~erg/s. We know less about low-luminosity GRBs mostly because they are dimmer, with a typical isotropic luminosity of about $10^{48}$~erg/s, and therefore are more difficult to observe. Low-luminosity GRBs have a Lorentz factor one order of magnitude less than high-luminosity ones.

GRBs produce high-energy neutrinos mostly through $p\gamma$ interactions~\cite{Waxman:1997ti,Guetta:2003wi,Waxman:2003vh,Meszaros:2012ye,Dai:2000dj}. The main reactions are
\begin{subequations}
\begin{eqnarray}
p+\gamma \rightarrow \Delta &\rightarrow& n + \pi^+ \ {\rm or } \  p +
\pi^0
\\[1ex]
p+\gamma &\rightarrow& K^+ + \Lambda/\Sigma
 \end{eqnarray}
 \end{subequations}
with the pions, muons, kaons and neutrons decaying to neutrinos of muon and electron flavor~\cite{Guetta:2003wi}, as described earlier.
Usually the injected photon energy distribution  is parametrized through a Band spectrum (broken power law) with a break energy defined as a function of the isotropic energy.
According to the fireball model~\cite{Piran:1999kx}, because the main neutrino production channel is through $p\gamma$ interactions and the proton spectrum is proportional to $E_p^{-2}$ (without breaks), the resultant neutrino spectrum will have a break corresponding to the break energy of the photon spectrum~\cite{Guetta:2003wi}. Above the first break, the neutrino spectrum should be the same as the proton spectrum. However, radiative processes (i.e., radiation losses through synchrotron, inverse Compton, bremsstrahlung, etc.) affect the observable neutrino spectrum and steepen it at higher energies \cite{Hummer:2011ms,Baerwald:2011ee,Tamborra:2015qza}.

The neutrino events thus far detected by IceCube are not in spatial or temporal correlation with known GRBs \cite{Aartsen:2016qcr,Aartsen:2017wea}; the neutrino non-observation from these sources places upper bounds on the neutrino emission that remains consistent with theoretical models. High-luminosity GRBs are also excluded as main sources of the diffuse high-energy neutrino flux
observed by IceCube \cite{Meszaros:2015krr}. However, low-luminosity or choked GRBs could produce high-energy neutrinos abundantly and partially explain the IceCube flux~\cite{Tamborra:2015qza,Senno:2015tsn,Murase:2008mr,Murase:2006mm}. A choked jet is a jet successful in accelerating particles, but such that the electromagnetic radiation
cannot escape \cite{Ando:2005xi,Razzaque:2004yv,Murase:2013ffa}. Choked jets have been invoked to explain the neutrino data in the low-energy tail of the spectrum in the
10--100~TeV range~\cite{Murase:2013ffa,Murase:2015xka,Senno:2015tsn}, although details of the modeling of the neutrino emission might produce results in tension with current
data~\cite{Denton:2018tdj}.

\begin{figure}
\includegraphics[width=0.98\columnwidth]{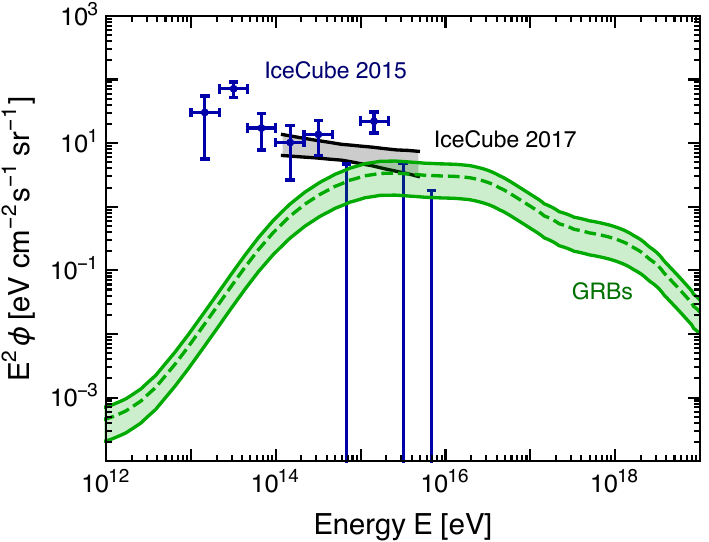}
\caption{Diffuse neutrino flux per flavor
  $\nu_\alpha+\overline{\nu}_\alpha$ from GRBs \cite{Denton:2017jwk}.
  The green band marked ``GRBs'' tracks the uncertainty
  of the local star-formation rate~\cite{Strolger:2015kra}. Also shown
  is the IceCube neutrino flux from two data sets as in Fig.~\ref{sfg}.}\label{grbs}
\end{figure}

Figure~\ref{grbs} shows the diffuse neutrino emission per flavor ($\nu_\alpha+\overline{\nu}_\alpha$) and mass eigenstates from long-duration GRBs. It has been obtained according to the advanced model presented in Fig.~5 of \textcite{Denton:2017jwk} which includes high-luminosity, low-luminosity, and choked GRBs. The astrophysical uncertainty is based on the error in the measurement of the local star-formation rate~\cite{Strolger:2015kra}.

Short GRBs have typical luminosities similar to the ones of long high-luminosity GRBs but originate from compact binary mergers.  The expected diffuse neutrino background from these sources is much smaller than the one from long-duration GRBs because of the merger distribution on cosmic scales~\cite{Tamborra:2015qza}. However, a sizable neutrino flux could be detected, if e.g.~one invokes a large fraction of magnetars connected to these bursts~\cite{Fang:2017tla} or by exploiting the GRB extended emission that can potentially provide a larger target photon field~\cite{Kimura:2017kan}.

\subsection{Blazars}

AGNs are mainly powered by mass accretion
onto supermassive black holes at the center of their host
galaxies \cite{Padovani:2017zpf}. AGNs are among the most luminous
sources of electromagnetic radiation and have been proposed as powerful high-energy cosmic accelerators~\cite{Murase:2015ndr}.

AGNs can be  divided in radio-quiet and radio-loud objects. The
latter are characterized by an emission from the jet and lobes that
is especially prominent at radio wavelengths, whereas
in radio-quiet objects the continuum emission comes from the core
regions and the jet-related emission is weak. Radio-loud AGNs are
promising cosmic accelerators and powerful neutrino sources.

Blazars are a special kind of loud AGNs with the jet pointing towards us. Blazars  are characterized by extreme variability and strong emission over the entire electromagnetic spectrum. Blazars are divided into BL Lacertae objects (BL-Lacs) and flat spectrum radio quasars (FSRQs). These two categories have different optical spectra, the latter showing strong and broad emission lines and the former characterized by optical spectra with weak emission lines.

In the following, we will focus on neutrino production from blazars as they are  expected to be rich neutrino factories.  However, radio-quiet AGNs may also contribute to the diffuse neutrino background~\cite{Murase:2015ndr}, although the neutrino production is affected by large uncertainties.

The photon spectrum of blazars is characterized by two broad bumps~\cite{Padovani:2017zpf}. The low-energy peak can occur at frequencies in the range 0.01--13~keV while the high-energy peak can be in the 0.4--400~MeV range. The low-energy emission of blazars comes from electron synchrotron radiation with the peak frequency being related to the maximum energy at which electrons can be accelerated. On the other hand, the origin of the high-energy emission is still under debate, it might originate from inverse Compton radiation or from the decay of pions generated by accelerated protons.

The electromagnetic spectrum evolves with the blazar luminosity, the so-called blazar sequence. The correspondent neutrino production occurs through  ${p\gamma}$ interactions \cite{Atoyan:2001ey,Protheroe:1996uu}; in fact high-energy protons are accelerated through diffusive shock acceleration or stochastic acceleration in the jet. Protons then interact with synchrotron photons coming from non-thermal electrons that are co-accelerated in the jets. Given their abundance and brightness, the detection of neutrinos from  stacked searches of blazars is statistically favored~\cite{Feyereisen:2016fzb,Mertsch:2016hcd,Murase:2016gly}.

BL-Lacs produce up to 40--70$\%$ of the gamma-ray diffuse background in the 0.1--10 GeV range. Assuming that neutrinos are produced  through $p\gamma$ interactions, the gamma-ray and neutrino luminosity from blazars may be connected through an efficiency factor $Y_{\nu\gamma}$ varying between 0.1 and 2, so that $L_\nu=Y_{\nu\gamma}L_\gamma$~\cite{Petropoulou:2015upa}.

To estimate the neutrino production from the blazar population, it is useful to rely on the blazar sequence. Although one can assume a distribution in the Lorentz factor of the
jet, $\Gamma=10$ is here assumed as a representative value during a typical variability time of $10^6$~s. Cosmic rays undergo Fermi acceleration and acquire a power-law energy distribution
$F_p(E_p)= E_p^{-2} \exp{(-E_p/E_{\mathrm{max}})}$ with $E_{\mathrm{max}}$ the maximum energy that cosmic rays have in the source. In $p\gamma$ interactions, neutrinos carry about $5\%$ of the energy of the primary proton.

The target photon field is determined according to the blazar sequence~\cite{Ghisellini:2017ico}. Beyond synchrotron and inverse Compton peaks present in the BL-Lac spectral energy distribution, FSRQs typically exhibit broad lines from atomic emission of the gas surrounding the accretion disk. By deriving the neutrino spectral energy distribution from the
gamma-ray one and by relying on the blazar distribution at
cosmological distances as from {\em Fermi}~\cite{Ajello:2013lka,Ajello:2011zi},  \textcite{Palladino:2018lov} estimated the neutrino diffuse emission from blazars by imposing bounds on the non-observation of neutrino events from dedicated stacking searches and by assuming that the baryonic loading varies with the luminosity as a power law. The neutrino production from FSRQs is estimated to be about 30$\%$ of the BL-Lac one~\cite{Padovani:2015mba}.

\begin{figure}
\includegraphics[width=1.0\columnwidth]{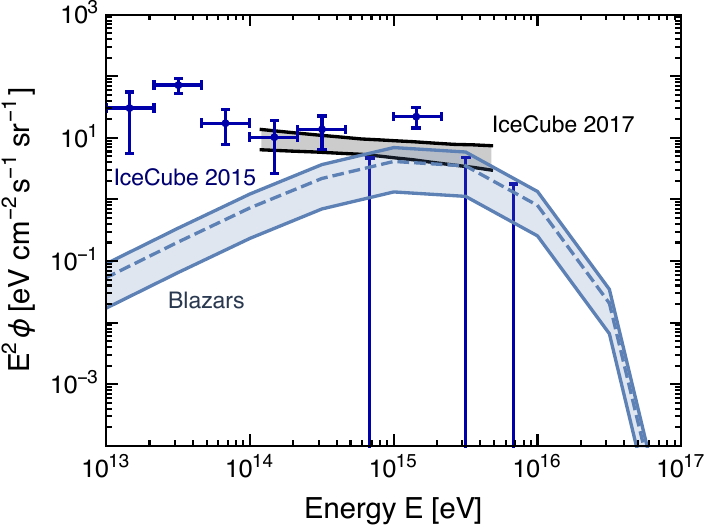}
\caption{Diffuse neutrino background per flavor
  $\nu_\alpha+\overline{\nu}_\alpha$ from blazars~\cite{Palladino:2018lov}. The
  blue band marked ``Blazars'' reproduces the possible variations ($1\sigma$) due to the
  uncertainties on the modeling of the neutrino emission.  Also shown
  is the IceCube neutrino flux from two data sets as in Fig.~\ref{sfg}. Thanks to A.~Palladino for providing the data used in this figure.}\label{blazars}
\end{figure}

Figure~\ref{blazars} shows the neutrino emission per flavor from
blazars. It was obtained from Scenario~3 of \textcite{Palladino:2018lov}; the $1\sigma$ uncertainty band includes all uncertainties due to the modeling of neutrino emission. Given the large uncertainties on the modeling of the diffuse neutrino emission from blazars, we refer the interested reader to \textcite{Murase:2014foa}, \textcite{Padovani:2015mba}, \textcite{Aartsen:2016lir}, \textcite{Murase:2015ndr}, and \textcite{Becker:2005ya} for examples of alternative independent estimations of the  neutrino emission.

In this Section, we focused on the diffuse neutrino emission from AGNs. However, IceCube recently reported hints for the detection of a dozen neutrino events along the direction of the blazar TXS 0506$+$056~\cite{IceCube:2018cha,IceCube:2018dnn}. The interpretation of the neutrino energy distribution is currently in tension with the correspondent electromagnetic observations; however, if confirmed, this would correspond to the first detection of high-energy neutrinos from a cosmic source.

\subsection{Cosmogenic neutrinos}

Ultra-high-energy cosmic rays (UHECRs) have energies up to $10^{20}$~eV; these are the particles with the highest energies observed in Nature~\cite{Anchordoqui:2018qom}. The sources producing them and the mechanisms behind their acceleration are unknown. Results from the Auger observatory suggest a light composition at 1~EeV which tends to become heavier with increasing energy~\cite{Aab:2014aea}.  Telescope Array (TA) data seem to confirm this trend,  suggesting a mixed composition~\cite{Abbasi:2018wlq}.

On their way to Earth, UHECRs interact with radiation, specifically with the cosmic microwave background (CMB) and the extragalactic background light (EBL), which is the cosmic population of photons, e.g.~in the infrared range. The energy spectrum of nucleons is mostly affected by the CMB because of pair production and photo-pion production, whereas the energy spectrum of heavier nuclei is affected by the EBL through pair production and photo-disintegration.
Photo-pion interactions occur when nucleons ($N$) with Lorentz factor $\Gamma \ge 10^{10}$ interact with the CMB and pions are produced ($N+\gamma \rightarrow N + \pi^{0,\pm}$). For lower $\Gamma$, the same process can take place with the EBL.
The strong flux suppression at high energies coming from the photo-pion production is responsible for the so-called Greisen-Zatsepin-Kuzmin (GZK) cutoff~\cite{1966PhRvL..16..748G,1966JETPL...4...78Z,2008PhRvL.100j1101A}.
Photodisintegration takes place when UHE nuclei are stripped by one or more nucleons by interacting with the CMB or EBL,
\begin{equation}
(A,Z)+\gamma \rightarrow (A-n, Z-n^\prime)+nN\ ,
\end{equation}
where $n$ and $n^\prime$ is the number of stripped nucleons and protons, respectively.
Mesons produced in these interactions quickly decay and produce a flux of cosmogenic neutrinos \cite{Beresinsky:1969qj,Heinze:2015hhp,Aloisio:2015ega,Ahlers:2012rz,Kotera:2010yn,Allard:2006mv}. The $\beta$-decay of nucleons and nuclei from photo-disintegration can also lead to neutrino
production. However, while neutrinos produced from pion decay have energies that are a few percent of the parent nucleus, those produced from $\beta$-decay carry less than one part per thousand of the parent nucleon's energy.

The cosmogenic neutrino spectrum is also sensitive to the maximum
UHECR energy and heavy composition at the source (or a weaker
evolution of cosmic-ray sources) tends to produce a significantly
lower cosmogenic neutrino flux~\cite{Aloisio:2015ega}. The largest
contribution is instead obtained if one assumes a proton source; this
is, however, currently disfavored by {\em Fermi\/} data~\cite{AlvesBatista:2018zui}.

Interestingly, while the cosmic-ray spectrum is dominated by nearby sources, the neutrino flux will receive contributions up to cosmological  scales. Moreover, the cosmogenic neutrino flux will also change according to the assumed source composition~\cite{Aloisio:2015ega,Ahlers:2012rz}.

\begin{figure}
\includegraphics[width=1.0\columnwidth]{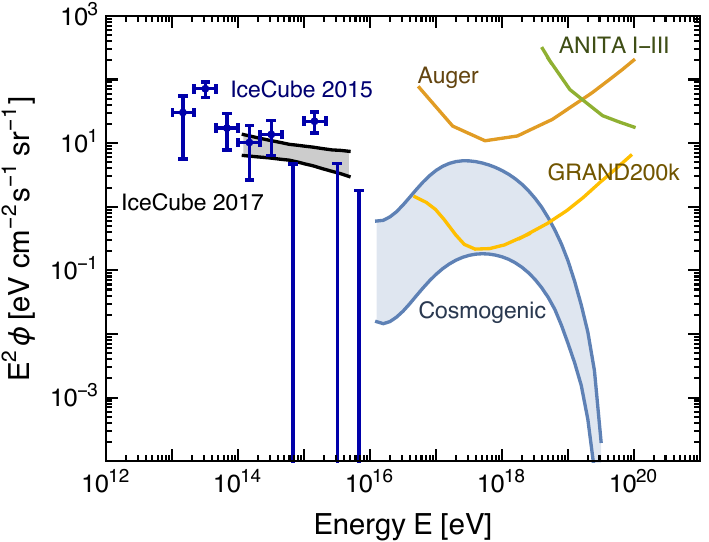}
\caption{Cosmogenic neutrino flux per flavor
  $\nu_\alpha+\overline{\nu}_\alpha$ \cite{Moller:2018isk}. The bands
  reproduce the largest possible variations due to the uncertainties
  on the ultra-high-energy cosmic ray composition and source redshift
  evolution. The exclusion measurements of Auger,
  ANITA phase I-III, and projected 3-year sensitivity for  GRAND (200,000-antenna
  array) are marked accordingly and shown respectively in  orange, green, and
  yellow~\cite{Ackermann:2019ows,Alvarez-Muniz:2018bhp}.  Also shown
  is the IceCube neutrino flux from two data sets as in Fig.~\ref{sfg}.}\label{cosmogenic}
\end{figure}

Cosmogenic neutrinos have not been detected yet and IceCube has recently
placed a new upper limit on this flux \cite{Aartsen:2016ngq}. This
non-observation disfavors sources with a cosmological evolution that
is stronger than the one predicted from the star-formation rate, such
as AGNs~\cite{Aartsen:2016ngq} if one assumes a proton composition at
the source. We show the predicted flux in Fig.~\ref{cosmogenic}, where
we reproduce the results reported in \textcite{Moller:2018isk}.
Note that a lower cosmogenic neutrino flux may also be
expected for mixed composition, see e.g.\ \textcite{AlvesBatista:2018zui} and \textcite{Kotera:2010yn}.  The
upper bounds obtained by ANITA and Auger are respectively shown in
green and orange~\cite{Ackermann:2019ows}. The next generation of
radio facilities, such as the Giant Radio Array for Neutrino Detection
(GRAND, see yellow curve in Fig.~\ref{cosmogenic} for its projected
sensitivity), and the Antarctic Ross Ice-Shelf ANtenna Neutrino Array
(ARIANNA), will also be able to detect this flux, which contributes to the
highest-energy range in the GUNS.

\subsection{Future detection prospects}

IceCube currently remains the only experiment that
detects high-energy neutrinos from astrophysical sources.
Considering further experimental efforts in the field of high-energy neutrino astronomy, IceCube-Gen2~\cite{Aartsen:2014njl} is currently under planning. Another upcoming detector is KM3NeT~\cite{Adrian-Martinez:2016fdl} which will have better sensitivity to Galactic  sources.

For neutrino energies above the PeV range, GRAND \cite{Fang:2017mhl} is currently being designed and developed. ARIANNA~\cite{Barwick:2014pca}, a hexagonal radio array, has already delivered first constraints on cosmogenic neutrinos. ARA, the Askaryan Radio Array, is currently being developed~\cite{Allison:2011wk}.  POEMMA~\cite{Olinto:2017xbi}  is being designed for the detection of cosmogenic tau neutrinos.

\section{Discussion and Outlook}
\label{sec:Conclusions}

In analogy to the seminal {\it Grand Unified Photon Spectrum} of \textcite{Ressell:1989rz}, we have presented the {\it Grand Unified Neutrino Spectrum} (GUNS). This is a complete overview of the
diffuse neutrino and antineutrino backgrounds at Earth ranging from the cosmic neutrino background
in the meV range to cosmogenic neutrinos reaching up to $10^{18}$~eV.

The lowest-energy neutrinos are those from the cosmic
neutrino background (CNB) and the $\bar\nu_e$'s from neutron and triton decay left over from big-bang nucleosynthesis. While these fluxes have not yet been detected, eventually the CNB may become experimentally accessible, depending on the actual neutrino mass spectrum.
The lowest-energy neutrinos on our plot that have ever been
observed are solar pp neutrinos down to 100~keV.

Neutrinos from nuclear reactions in the Sun and atmospheric neutrinos are the best measured
and theoretically best understood sources. Most importantly,
they have played a crucial role in detecting and
exploring neutrino flavor conversion.
Neutrinos from these sources continue to contribute to global fits of neutrino
mixing parameters and in the search for possible non-standard effects in the neutrino sector.

In the few-MeV range, ``antineutrino astronomy'' is a field
encompassing geological $\bar\nu_e$ sources as well as nuclear power
reactors. Geoneutrinos have been observed for more than a decade, but
thus far with somewhat limited statistics. It will take
larger detectors to begin neutrino geology in earnest. In the long run,
geoneutrino research will have practical
implications, e.g.~it could be employed for verification in the context of
nuclear non-proliferation. Reactor
neutrinos remain crucial sources for investigating neutrino mixing
parameters or to study coherent neutrino scattering.

Geoneutrinos and neutrinos from reactors are relevant in the same energy range as the diffuse supernova neutrino background (DSNB). Likely the latter will be measured for the first time by the upcoming gadolinium-enhanced Super-Kamiokande detector and the JUNO scintillator detector, advancing the frontiers of neutrino astronomy to cosmological distances.

Atmospheric neutrinos partly overlap with the DSNB signal. Atmospheric neutrinos were crucial for the discovery of neutrino flavor conversions and to measure the neutrino mixing parameters. In the next future, a precise determination of the atmospheric neutrino flux will be fundamental for the
detection of the DSNB to extract better constraints on the supernova population,  as well to better discriminate the neutrinos of astrophysical origin in the range $1$--$50$~TeV.

Above some TeV, the neutrino backgrounds are far less explored. The IceCube neutrino telescope detects a flux of astrophysical neutrinos up to a few~PeV, whose sources remain to
be unveiled. At even higher energies, the detection of cosmogenic neutrinos will open a new window on the ultra-high-energy sky. Future experimental progress in this energy range will depend on
increased statistics based on larger detection volumes as well as new detector technologies. An
improved source identification in the context of multi-messenger
studies will also contribute to better explore this part of the GUNS.

The GUNS plot reveals the neutrino potential of charting an extremely wide energy range.
While astrophysical neutrinos have been instrumental for detecting flavor conversion, the
mass and mixing parameters are now becoming a question of precision
determination in laboratory experiments and global fits. The most exciting perspectives
to learn about neutrino properties as well as their sources sit at the low- and high-energy tails
of the GUNS. In particular, the branch of high and ultra-high energy
neutrino astronomy is only in its infancy.
The high-energy tail of the GUNS could unlock the secrets of the most energetic events occurring in our Universe,  shed light on
the origin of cosmic rays, and constrain standard and non-standard neutrino properties.

While we hope that our GUNS plot provides a useful overview of the
global neutrino flux at Earth, we also anticipate that it will continue to
require frequent updating both from new observations and new
theoretical ideas and insights.

\section*{Acknowledgments}

In Munich, we acknowledge support by the Deutsche Forschungsgemeinschaft through Grant No.\ SFB 1258 (Collaborative Research Center ``Neutrinos, Dark Matter, Messengers'') and the
European Union through Grant No.\ H2020-MSCA-ITN-2015/674896
(Innovative Training Network ``Elusives'').  In Copenhagen, this
project was supported by the Villum Foundation (Project No.\ 13164),
the Carlsberg Foundation (Grant No.\ CF18-0183), the Danmarks Frie Forskningsfonds (Grant No.\ 8049-00038B), and the Knud H{\o}jgaard Foundation.

\appendix\section{Units and dimensions}
\label{sec:Units}

We use natural units with $\hbar=c=k_{\rm B}=1$.  The neutrino flux at
Earth is shown integrated over all angles for all types of sources
(point-like sources such as the Sun, diffuse such as geoneutrinos, or
isotropic such as the DSNB), for example in units ${\rm cm}^{-2}~{\rm
  s}^{-1}~{\rm MeV}^{-1}$. On the other hand, in Secs.~\ref{sec:ATM}
  and \ref{sec:HE} we follow the convention usually adopted in
  the neutrino-astronomy
  literature and show the fluxes in units ${\rm cm}^{-2}~{\rm
  s}^{-1}~{\rm GeV}^{-1}~{\rm sr}^{-1}$; i.e.~the fluxes have been obtained by
  integrating over all angles and dividing by $4\pi$.

Multiplying the angle-integrated neutrino flux with the cross
section of a target particle provides a differential detection rate
per target particle $d\dot{n}/dE$. To take advantage of directional
capabilities in some detectors one needs to restore, of course, the
angular characteristic of the neutrino flux.  Dividing our
$4\pi$-integrated flux by the speed of light provides the number
density at Earth per energy interval $dn/dE$, for example in units of
${\rm cm}^{-3}~{\rm MeV}^{-1}$, with the exception of nonrelativistic
CNB neutrinos where the appropriate velocity has to be used.  We
mention these seemingly trivial details because sometimes one finds
incorrect factors $4\pi$ in plots showing fluxes from both diffuse and
point-like sources.

There is no ambiguity about the local number density, which may or may not
have a nontrivial angular distribution. The $4\pi$-integrated flux can
be interpreted as the number of neutrinos passing through a sphere of
$1~{\rm cm}^2$ cross-sectional area per second. Other authors have
used the picture of neutrinos passing from one side through a disk of
$1~{\rm cm}^2$ area per second, which is a factor 1/4 smaller for an
isotropic distribution such as the CNB. Such a definition would be
appropriate, for example, for the detection of CMB photons by a horn
antenna where we could only count the photons passing through the
entrance cross section of the horn.  Likewise, the emitted flux from a
blackbody surface, as expressed by the Stefan-Boltzmann Law, is a
factor 1/4 smaller than the energy density of isotropic blackbody
radiation.

\begin{figure*}
  \includegraphics[width=0.40\textwidth]{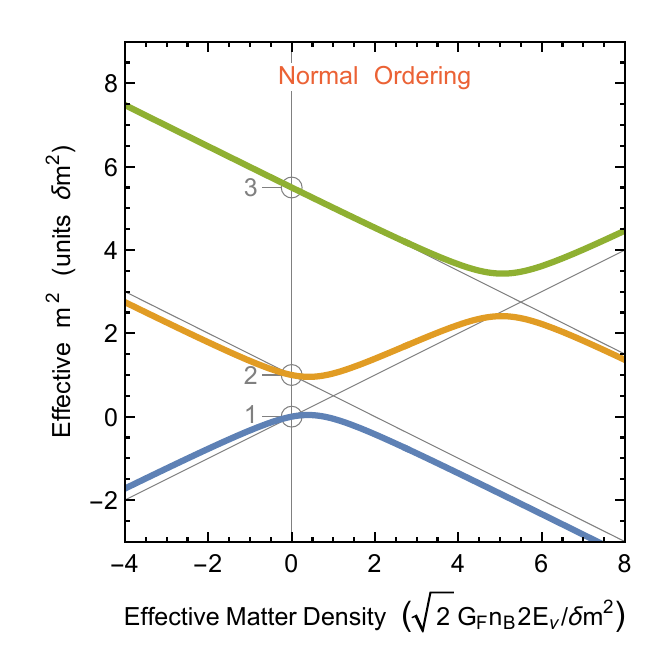}\hskip24pt
  \includegraphics[width=0.40\textwidth]{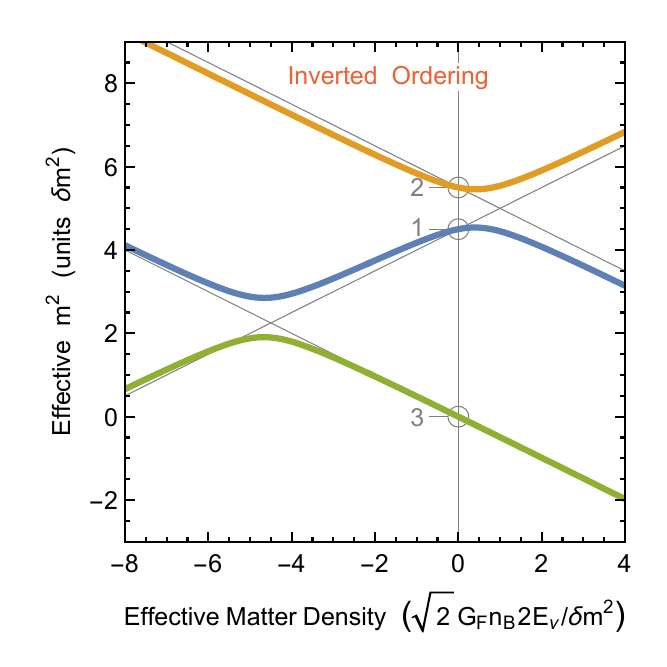}
  \caption{Effective neutrino masses in a medium in units of the solar
  mass difference $\delta m$. For this schematic plot,  $m_1=0$,
  the atmospheric mass difference was chosen
  as $\Delta m^2=5\,\delta m^2$, and the mixing angles as $\sin^2\theta_{12}=0.30$ and
  $\sin^2\theta_{13}=0.01$. The electron and neutron densities were taken to be equal
  ($n_e=n_n=n_B/2$)
  appropriate for a medium consisting of $^4$He or $^{12}$C.
  A~negative density is to be interpreted as a positive density for
  the energy levels of antineutrinos. At zero density the levels are
  the squared vacuum masses. The thin (gray)
  lines are the energy levels for vanishing mixing angles.}\label{fig:masslevels}
\end{figure*}

\section{Neutrino Mass Matrix}
\label{sec:MassMatrix}

Neutrino fluxes from practically any source depend on flavor so that
what arrives at the detector depends on flavor oscillations driven by
neutrino masses and mixing. We restrict ourselves to a minimal
scenario that includes only the three known species.

The weak-interaction neutrino fields $\nu_\alpha$ with $\alpha=e$,
$\mu$ or $\tau$ are given in terms of fields with definite masses
$\nu_i$ by a unitary transformation \smash{$\nu_\alpha=\sum_{i=1}^3
  {\sf U}_{\alpha i}\nu_i$}, implying
\begin{equation}\label{eq:Umixing-states}
  |\nu_\alpha\rangle=\sum_{i=1}^3 {\sf U}^*_{\alpha i}|\nu_i\rangle
  \quad\hbox{and}\quad
  |\overline\nu_\alpha\rangle=\sum_{i=1}^3 {\sf U}_{\alpha i}|\overline\nu_i\rangle
\end{equation}
for neutrino and antineutrino single-particle
states \cite{Giunti:2007ry}. The mixing
matrix is conventionally expressed in terms of three two-flavor mixing
angles $0\leq\theta_{ij}<\pi/2$ and a CP-violating phase
$0\leq\delta<2\pi$ in the form
\begin{eqnarray}\label{eq:Uvac}
  &{\sf U}&=
  \hbox{\footnotesize$\begin{pmatrix}1&0&0\\0&c_{23}&s_{23}\\0&-s_{23}&c_{23}\end{pmatrix}
  \begin{pmatrix}c_{13}&0&s_{13}e^{-i\delta}\\0&1&0\\-s_{13}e^{i\delta}&0&c_{12}\end{pmatrix}
  \begin{pmatrix}c_{12}&s_{12}&0\\-s_{12}&c_{12}&0\\0&0&1\end{pmatrix}$}
\nonumber\\[1.5ex]
&=&\hbox{\footnotesize$
\begin{pmatrix}c_{12}c_{13}&s_{12}c_{13}&s_{13}e^{-i\delta}\\
-s_{12}c_{23}-c_{12}s_{23}s_{13}e^{i\delta}&c_{12}c_{23}-s_{12}s_{23}s_{13}e^{i\delta}&s_{23}c_{13}\\
s_{12}s_{23}-c_{12}c_{23}s_{13}e^{i\delta}&-c_{12}s_{23}-s_{12}c_{23}s_{13}e^{i\delta}&c_{23}c_{13}
\end{pmatrix}$},
\nonumber\\
\end{eqnarray}
where $c_{ij}=\cos\theta_{ij}$ and $s_{ij}=\sin\theta_{ij}$. We have
left out a factor
\smash{${\rm diag}(1,e^{i\alpha_{21}/2},e^{i\alpha_{31}/2})$}
of Majorana phases that are important, for example, in neutrinoless
double-beta decay, but not for flavor oscillations.

\begin{table}[b]
  \caption{Neutrino mixing angles according
  to \textcite{Capozzi:2018ubv}, very similar to those of
  \textcite{deSalas:2017kay} and \textcite{Esteban:2016qun}.}\label{table:mixing-angles}
\begin{tabular*}{\columnwidth}{@{\extracolsep{\fill}}lllll}
\hline
\hline
&\multicolumn{2}{l}{Normal Ordering}&\multicolumn{2}{l}{Inverted Ordering}\\
&Best fit&1$\sigma$ range&Best fit&1$\sigma$ range\\
\hline
$\sin^2\theta_{12}$&0.304&0.291--0.318 &0.303&0.290--0.317\\
$\sin^2\theta_{13}$&0.0214&0.0207--0.0223 &0.0218&0.0211--0.0226\\
$\sin^2\theta_{23}$&0.551&0.481--0.570 &0.557&0.533--0.574\\
$\delta/\pi$&1.32&1.14--1.55&1.52&1.37--1.66\\
\hline
\end{tabular*}
\end{table}

The best-fit mixing angles determined from global fits of all flavor
oscillation data are given in Table~\ref{table:mixing-angles}.
Within uncertainties, the octant of $\theta_{23}$ remains unknown,
i.e., if $\sin^2\theta_{23}$ is larger or smaller than~$1/2$.  CP
violation is favored, but the range of allowed $\delta$
remains large.

With the results of Table~\ref{table:mixing-angles}
for normal mass ordering one finds the matrix of mixing
probabilities, which is the same for $\nu$ and $\overline\nu$,
\begin{equation}\label{eq:probability-matrix}
  \bigl(|{\sf U}_{\alpha i}|^2\bigr)=
  \begin{pmatrix}
    0.681^{+0.013}_{-0.014} & 0.297^{+0.014}_{-0.013} & 0.0214^{+0.0009}_{-0.0007}\\[1ex]
    0.109^{+0.074}_{-0.035} & 0.352^{+0.080}_{-0.065} & 0.539^{+0.019}_{-0.069} \\[1ex]
    0.210^{+0.040}_{-0.073} & 0.351^{+0.067}_{-0.082} & 0.439^{+0.069}_{-0.019}
  \end{pmatrix}.
\end{equation}
The uncertainties correspond to the maximal and minimal values within
the $1\sigma$ ranges shown in
Table~\ref{table:mixing-angles}. Of course, the rows and columns of
this matrix of probabilities always have to add up to~1.
The first row, for example, means that a produced $\nu_e$ has a 68\%
chance to be a $\nu_1$, 30\% to be $\nu_2$, and 2\% to be $\nu_3$.
The mass eigenstates are conventionally numbered such that the
probabilities in the first row appear in declining order, i.e.,
according to the $\nu_i$ admixtures to $\nu_e$.

The matrix ${\sf U}$ being unitary, its inverse ${\sf U}^{-1}$, which
allows us to express mass states in terms of flavor states,
is identical with its conjugate transpose
${\sf U}^\dagger$. Therefore, the probabilities for finding a given mass
eigenstates in any of the flavors correspond to the columns of
Eq.~(\ref{eq:probability-matrix}).  For example, the last column tells
us that a $\nu_3$, for example in the cosmic neutrino background, has
a 2\% chance of being $\nu_e$, a 54\% chance of being $\nu_\mu$, and
44\% of being $\nu_\tau$, and analogous for the other columns.

The numbering convention of mass states leaves open the ordering
of the mass values.  The matter effect on flavor conversion in the Sun
implies $m_1<m_2$. The atmospheric ordering may be normal with
$m_1<m_2<m_3$ or inverted with $m_3<m_1<m_2$.  Global fits somewhat
prefer normal ordering. The probability matrix for inverted ordering
is similar to Eq.~\eqref{eq:probability-matrix} within the shown
uncertainties.

Flavor oscillations of relativistic neutrinos are driven by the
squared-mass differences. We express the mass spectrum in terms of the
parameters \cite{Capozzi:2018ubv}
\begin{subequations}
\begin{eqnarray}
\kern-2em
\delta m^2&=&m_2^2-m_1^2
\kern2.7em{}=73.4~{\rm meV}^2\,,
\\[1ex]
\kern-2em
\Delta m^2&=&m_3^2-\frac{m_1^2+m_2^2}{2}
=\pm2.45\times10^{3}~{\rm meV}^2\,,
\end{eqnarray}
\end{subequations}
where normal ordering corresponds to $\Delta m^2>0$, inverted
ordering to $\Delta m^2<0$.  The nominal $1\sigma$ range of the
measured values is 1.4 and 2.2\%, respectively.  The small mass
splitting $\delta m^2$ is also called the solar mass difference
because it drives solar neutrino conversion, whereas $\Delta m^2$ is
the atmospheric one. Often the atmospheric splitting
is instead identified with
either $m_3^2-m_2^2$ or $m_1^2-m_3^2$, depending on the mass ordering,
which however is a less practical convention.

Direct laboratory limits on the unknown overall mass scale of
approximately 2~eV derive from the electron endpoint spectrum in
tritium $\beta$ decay \cite{Tanabashi:2018}. The KATRIN experiment is
expected to improve the sensitivity to approximately 0.2~eV in the
near future \cite{Arenz:2018kma}.

Cosmological data constrain the fraction of hot dark matter, implying
95\% C.L.\ limits $\sum m_\nu< 0.11$--$0.68~{\rm eV}$,
depending on the used data and cosmological model
\cite{Ade:2015xua, Aghanim:2018eyx, Lesgourgoues:2017}.  Near-future
surveys should be able to set a lower limit, i.e., provide a
neutrino-mass detection \cite{Lesgourgoues:2017}.  Of course, these
results have to be interpreted with the usual caveats about
cosmological assumptions and possible unrecognized systematics.

The neutrino signal from the next nearby supernova can provide a 95\%
C.L.\ time-of-flight limit of 0.14~eV if the emission shows
few-millisecond time variations caused by hydrodynamic instabilities
as suggested by 2D and 3D simulations \cite{Ellis:2012ji}.

Searches for neutrinoless double beta decay are only sensitive to
Majorana masses, and specifically to the combination $\langle
m_\nu\rangle=\bigl|\sum_{i=1}^{3}{\sf U}_{ei}^2 m_i\bigr|$.  Current limits
are on the level of 0.11--0.52~eV, depending on isotope and on
uncertainties of the nuclear matrix elements \cite{Tanabashi:2018}.

\section{Neutrino Mixing in Matter}
\label{sec:MatterMixing}

When they propagate in matter, neutrinos experience a flavor-dependent
potential caused by the electroweak interaction. In a normal medium,
consisting of nuclei and electrons, it is
\begin{equation}\label{eq:weak-potential}
  V_{\rm weak}=\pm\sqrt{2} G_{\rm F} n_B\times
  \begin{cases}
    Y_e-Y_n/2, & \mbox{for $\nu_e$},\\
    -Y_n/2,      & \mbox{for $\nu_{\mu,\tau}$},
  \end{cases}
\end{equation}
where $n_B$ is the baryon density, $Y_e=n_e/n_B$ the net electron fraction per baryon
(electrons minus positrons),
and $Y_n=n_n/n_B$ the neutron fraction. The upper sign is for $\nu$,
the lower sign for $\overline\nu$. Equivalently, we can use a nominally negative
baryon density to denote the $\overline\nu$ potential. Radiative corrections actually provide
a small difference between the $\nu_\mu$ and $\nu_\tau$ potentials
\cite{Botella:1986wy, Mirizzi:2009td}, as does
the possible presence of muons in a supernova core \cite{Bollig:2017lki}.
We also ignore background neutrinos which complicate neutrino
propagation in the form of collective flavor evolution \cite{Duan:2010bg}.

The flavor of a neutrino of fixed energy $E$ evolves
as a function of distance $z$ as
$i\partial_z\Psi=({\sf H}_0+{\sf V})\Psi$, where
$\Psi$ is a three-vector
of flavor amplitudes, whereas antineutrinos evolve as
$i\partial_z\overline\Psi=({\sf H}_0^*-{\sf V})\overline\Psi$.
In the ultrarelativistic limit, the mass contribution
in the flavor basis is
\begin{equation}
  {\sf H}_0=\frac{1}{2E}{\sf U}\begin{pmatrix}
              m_1^2 & 0 & 0 \\
              0 & m_2^2 & 0 \\
              0 & 0 & m_3^2
            \end{pmatrix}{\sf U}^\dagger
\end{equation}
and the matrix of potential energies is
\begin{equation}
 {\sf V}=\sqrt{2}G_{\rm F}\begin{pmatrix}
              n_e-n_n/2 & 0 & 0 \\
              0 & -n_n/2 & 0 \\
              0 & 0 & -n_n/2
            \end{pmatrix}.
\end{equation}
Without flavor mixing, the in-medium dispersion relation in the
relativistic limit is given by the effective masses
$m_{\rm eff}^2=m^2+V_{\rm weak}2E$, shown as thin gray lines in
Fig.~\ref{fig:masslevels} for a schematic choice of mass and mixing
parameters. A nominally negative density is used to show the energy
levels for antineutrinos.  The background medium is taken to have
equal densities of electrons and neutrons as would be the case for
$^4$He or $^{12}$C. For a different composition, the lines acquire a
different slope caused by the common neutral-current potential for all
flavors.

For nonvanishing mixing angles, the effective
masses are obtained by diagonalizing ${\sf H}_0+{\sf V}$,
which is achieved by a unitary matrix ${\sf U}_{\rm M}$ such that
\begin{eqnarray}
{\sf M}_{\rm eff}^2&=&
\begin{pmatrix}
  m_{1,\rm eff}^2 & 0 & 0 \\
  0 & m_{2,\rm eff}^2 & 0 \\
  0 & 0 & m_{3,\rm eff}^2
\end{pmatrix}\nonumber\\[1ex]
&=&{\sf U}^\dagger_{\rm M}\big({\sf U}{\sf M}^2{\sf U}^\dagger
+2E\,{\sf V}\big){\sf U}_{\rm M}.
\end{eqnarray}
For antineutrinos, one substitutes ${\sf V}\to-{\sf V}$ and
$\delta\to-\delta$, the latter equivalent to ${\sf U}\to{\sf
  U}^*$. Notice that $m_{i,\rm eff}^2$ can be negative because it is
just a formal way for writing the in-medium energy levels.  In
Fig.~\ref{fig:masslevels}, the $m_{i,\rm eff}^2$ are shown as thick
colored lines.  Notice that $\theta_{23}$ and $\delta$ do not enter if
the $\nu_\mu$ and $\nu_\tau$ potentials are equal --- otherwise there
will be a third level crossing. Notice also that asymptotically the
colored lines have a nonvanishing offset relative to the gray
lines.\footnote{Similar plots in the context of supernova neutrinos
  \cite{Dighe:1999bi} show in-medium curves asymptotically approaching
  the zero-mixing lines. This behavior is caused by the transition
  from their Eq.~(43) to (44) where one should expand consistently to lowest
  order in all $m^2$.}

Of particular interest is the case of neutrinos produced at high
density in the interior of a star which then propagate all the
way to the surface. If the
propagation is adiabatic (and this is the case for
solar and supernova neutrinos), a state originally in a
propagation eigenstate emerges as such. So we
should decompose the flavor states at the point of production
into local propagation states which then connect to vacuum mass
states at the stellar surface (MSW effect). In sufficiently dense
matter, the propagation eigenstates coincide with interaction
eigenstates. In Fig.~\ref{fig:masslevels} (normal ordering), a $\nu_e$
produced at very high density corresponds to the largest $m_{\rm
  eff}^2$, i.e., the thick green line. Following this line to zero
density (vacuum), we see that a produced $\nu_e$ will emerge as the
mass eigenstate $\nu_3$. Conversely, a $\overline\nu_e$ (large negative
density) is on the blue line and thus emerges as $\nu_1$. A detailed
discussion of all such cases, relevant in the supernova context, was
provided by \textcite{Dighe:1999bi}.

\begin{figure}[!thbp]
  \vskip3pt
  \includegraphics[width=0.88\columnwidth]{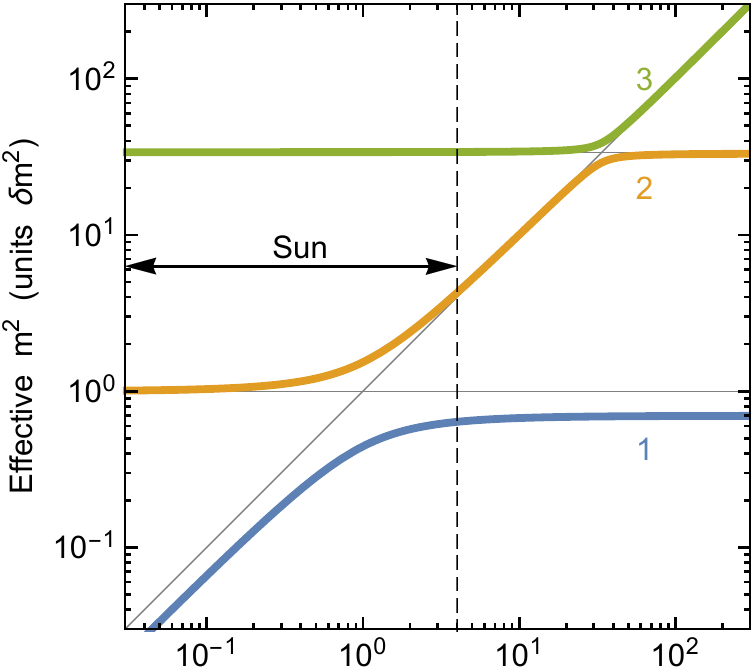}
  \includegraphics[width=0.88\columnwidth]{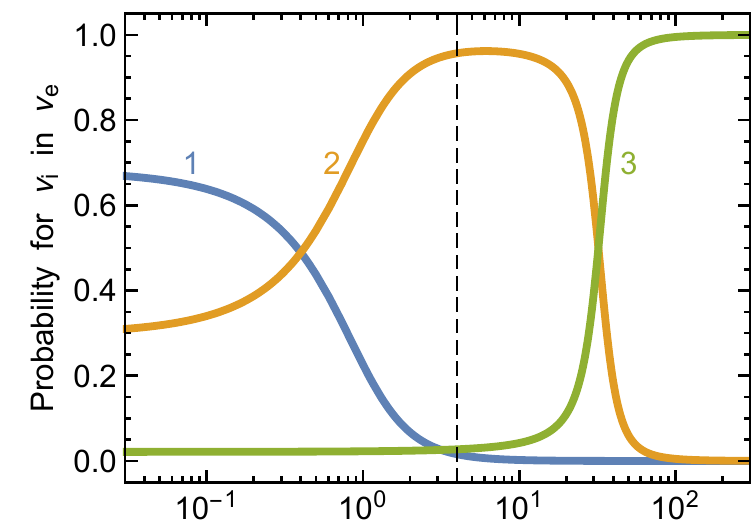}
  \includegraphics[width=0.88\columnwidth]{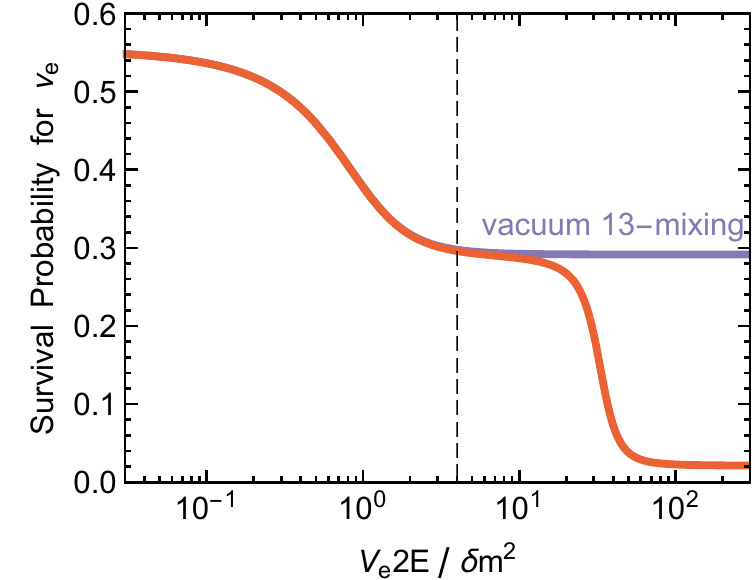}
  \caption{Neutrino mixing in matter.
  {\em Top:\/} Effective masses-squared,
    $m_{i,{\rm eff}}^2$ with $i=1$, 2 or 3, in units of the solar
    value $\delta m^2$. The neutron contribution, which is flavor
    diagonal, has been ignored. The mixing parameters are the best-fit
    values in normal ordering from Table~\ref{table:mixing-angles}. In
    the Sun, the maximum on the horizontal axis is 4.0 (vertical
    dashed line), corresponding to $n_e=6.14\times10^{25}~{\rm cm}^{-3}$ at the solar center
    and the largest $\nu_e$ energy of $E=18.8$~MeV.
    {\em Middle:\/} Probability of a produced $\nu_e$ to be in the
    propagation eigenstates 1, 2 or~3, corresponding to
    $|{\sf U}^{\rm M}_{ei}|^2$. {\em Bottom:\/} Probability of
    a produced $\nu_e$, after adiabatic propagation, to be measured as
    a $\nu_e$ according to Eq.~\eqref{eq:survival-prob}.
    For solar conditions (left of the dashed line), ignoring the
    matter effect on 13--mixing yields an excellent
    approximation.}\label{fig:masslevels-sun}
\end{figure}

Often the flavor-diagonal contribution to ${\sf V}$ provided by
neutrons is not included because it drops out of the oscillation
equation. In this case, and using the best-fit mixing parameters in
normal ordering from Table~\ref{table:mixing-angles}, the same plot of
$m_{i,\rm eff}^2$ is shown in Fig.~\ref{fig:masslevels-sun} (top).  In
the Sun, the central density is $150~{\rm g}~{\rm cm}^{-3}$ with
$Y_e=0.681$, corresponding to $n_e=6.14\times10^{25}~{\rm cm}^{-3}
=4.72\times10^{11}~{\rm eV}^3$ and thus to
$V_e=7.8\times10^{-12}~{\rm eV}$, where
\begin{equation}\label{eq:Ve}
  V_e=\sqrt{2}G_{\rm F} n_e.
\end{equation}
With $E=18.8~{\rm MeV}$, near the highest solar $\nu_e$ energy,
one finds
$V_e 2E<233~{\rm meV}^2=4.0~\delta m^2$, indicated by a
vertical dashed line in Fig.~\ref{fig:masslevels-sun}.

The probability for a $\nu_e$ that was produced in the medium to be
found in any of the propagation eigenstates $i$ is $P_{ei}^{\rm
  M}=|{\sf U}_{ei}^{\rm M}|^2$, shown in Fig.~\ref{fig:masslevels-sun}
(middle) from a numerical solution for ${\sf U}_{\rm M}$ using the
best-fit mixing parameters. At zero density, the $P_{ei}$ correspond
to the top row in the matrix of Eq.~\eqref{eq:probability-matrix}. At
very high density, $\nu_e$ essentially coincides with $\nu_3$, so
after adiabatic propagation it would emerge in the third mass
eigenstate as mentioned earlier.

Neutrinos propagating from a distant source decohere into mass
eigenstates, so for example the $\nu_e$ produced in the Sun arrive
with probabilities $P_{ei}$ in the different mass eigenstates,
depending on the exact point of production and depending on their
energy.  A detector which is only sensitive to $\nu_e$ projects from
each of the $\nu_i$ fluxes the $\nu_e$ component, corresponding to the
probability $|{\sf U}_{ei}|^2$, so the $\nu_e$ survival probability is
\begin{equation}\label{eq:survival-prob}
  P_{ee}=\sum_{i=1}^{3} \bigl|{\sf U}_{ei}^{\rm M}\bigr|^2\,
  \bigl|{\sf U}_{ei}\bigr|^2,
\end{equation}
shown as a red line in Fig.~\ref{fig:masslevels-sun} (bottom). For
neutrinos produced at very low density and/or with very low energies,
this is, using Eq.~\eqref{eq:Uvac}
\begin{equation}
  P_{ee}^{\rm vac}=(c_{12}^4+s_{12}^4)c_{13}^4+s_{13}^4=
0.553,
\end{equation}
where the numerical value is for the best-fit mixing angles in normal
mass ordering.

The mass differences are hierarchical,
$\delta m^2\ll \Delta m^2$, allowing for an approximate determination of
${\sf U}_{\rm M}$ \cite{Denton:2016wmg,Ioannisian:2018qwl}. Writing it in the form
of Eq.~\eqref{eq:Uvac}, one finds for the
in-medium mixing angles $\theta_{23}^{\rm M}=\theta_{23}$,
$\delta^{\rm M}=\delta$, and
\begin{subequations}
\begin{eqnarray}
  2\theta_{12}^{\rm M} &=&
  {\rm ArcTan}\left(\cos2\theta_{12}{-}\epsilon_\odot,
  \cos\theta_{13}'\sin2\theta_{12}\right),\\
  2\theta_{13}^{\rm M} &=&
  {\rm ArcTan}\left(\cos2\theta_{13}{-}\epsilon_{\rm a},\sin2\theta_{13}\right),
\end{eqnarray}
\end{subequations}
where $\theta_{13}'=\theta_{13}^{\rm M}-\theta_{13}$. Here,
$\alpha={\rm ArcTan}(x,y)$ is an angle such that
$\sin\alpha=y/\sqrt{x^2+y^2}$ and $\cos\alpha=x/\sqrt{x^2+y^2}$. Further,
\begin{subequations}
\begin{eqnarray}
  \epsilon_\odot &=&\frac{2E V_e}{\delta m^2}\,
  \left(\cos^2\theta_{13}^{\rm M}+\frac{\sin^2\theta_{13}'}{\epsilon_{\rm a}}\right),
  \\
  \epsilon_{\rm a}&=&\frac{2E V_e}{m_3^2-m_1^2-\delta m^2\sin^2\theta_{12}},
\end{eqnarray}
\end{subequations}
where  $\epsilon_{\rm a}<0$ for inverted mass ordering ($m_3^2<m_1^2$).
The approximate analytic
probabilities $|U^{\rm M}_{ei}|^2$ agree very well with the
numerical values shown in the middle panel of
Fig.~\ref{fig:masslevels-sun}. The agreement is better than $10^{-3}$
except for $|U^{\rm M}_{e1}|^2$ where above 10 on the horizontal axis
the analytic probability falls off faster than the numerical
one. The differences between analytic and numerical solutions
are essentially irrelevant on a level of precision where we have
ignored radiative corrections to the weak potential.

The maximum value of $2EV_e$ in the Sun is small compared with
$m_3^2-m_1^2$, so for solar conditions we may neglect the matter effect
on $\theta_{13}$. In this case ${\sf U}_{\rm M}$ is given in terms of the
vacuum mixing angles except for
\begin{equation}
  2\theta_{12}^{\rm M}=
  {\rm ArcTan}\left(\cos2\theta_{12}{-}\epsilon_\odot,\sin2\theta_{12}\right)
\end{equation}
with
\begin{equation}
  \epsilon_\odot=
  \frac{2E V_e}{\delta m^2}\,\cos^2\theta_{13}\,.
\end{equation}
In this case, the probability for a produced $\nu_e$ to be found in any of
the three propagation eigenstates is
\begin{subequations}\label{eq:two-flavor-probabilities}
\begin{eqnarray}
  P_{e1}&=&\cos^2\theta_{13}\,\cos^2\theta_{12}^{\rm M},\\
  P_{e2}&=&\cos^2\theta_{13}\,\sin^2\theta_{12}^{\rm M},\\
  P_{e3}&=&\sin^2\theta_{13}\,.
 \end{eqnarray}
\end{subequations}
The $\nu_e$ survival probability is
\begin{equation}\label{eq:survival-probability}
  P_{ee}=\frac{1+\cos2\theta_{12}\cos2\theta_{12}^{\rm M}}{2}\,
 \cos^4\theta_{13}+\sin^4\theta_{13}
\end{equation}
marked as ``vacuum 13--mixing'' in
Fig.~\ref{fig:masslevels-sun} (bottom).
The best-fit value $\sin^2\theta_{13}=0.0214$ implies that
we can safely neglect $\sin^4\theta_{13}$, whereas
$\cos^4\theta_{13}=0.958$ deviates significantly from~1.
\section{Constructing the GUNS plot}
\label{sec:GUNS}

For those wishing to construct their own version of the GUNS plot of
Fig.~\ref{fig:GUNS0} we here provide the exact input information that
we have used. We also provide numerical tables, enclosed as ancillary
files in the arXiv repository, that can be used for this purpose.

The GUNS plot shows the sum over all flavor or mass eigenstates for neutrinos
(solid lines) and antineutrino (dashed lines). Notice that the
uncertainty bands for atmospheric, IceCube, and cosmogenic neutrinos
are only indicative because in the literature one finds $E^2\phi_\nu$
and its uncertainty, so the uncertainty of $\phi_\nu$ also depends on
the energy uncertainty in a steeply falling spectrum.

The IceCube and cosmogenic fluxes in Fig.~\ref{fig:GUNS0} can be
compared to the corresponding figures in the main text by multiplying
the curves in Fig.~\ref{fig:GUNS0} with ${E^2/4\pi \times 2/3}$. In
Fig.~\ref{fig:GUNS0} we show an angle-integrated flux, hence the
factor $1/4\pi$ to arrive the traditional fluxes per solid angle shown
in high-energy neutrino astronomy. The factor 1/3 translates the sum
over all flavors to a single-flavor flux, assuming flavor
equipartition arriving at Earth. The factor 2 sums over neutrinos plus
antineutrinos.

\subsection{Cosmic neutrino background}

For the CNB we assume the neutrino masses of
Eq.~\eqref{eq:mass-spectrum}. We show blackbody radiation for one mass
eigenstate, following Eq.~\eqref{eq:E-flux} and listed as a numerical
table in the file \texttt{CNB.dat}. In addition there are two line
sources (expressed in units of $\rm cm^{-2}\,s^{-1}$ in the upper panel
and in units of $\rm eV \,cm^{-2}\,s^{-1}$ in the lower panel),
corresponding to the $m_2$ and $m_3$ mass eigenstates. The normalization
is given by the integral of Eq.~\eqref{eq:mass-spectrum}. The line
fluxes can be found in the file \texttt{CNB-lines.dat}.

\subsection{Neutrinos from big-bang nucleosynthesis}

For neutrinos produced by the decay of neutrons and tritium during
big-bang nucleosynthesis we adopt the fluxes found in
\textcite{Ivanchik:2018fxy}. The tables \texttt{BBN-neutron.dat} and
\texttt{BBN-tritium.dat} are courtesy of these authors.

\subsection{Thermal neutrinos from the Sun}

For neutrinos produced by thermal processes in the Sun we use the flux
computed in \textcite{Vitagliano:2017odj}, see the table
\texttt{Sun-thermal.dat}.

\subsection{Solar neutrinos from nuclear reactions}

The solar neutrino flux is equivalent to Fig.~\ref{sunnuclear}. The
flux from the pp chains and CNO cycle is given (except for hep) by the
measurements shown in Table~\ref{table:sun}. For the CNO and hep
fluxes the range is bracketed by the lowest AGSS09 and highest GS98
predictions. The flux is given by the sum of all the processes
contributing to the solar neutrino flux, which can be found in
\texttt{Sun-nuclear-B8.dat}, \texttt{Sun-nuclear-N13.dat}, \texttt{Sun-nuclear-O15.dat}, \texttt{Sun-nuclear-hep.dat}, and \texttt{Sun-nuclear-pp.dat}.
The lines due to electron capture are provided in \texttt{Sun-lines.dat}.

\subsection{Geoneutrinos}

The average geoneutrino flux is the sum over the different processes
shown in Fig.~\ref{geonus}. We have used the publicly available data
of \textcite{Enomoto:2005}. The flux is tabulated in
\texttt{Geoneutrinos.dat}.

\subsection{Reactor neutrinos}

The reactor $\overline{\nu}_e$ flux is tabulated in
\texttt{Reactor.dat} for the example of 140~MW thermal power and a
detector at the close distance of 81~m. We assume that each reaction
release $205 \rm \, MeV$ and the number of neutrinos per reaction per
unit energy is obtained as in Fig.~\ref{fig:reac} (see also main
text). Notice that this is not equivalent to the JOYO detector example
shown in Fig.~\ref{fig:reacflux}. The choice of this reactor-detector
system is arbitrary, and can be rescaled accordingly.
Of course, what to show on the GUNS plot is somewhat arbitrary because
reactor fluxes depend most sensitively on location of all GUNS components.

\subsection{Diffuse supernovae neutrino background}

The DSNB neutrino and antineutrino fluxes can be found in
\texttt{DSNB.dat}. The table has five entries: energy, lower and upper
bound for the neutrino flux, and lower and upper bound for the
antineutrino one. Each flux was calculated as
${\phi_{\nu_e}+2\phi_{\nu_x}}$. The uncertainty band was obtained
considering the simplified scenario discussed in the main text for the
supernova masses (9.6, 27, 40) $M_\odot$ with (50,20,21)\% and
(59,32,9)\% for the supernova models.

\subsection{Atmospheric neutrinos}

Atmospheric neutrino and antineutrino fluxes can be found in
\texttt{Atmospheric.dat}, in each case the sum of the electron and
muon flavored fluxes.  The low-energy points ($\lesssim 100\,
\text{MeV}$) are from the tables of \textcite{Battistoni:2005pd},
while the energy range ($100 \, \text{MeV}\lesssim E \lesssim 1\rm\,
TeV$) is from the publicly available results of
\textcite{Honda:2015fha}. We show the flux for average
solar activity at the Kamioka site. The high-energy ($E \gtrsim 1\rm\,
TeV$) flux is taken from \textcite{Richard:2015aua} and \textcite{Aartsen:2014qna}.

\subsection{IceCube neutrinos}

The high-energy astrophysical neutrino flux (in units $10^{-20}\,\rm
  eV^{-1}\,cm^{-2}\,s^{-1}$), as measured by IceCube, is tabulated in
\texttt{IceCube.dat} with an upper and lower bound of the expected
cosmogenic flux.  These data are taken from the high-energy events
analysis of \textcite{Aartsen:2017mau}.

\subsection{Cosmogenic neutrinos}

Cosmogenic neutrinos are tabulated in three columns in
\texttt{Cosmogenic.dat}, i.e., energy and fluxes (in units
$10^{-30}\,\rm eV^{-1}\,cm^{-2}\,s^{-1}$) for two models of the
primary cosmic-ray composition, taken to represent an upper and a
lower bound. The data used in the figure have been estimated in
\textcite{Moller:2018isk}.

\subsection{Overall GUNS plot}

As an example for constructing the GUNS plot from these data files we
also include a {\sc Mathematica} notebook named
\texttt{Produce-your-GUNS.nb} that can be used to create other
variations of this plot.

\raggedright

\bibliography{Bibliography}

\end{document}